\documentclass[useAMS,usenatbib]{mn2e}
\usepackage{epsfig}
\usepackage{lscape}
\usepackage{graphicx}
\usepackage{rotating}
\usepackage{times}

\def\cm3{cm$^{-3}$}
\def\kms{km~s$^{-1}$}

\def\lsun{L$_{\odot}$}
\def\rsun{R$_{\odot}$}

\def\msun{M$_{\odot}$}

\def\beq{\begin{equation}}
\def\eeq{\end{equation}}

\def\lesssim{\mathrel{\hbox{\rlap{\hbox{\lower4pt\hbox{$\sim$}}}\hbox{$<$}}}}
\def\gtrsim{\mathrel{\hbox{\rlap{\hbox{\lower4pt\hbox{$\sim$}}}\hbox{$>$}}}}

\def\aj{AJ}

\def\apj{ApJ}
\def\apjs{ApJS}
\def\apjl{ApJL}
\def\aap{A\&A}
\def\araa{ARA\&A}
\def\aaps{A\&AS}
\def\mnras{MNRAS}
\def\nat{Nature}
\def\iaucirc{IAU~Circ.}
\def\solphys{Sol.~Phys.}
\title[Non-LTE time-dependent radiative transfer]{Supernova radiative-transfer
modeling: A new approach using non-LTE and full time dependence}

\author[Luc Dessart and D. John Hillier]
{
Luc Dessart$^{1}$\thanks{E-mail: Luc.Dessart@oamp.fr},
D. John Hillier$^{2}$\\
$^{1}$ Laboratoire d'Astrophysique de Marseille, Universit\'e de Provence,
CNRS, 38 rue Fr\'ed\'eric Joliot-Curie, F-13388 Marseille Cedex 13, France \\
$^{2}$ Department of Physics and Astronomy, University of Pittsburgh, USA
}

\begin{document}

\date{Accepted 2010 March 2.  Received 2010 February 22; in original form 2009 December 8}

\pagerange{\pageref{firstpage}--\pageref{lastpage}} \pubyear{2009}

\maketitle

\label{firstpage}

\begin{abstract}
We discuss a new one-dimensional non-LTE time-dependent radiative-transfer
technique for the simulation of supernova (SN) spectra and light curves.
Starting from a hydrodynamical input characterizing the homologously-expanding ejecta
at a chosen post-explosion time, we model the evolution
of the {\it entire} ejecta, including gas and radiation. The boundary constraints
for this time-, frequency-, space-, and angle-dependent problem are the adopted
initial ejecta, a zero-flux inner boundary and a free-streaming outer boundary.
This relaxes the often unsuitable assumption of a diffusive inner boundary, but will also
allow for a smooth transition from photospheric to nebular conditions.
Non-LTE, which holds in all regions at and above the photosphere, is
accounted for. The effects of line blanketing on the radiation field are explicitly included,
using complex model atoms and solving for all ion level populations appearing
in the statistical-equilibrium equations.
Here, we present results for SN1987A, evolving the model ``lm18a7Ad'' of Woosley
from 0.27 to 20.8\,d. The fastest evolution occurs prior to day 1, with a spectral
energy distribution peaking in the range $\sim$300-2000\AA, subject to line
blanketing from highly ionized metal and CNO species. After day 1, our synthetic
multi-band light curve and spectra reproduce the observations to within 10-20\%
in flux in the optical, with a greater mismatch for the faint UV flux.
We do not encounter any of the former discrepancies associated with the He\,{\sc i}
and H\,{\sc i} lines in the optical, which can be fitted well with a standard Blue-supergiant-star
surface composition and no contribution from radioactive decay.
The effects of time dependence on the ionization structure, discussed in Dessart \& Hillier,
are recovered, and thus nicely integrated in this new scheme.
Despite the 1D nature of our approach, its high physical consistency
and accuracy will allow reliable inferences to be made
on explosion properties and pre-SN star evolution.
\end{abstract}

\begin{keywords} radiative transfer -- stars: atmospheres -- stars:
supernovae - stars: supernovae: individual: SN 1987A
\end{keywords}

\section{Introduction}

 Core-collapse supernovae (SNe) are extraordinary events situated at the crossroads
 of many fields of astrophysics. They mark the birth of compact objects, either neutron
 stars or stellar-mass black holes which, owing to their high compactness, are often the site of tremendous
 magnetic fields, as in magnetars, or the site of tremendous rotation rates, as in millisecond-period pulsars.
 Their ejecta make a significant contribution to the chemical enrichment of galaxies, while affecting
 their dynamics and energetics. The connection to $\gamma$-ray bursts, for a subset of these SNe,
 may turn them into excellent probes of the early Universe.
 Our focus, however, is to characterize the SN ejecta itself
 and extract information that can help us understand the  properties of the explosion and
of the progenitor star. Such inferences are based on the analysis
 of the SN light, using photometric, spectroscopic, or spectropolarimetric data. Hence, developing
 accurate radiative-transfer tools capturing the key physics controlling the interaction of light and matter
is of prime importance.

 Of all SNe, only core-collapse, and in particular Type II SNe, have well-identified
 progenitors.  Numerous Type II-Plateau (II-P) SNe have now been associated with
 the explosion of low-mass Red-Supergiant (RSG) stars \citep{smartt_09}, while
 the progenitor of the Type II-peculiar SN1987A was a Blue-Supergiant (BSG) star, named Sk -69 202
 (for a review, see,  e.g., \citealt{ABK89_rev}).
 The excellent quality of the observational data for SN1987A makes it ideal
 for detailed modeling. Numerous radiative-transfer studies of the early-time spectra of SN1987A have been
 done, with some success \citep{EK89_87A,hoeflich_87,lucy_87,hoeflich_88, lucy_87,SAR90_87A,mazzali_etal_92}.
However, these studies are now twenty years old.
They generally assumed Local Thermodynamic Equilibrium (LTE; at best treating only a fraction of the species/levels in non-LTE)
and/or steady-state for the radiative transfer, and employed small model atoms.
\citet{lucy_87} and \citet{SAR90_87A} have emphasized the effects of line blanketing in the UV.
One explicit question raised by these studies was the problematic
observation of He\,{\sc i} lines in optical spectra, which required unacceptable helium enrichments.
More recently, \citet{MBB01_mixing,MBB02_87A} argued for significant mixing of $^{56}$Ni beyond 5000\,\kms\
in order to reproduce Balmer line profiles during the first weeks after explosion.

The Balmer line strength problem has now been associated with the erroneous neglect of important
time-dependent terms that appear in the energy and statistical-equilibrium equations
(\citealt{UC05_time_dep}; \citealt{DH08_time}), while reproducing the characteristics of
He\,{\sc i} lines seems to require a non-LTE treatment.
Hence, some of these discrepancies may simply reflect the shortcomings of the radiative-transfer tools employed
rather than a genuine peculiarity of the progenitor. Taking a new look at this dataset therefore seems warranted.
Furthermore, studying SN1987A is a good exercise to gauge the level of accuracy of radiative-transfer codes
and assumptions, facilitated by the considerable advances in computer technology over the last twenty years.
Because atomic data represent a fundamental and essential ingredient of radiative-transfer computations,
the considerable improvements in that domain over that period make such calculations more accurate
than calculations done when SN1987A went off.

  In this paper, we present a new approach for non-LTE time-dependent radiative transfer
  modeling of SN ejecta using the code {\sc cmfgen} \citep{HM98_lb,DH05_qs_SN, DH08_time}.
  We discuss the conceptual aspects of the method and emphasize the key equations
  that are solved, delaying a more comprehensive presentation of the technical details to a forthcoming paper
  (Hillier \& Dessart 2010, in preparation).
  We illustrate this new capability with results obtained for SN1987A, starting our
  time evolution from an hydrodynamical input of an exploded BSG star (Woosley, priv. comm.;
  model ``lm18a7Ad").
  We do not present an in-depth study of SN1987A, but merely use this well-observed
  SN to check and confront our model results.
  A preliminary set of results were presented in \citet{DH09_boulder}.
  In the following section, we discuss the various approaches we have used in the recent years
  for SN spectroscopic modeling, emphasizing their merits and limitations.
  We then describe in \S\ref{sect_setup} the hydrodynamical model we employ
  as a basis for this calculation, summarizing its various
  properties, before presenting our setup for the radiative-transfer calculation.
  In \S\ref{sect_results}, we describe the ejecta evolution, such as temperature and ionization structure,
  and present checks on our numerical technique.
  We then discuss in \S\ref{sect_synthetic_rad} the radiative properties of our
  non-LTE time-dependent models, covering in turn synthetic spectra and the bolometric light curve,
  and detailing in particular the sources of line blanketing at various epochs.
  In \S\ref{sect_comp_to_obs}, we compare our theoretical predictions to UV and optical observations
  of SN1987A.
  Finally we present our conclusions in \S\ref{sect_ccl}, and lay out the various projects ahead.

\section{The Radiative Transfer Problem and the approaches to its solution}
\label{sect_rad_trans}

Radiative-transfer modeling of SN ejecta can be performed in a variety of ways and with different
accuracies and consistencies: the interaction between the gas and the radiation
may be done in LTE or in non-LTE; the problem may be simplified to its steady-state form
($c \rightarrow \infty$ and $\partial/\partial t =$0); the relative importance
of absorption versus scattering opacity may be imposed or calculated.
The latter is controlled by transition rates, level
populations, and the ionization state, which are all connected to the radiation field and thus the whole
problem is tightly coupled.

   Our fundamental aim is to model the time evolution of the gas and radiation field from near the onset of
the explosion (i.e., shock breakout in core-collapse SNe) when the ejecta are hot, dense, and possess
a large optical depth, until the nebular phase when expansion and cooling have made the ejecta
cool, tenuous, and optically thin at all wavelengths. To do so, we need to track the evolution of the following:

1) The specific intensity $I$ as a function of radius, time, frequency, and angle, i.e., $I \equiv I(r,t,\nu,\mu)$, as well as its moments $J$, $H$, and $K$.

 2) The properties of the gas, such as electron density $n_{\rm e}$
and ion density $n_{\rm i}$, temperature, energy. This requires accounting for changes in excitation and ionization, as well
as for contributions from radioactive decay of unstable isotopes.

3)  The coupling between the radiation and the gas, i.e., opacity and emissivity. This represents a considerable task
in SN ejecta because of departures from LTE conditions and the importance of scattering.

4) The changes in composition of unstable isotopes.

\noindent
Finally, all this work has to be done using a reliable and comprehensive atomic dataset
for a wide range of species and ions.

  Schematically, assuming homologous expansion and retaining only the terms of order $v/c$,
the combined set of equations we wish to solve are:

\begin{itemize}

  \item The statistical equilibrium equations (one for each level $i$ of each ion treated):

    \begin{equation}
      \rho {D n_i/\rho \over Dt} = {1 \over r^3} {D(r^3 n_i)  \over Dt} =  \sum_j \left(n_j R_{ji}  - n_i R_{ij}\right)\,;
    \end{equation}

  \item The energy equation:
    \begin{equation}
    \rho {De \over Dt} - {P \over \rho} {D\rho \over Dt}= 4\pi \int d\nu (\chi_\nu J_\nu  - \eta_\nu) + {De_{\rm decay} \over Dt} \,;
     \label{eq_energy}
     \end{equation}

\item The time dependent zeroth and first moment of the radiative transfer equation:

\begin{equation}
  {1 \over cr^3}  {D(r^3 J{_\nu})  \over Dt} + {1 \over r^2} {\partial (r^2 H_\nu)  \over \partial r}
  - {\nu V \over rc} { \partial J_\nu \over \partial \nu } = \eta_\nu - \chi_\nu J_\nu
 \label{eq_zero_mom}
 \end{equation}

\noindent
and

\begin{eqnarray}
  {1 \over cr^3}  {D(r^3 H_\nu)  \over Dt} + {1 \over r^2} { \partial(r^2 K_\nu)  \over \partial r}
  &+& {K_\nu - J_\nu \over r} \\
  &-& {\nu V \over rc}{ \partial H_\nu \over \partial \nu } = - \chi_\nu H_\nu \,\,.
 \end{eqnarray}

\noindent
In the above equations $D/Dt$ is the Lagrangian derivative, $n_i$ the number density of state $i$, $R_{ij}$ is a term that represents the rates connecting the level $i$ to the level $j$, $e$ is the internal energy per unit mass, $\nu$ the frequency, $\chi_\nu$
the opacity, $\eta_\nu$ the emissivity, $J_\nu$ the mean intensity, and
$e_{\rm decay}$ is the specific energy associated with the decay of unstable nuclei.
The internal energy can be written in the form

      \begin{equation}
      e = e_K +e_I\,,
      \end{equation}

\noindent where

      \begin{equation}
       e_K={3 kT(n+n_e) \over 2 \mu m n }\,,
      \end{equation}

      \begin{equation}
      e_I=  \sum_i {n_i E_i \over \mu m n}\,.
      \end{equation}

\noindent
$n$ is the total particle density (excluding electrons), $m$ is the atomic
mass unit, $\mu$ is the mean atomic weight, and $E_i$ is the total energy (excitation and ionization)
of  state $i$.

\end{itemize}

   In a steady-state approach, we drop all terms involving partial-derivatives with respect to time,
in the statistical-equilibrium/energy-equations and/or in the moments of the transfer equation.
By neglecting these in both, the approach allows for departures from LTE and assumes steady-state (\S\ref{sect_noddt}).
Retaining them in the first but neglecting them in the second, one grasps the time-dependent effects
important for the ionization structure, while the radiation field is still assumed steady-state (\S\ref{sect_ddt}).
By retaining these terms in all equations above, one accounts for the full time dependence of the problem
and thereby reaches a much higher level of physical consistency (\S\ref{sect_djdt}).
These various approaches correspond chronologically to what we have followed so far. We now present them
in more details.

   \subsection{Non-LTE and steady-state approach}
\label{sect_noddt}

    Because Type II-Plateau (II-P) SNe, associated with the explosion of RSGs, are
characterized by a massive quasi-homogeneous hydrogen-rich homologously-expanding ejectum during
their photospheric phase, their photospheric properties are weakly sensitive to the details of the explosion
mechanism or morphology at early times. In particular, the hydrogen-rich region of
Type II-P SN ejecta seems to remain quasi-spherical even when there is strong evidence
for an aspherical explosion, as revealed by the growing linear polarization of continuum
photons at the end of the plateau phase \citep{leonard_etal_06}.
This property gives impetus to model atmosphere calculations in which multi-dimensional effects
are ignored, but which include a much more accurate treatment of non-LTE and line-blanketing effects
(both of which are very computationally costly).
Being free of these various uncertainties, Type II-P SNe are ideal for benchmarking purposes,
and we therefore focused our initial investigations on these objects.

   Our initial perspective was to consider the photospheric layers of the SN ejecta to be just like those of
a stellar atmosphere, characterized here by a homologous and fast expansion, a steep density distribution,
and spherical symmetry. In this approach, we neglected the Lagrangian derivative terms
in the radiative-transfer, statistical-equilibrium,
and energy equations and solved the radiative-transfer problem in 1D
(see \citealt{HM98_lb,DH05_qs_SN}).\footnote{If we neglect only $\partial J/\partial t$ we
get extra terms of order $v/c$ in the transfer equation which is generically referred to
as the relativistic form of the transfer equation. These extra terms are unimportant in O and Wolf-Rayet stars,
but are of crucial importance for the temperature structure in SN, particularly
below the photosphere \citep{MM84_RH}.
Various approximations to the transfer equation have been suggested to avoid the explicit-time dependence
in the transfer equation, but these are ultimately unphysical. \citet{PE00_Ia_anal} concluded
``that a fully time-dependent solution to the transport problem is needed in order to compute
SNe Ia light curves and spectra accurate enough to distinguish subtle differences of various
explosion models'', and we believe the same must be true for Type II SNe.}
Observed spectra are computed
in the observer's frame using the code of \citet{BH05_2D}, with relativistic terms taken into
account but ignoring time dependence (see Hillier \& Dessart 2010 for details).

The radiative-transfer grid is set to
cover the vicinity of the photospheric region, extending inwards to a Rosseland mean optical depth of $\sim$100
and extending outwards to distances where radiation is in a free-streaming regime.
In between the boundaries, we adopt an analytical distribution for the SN ejecta density.
We impose a uniform composition  (i.e., the modeled ejecta layers are chemically homogeneous),
so that changes in ion densities reflect exclusively changes in ionization.
The temperature structure of the ejecta is controlled through the imposed diffusive inner boundary
flux and constrained through fits to the spectral energy distribution (SED) as well as line profile strength
and shape.  An independently inferred distance to the Galaxy host and/or the estimated time of explosion allow one
to approximate the radius at a given epoch, thereby fully constraining the problem.

Equipped with the assumptions outlined above, we analyzed the photospheric phase of SNe 1999em,
2005cs, and 2006bp \citep{DH06_SN1999em,BDH07_SN2005cs, DBB08_SN2005cs}.
We discussed the importance of line blanketing on the SED, the importance of non-LTE effects
for line formation, and the importance of line overlap for proper identifications of spectral features.
We found that at early times, prior to
the ejecta recombining to a neutral state, our approximations produced very satisfactory fits to
observations, with ejecta properties compatible with the RSG nature of the progenitor star.
Based on these quantitative analyzes, we also determined distances to these SNe, obtaining
values in agreement with those obtained with independent methods (e.g., Cepheids).

   The advantage of this method is that it is self-contained, so that one can compute a model without
considering the previous evolution or the ejecta conditions below the inner grid boundary.
But this approach becomes inaccurate when the time-dependent terms appearing in the equations
above become non-negligible, or when the inner boundary flux cannot be accurately described
by a steady-state diffusion solution.

     \subsection{Non-LTE radiative transfer with steady-state radiation-field but time-dependent
terms in the rate and energy equations}
\label{sect_ddt}

    The steady-state assumption is a simplification of the radiative transfer problem, one
that holds safely in the context of hot star atmospheres for which {\sc cmfgen} was originally designed.
However, in SNe, low density and fast expansion conspire to make the expansion and the recombination
timescales eventually comparable. Exacerbated further by optical-depth effects, time dependence can lead to an
ejecta ionization freeze-out, modifying the electron and ion densities, line and continuum
optical depths, and consequently the emergent flux globally.
This was demonstrated by \citet{UC05_time_dep} in the case of SN 1987A,
with primary focus on H$\alpha$ and Ba\,{\sc ii}\,6142\AA. By retaining the time-dependent terms in the
statistical equilibrium and energy equations in {\sc cmfgen} (while keeping the same approach otherwise),
we confirmed and generalized the strong time-dependent effects on
Type II SN ejecta ionization, and extended the implications to the entire spectrum, including continuum and
lines \citep{DH08_time}. In particular, we demonstrated the importance of these time-dependent terms for
reproducing the strength and shape of Balmer line profiles, although we emphasized that all lines and all ions are
to some extent affected. Even at early times when
hydrogen is fully ionized, time-dependent effects on the helium ionization are noticeable, leading to
stronger and broader He\,{\sc i} lines compared to equivalent steady-state models.

    While such important time-dependent terms were included in \citet{DH08_time}, the approach focused
    only on a portion of
    the SN ejecta, characterized by a prescribed density distribution and chemical homogeneity, with an imposed
    flux at the inner, diffusive, boundary. While suitable for general studies of the radiative-transfer problem,
    and in particular for gauging the importance of time dependence on the ejecta ionization structure,
    this setup is not very amenable for comparing models with observations, or for assessing the adequacy of a given
    explosion model.

     \subsection{Non-LTE approach with full time dependence}
\label{sect_djdt}

   The recent improvements to the code retain all the assets of the previous version, namely non-LTE and time  dependence in
   the statistical-equilibrium and energy equations.
   However, in addition to the derivatives with respect to space, angle, and frequency,
   we now treat time derivatives of the moments of the mean intensity that control
   the transport of radiation through the ejecta.
   For simplicity, we still ignore time derivatives of the specific intensity.
   Importantly, rather than focusing merely on a portion of the ejecta, we start from hydrodynamical inputs
   describing a given explosion and we map the {\it entire ejecta} onto the radiative-transfer grid.
   We then follow explicitly the evolution of both the radiation and the gas at all depths.
   Hence, with such a ``full-ejecta''
    approach, we now retain all time-, depth-, angle-, and frequency-dependent terms in the radiative-transfer
    equations, the energy equation, and the statistical-equilibrium equations. Importantly,
this approach will allow us to model the full non-LTE spectral evolution as a function of time,
which in turn will allow us to accurately model the flux and color evolution of the hydrodynamical
model of the SN.

    The treatment of boundaries is more straightforward in this approach.
    The outer boundary is positioned at large distances (corresponding to a maximum velocity as high as
    $\sim$60000\,\kms), where the radiation is freely streaming at all wavelengths apart from, perhaps,
    the He\,{\sc i}--He\,{\sc ii} continua - this matters little since few (if any) photons are injected at such
    wavelengths at any time.
    The inner boundary flux is constrained by imposing a zero-flux condition.
    This is a reasonable approximation when the entire envelope is optically thick --- it ignores effects due
    to Doppler shifts at the inner boundary and the finite light travel time. At later times we  use a  a nebular approximation
    (the incoming and outgoing specific intensities are equal, modulo Doppler shifts which we correct for).
    Diffusive energy emerging from large optical depths as time proceeds is naturally taken into account.
    The parameters describing the
    model are entirely determined at the first time step, and we just let the ejecta evolve under the various
    conservation laws for level-population rates, charge and gas/radiation energy. This is
    primarily an initial-value problem, with straightforward boundary conditions at spatial boundaries.
    Full details of the solution technique, and approximations, will be provided by Hillier \& Dessart
 (2010, in preparation).

    Using a hydrodynamical model as input for our radiative-transfer calculations represents
    a big improvement over
    our former approaches, which required
    the setting of all ejecta conditions (density, radius, velocity and boundary conditions). The past freedom
    allowed the tuning of model parameters but lacked physical consistency.
    Although it will now be harder to fit a given observation, this new approach allows a more direct confrontation
    of synthetic observables with ejecta and progenitor properties, providing constraints on the progenitor
    star and its pre-SN evolution, and the characteristics of the explosion.

    To track composition changes with depth, we allow for a depth-dependent
    composition. In addition, we follow the change in composition and the associated energy release following the decay
    of unstable nuclei. Since most explosion models give $^{56}$Ni as the most significant radioactive  species, we    follow only one decay chain.
    We assume either pure local energy deposition or solve for the non-local deposition
    using a Monte-Carlo approach for $\gamma$-ray transport (this code will be presented in a separate paper).
    In the present calculations on the early-time evolution of
    SN 1987A, the typical $\gamma$-ray photon mean-free path is much smaller than the scale of the ejecta
    so that the energy deposition is purely local.

    In the following section, we illustrate the results from our non-LTE time-dependent radiative-transfer modeling
    for SN1987A, based on the model ``lm18a7Ad'' (Woosley,  priv. comm.). With our approach, we can now
    compute both spectra and light curves in full non-LTE and allow for time dependence, with explicit account of
    line blanketing. Such an approach has never been followed in the past and constitutes a key asset of our work.
    In the present paper, we wish to highlight the key properties of the SN1987A ejecta and show the match
    of our synthetic spectra to those observed. Aiming at validating our code against the excellent
    observations of SN1987A, we do not address specifically the issues concerning the explosion, the ejecta
    geometry, or the progenitor identity of SN1987A. Such an in-depth study will be performed later,
    allowing for variations in the properties of the progenitor and of the explosion, and gauging the
    impact on observables.

\begin{figure}
\epsfig{file=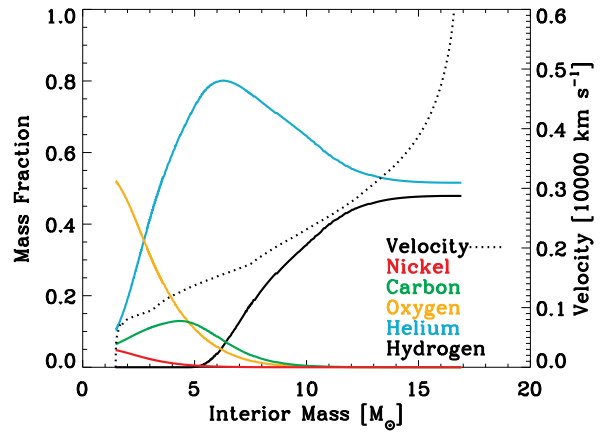,width=8.5cm}
\caption{Elemental mass fractions as a function of Lagrangian mass for the hydrodynamical input model
``lm18a7Ad''  used as the starting ejecta conditions for our radiative-transfer study. We show
hydrogen (black), helium (blue), oxygen (orange), carbon (green), and $^{56}$Ni (red).
We also overplot the velocity as a dotted black line.
Note that there is $\sim$1.1\,\msun\, of ejecta material traveling faster than 4500\,\kms,
which is the photospheric velocity at the end of our time-sequence at 20.8\,d.
The whole spectroscopic and photometric evolution presented in this paper thus corresponds
to a constant photospheric composition, i.e., that of the progenitor surface.
\label{fig_comp_87A_prog}
}
\end{figure}

\section{Initial hydrodynamical model and setup}
\label{sect_setup}

The hydrodynamical model we use for the calculations presented in this paper was produced
in two steps by Stan Woosley (priv. comm.), using the code {\sc kepler} \citep{weaver_etal_78}.
First, an 18\,\msun\, star on the main sequence
was evolved with allowance for rotation, magnetic torques, and mass loss.
The angular momentum halfway through hydrogen-core burning on the main sequence
is 2.9$\times$10$^{52}$\,erg$\cdot$s. At the stellar surface, the rotation speed at that time is 240\,\kms\
on the equator, which corresponds to $\sim$33\% of the local Keplerian (critical) velocity.
An LMC metallicity $Z$ was adopted, with $Z=0.4 Z_{\odot}$.
No convective overshooting was used, but a modest semiconvection was allowed for.

At the onset of collapse, the star is a blue-supergiant (BSG) with a radius of
3.28$\times$10$^{12}$\,cm ($\sim$47\,\rsun),
a luminosity of 8$\times$10$^{38}$\,erg ($\sim$2.1$\times$10$^8$\,\lsun),
an effective temperature of 18000\,K, and has an age of 14.7\,Myr.

The pre-SN envelope above a mass cut of 1.47\,\msun\, (the proto-neutron star which
forms a pulsar with a rotational period of $\sim$12\,ms)
is then exploded with a piston.
Its trajectory is characterized so that it yields an ejecta mass of 15.45\,\msun\
with an asymptotic kinetic energy of 1.2$\times$10$^{51}$\,erg (for details, see \S 2.3 of \citealt{woosley_weaver_95}).
The ejecta composition we inherit, when we start our radiative-transfer computations,
is given in Table~\ref{tab_comp}, while the distribution of the species with mass and velocity
is shown in Fig.~\ref{fig_comp_87A_prog}.
The model provided did not include a comprehensive description of isotopes, and lacked
some elements with low abundances (e.g., sodium). For these, we adopted the corresponding
LMC-metallicity value (given by scaling the solar metallicity value by 0.4)
and set it to be constant with depth (we expect an abundance variation
at the photosphere as it recedes in the ejecta towards deeper mass shells,
but not over the $\sim$21 days covered by our time sequence so this expedient
should be adequate). In practice, we use a depth-independent mass fraction of
1.35$\times$10$^{-5}$ for sodium. Similarly, although $^{44}$Ti is present and quite abundant
at depth in the ejecta, it is absent at the surface and so we enforce a floor value of 1.32$\times$10$^{-6}$
for titanium to reflect the LMC metallicity value for that species.
Barium and Scandium, which are both strongly under-abundant, are ignored
in this first set of simulations, and hence their associated line features are absent from our synthetic spectra.
Note that some chemical mixing has been done to mimic the effects of multi-dimensional fluid
instabilities (e.g., Rayleigh-Taylor).
The moderate mixing softens the composition gradients but preserves the strong ejecta stratification,
hydrogen being absent in the inner 3 solar masses of the ejecta (below 1500\,\kms).
The $^{56}$Ni mass fraction is sizable only below 1000--2000\,\kms\ but owing to mixing, shows a non-zero
mass fraction even at the outer edge of the ejecta. This is unphysical but tests we performed show that
such a low  $^{56}$Ni abundance does not impact the gas, nor the radiation, properties in regions exterior to
a velocity of 4500\,\kms (the ejecta photospheric velocity at the end of our time sequence) during the 21-day
timespan we cover.

\begin{table}
\begin{center}
\caption[]{Ejecta composition, given in solar masses, of the ``lm18a7Ad'' model used as input
for our radiative-transfer computations. Note that all metal mass-fractions have a floor value
corresponding to the LMC metallicity. The total ejecta mass is approximately 15.6\,\msun}
\label{tab_comp}
\begin{tabular}{cccc}
\hline
 Element        &  Mass [\,\msun]  &  Element        &  Mass [\,\msun]\\
\hline
H  &  3.94 & Si &  0.08  \\
He &  9.04  & S  &  0.03  \\
C  &  0.61 & Ar & 0.005  \\
N  &  0.027  & Ca & 0.004  \\
O  &  1.2  & Ti & 6.5$\times$10$^{-5}$  \\
Ne &  0.34 & Fe & 0.01  \\
Mg &  0.06 & $^{56}$Ni &  0.084  \\
\hline
\end{tabular}
\end{center}
\end{table}

 Before starting our radiative-transfer calculations, we need to make a number of adjustments
to this input hydrodynamical model. The innermost part of the ejecta is infalling at moderate
speeds and shows a pronounced
broad peak in both temperature and density. We thus start our grid just outside
of that region, at a radius of 1.3$\times$10$^{12}$\,cm corresponding to a velocity cut of 500\,\kms.
This excision, which amounts to $\sim$0.02\,\msun, is very small. As the outer regions of the
ejecta are described by only a handful of points beyond $\sim$10000\,\kms,
and reach a maximum velocity of a third of the speed of light, we trim this poorly-resolved region
at $\sim$8000\,\kms, and stitch an outer region in homologous expansion, constant temperature
and a power-law density distribution with an exponent $-20$, extending to a maximum radius of
1.3$\times$10$^{14}$\,cm and a maximum velocity of 55000\,\kms.
Although at the time of 0.27\,d in this initial model homologous expansion is pretty
well established (apart from the inner regions which we have trimmed, $r/v$ varies by about 15\%
between the inner and the outer ejecta), we enforce
strict homologous expansion (exact numerically, i.e. $r/v$=constant), using the values of velocity and radius halfway
through the ejecta. This is done because our time-dependent radiative-transfer solver requires
$dv/dr=v/r$, but this tinkering does not alter the corresponding
time of explosion by more than a few percent in the surface layers we model here.
That homology does not hold exactly in the ejecta
at 0.27 days results primarily from the finite size of the progenitor star, whose surface radius represents
10\% of the SN photosphere radius at 0.27\,d.
These edits to the initial model are merely for starting our time-sequence. After that, we just let the ejecta
evolve under the constraints set by radiation transport and energy/charge conservation.

To limit the size of the numerical problem to solve, we exclude some species/ions that
do not contribute sizably to the problem, e.g., species that are not identified
conspicuously in spectra at any time (e.g., neon or sulfur), or ionization stages that are not needed
for the accuracy of the problem (e.g., high ionization stages of sodium, magnesium, or calcium).
This expedient is justified at this exploratory stage, in particular given
the brute-force approach followed. In the future, we will employ a more complete atomic model.
In Table~\ref{tab_atom}, we provide details of the model atom used in the present computations
(the same one is used throughout the sequence), including for each ion species the number of super
and full levels, the number of transitions, and the uppermost level included.
Overall, we identify only one inadequate setting in the present calculation,
which has to do with the neglect of Ni\,{\sc ii}, Ni\,{\sc iii}, and Ni\,{\sc iv} in the model atom.

The basic atomic processes, atomic data, and its utilization, are discussed by \citet{HM98_lb}.
In the code we include the following processes: bound-free, bound-bound, free-free, dielectronic
recombination, electron scattering, Rayleigh scattering by hydrogen, inner-shell ionization by X-rays,
collisional ionization/recombination, collisional excitation/de-excitation, and charge exchange
reactions with H and He. To simplify the atomic model atoms we utilize super-levels,
although all transitions are still treated at their correct wavelength.
Photoionization cross sections are smoothed (typically with a Gaussian of width 3000\,\kms)
with recombination treated as the inverse of photoionization.
At present we do not include ionization by non-thermal electrons which are particularly important at later times.

For most ions there exist states that lie above the ground state of the ion,
and these can be treated in a variety of ways. Generally, for states permitted to autoionize
in LS coupling, they are treated as resonances in the photoionization cross-section.
The Opacity Project \citep{Sea87_OP} data explicitly contain these resonances while older calculations may not.
The treatment of states not permitted to autoionize in LS coupling varies -- in some cases
(e.g., C\,{\sc iii}) these states are expected to autoionize and are added to the photoionization data.
In other cases the levels are ignored although for many CNO atoms we now have accurate autoionizing
rates \citep{ARS_auto}, and it is possible to include the states explicitly in the statistical equilibrium
equations (at the expense of potentially very large model atoms).

The sources of atomic data are varied, and in many cases multiple data sets for a given ion are available.
In some cases these multiple data sets represent an evolution in data quality and/or quantity, while in
other cases they represent different sources and/or computational methods. Comparisons of models
calculated with different data sets and atomic models potentially provide insights into the sensitivity
of model results to the adopted atomic atoms and models (although such calculations have yet to be
undertaken for SN). Oscillator strengths for CNO elements were originally taken from
\citet{NS83_LTDR, NS84_CNO_LTDR}. These authors also provide transition probabilities to states in the ion
continuum. The largest source of oscillator data is from \citet{Kur09_ATD,Kur_web}; its principal advantage
over many other sources (e.g., Opacity Project) is that LS coupling is not assumed. More recently, non-LS
oscillator strengths have become available through the Iron Project \citep{HBE93_IP}, and work done by
the atomic-data group at Ohio State University \citep{Nahar_OSU}. Other important
sources of radiative data for Fe include \citet{BB92_FeV,  BB95_FeVI, BB95_FeIV}, \cite{Nahar95_FeII}.
Energy levels have generally been obtained from NIST. Collisional data is sparse, particularly for
states far from the ground state. The principal source for collisional data among low lying states
for a variety of species is the tabulation by \citet{Men83_col}; other sources include
\citet{BBD85_col}, \citet{LDH85_CII_col}, \citet{LB94_N2}, \citet{SL74},
\citet{T97_SII_col,T97_SIII_col}, Zhang \& Pradhan (\citeyear{ZP95_FeII_col,ZP95_FeIII_col,ZP97_FeIV_col}).
Photoionization data is taken from the Opacity Project \citep{Sea87_OP} and the  Iron Project
\citep{HBE93_IP}. Unfortunately Ni and Co photoionization data is generally unavailable,
and we have utilized crude approximations. Charge exchange cross-sections are from the tabulation
by \citet{KF96_chg}.  


\begin{table}
\begin{center}
\caption[]{Summary of the model atom used in our radiative-transfer calculations.
The source of the atomic datasets is given in text.
N$_{\sc f}$ refers to the number of full levels, N$_{\sc s}$
to the number of super levels, and N$_{\sc trans}$ to the corresponding number of
bound-bound transitions. The last column refers to the upper level of a given ion
included in our treatment. At the bottom of the table, we give the total number
of full levels treated, and the corresponding number of transitions explicitly included.}
\label{tab_atom}
\begin{tabular}{l@{\hspace{0.5mm}}r@{\hspace{2mm}}r@{\hspace{2mm}}r@{\hspace{4mm}}l}
\hline
 Species        &  N$_{\sc f}$  &  N$_{\sc s}$ & N$_{\sc trans}$ & Upper Level \\
\hline
H\,{\sc i}      & 30  &  20   &     435    & $n\le$30\\
He\,{\sc i}     & 51  &  40   &     374    & $n\le$11\\
He\,{\sc ii}    & 30  &  13   &     435    & $n\le$30\\
C\,{\sc i}      & 26  &  14   &     120    & $n\le$2s2p$^3$$\,^3$P$_{0}$\\
C\,{\sc ii}     & 26  &  14   &      87    & $n\le$2s2s4d$^2$D$_{5/2}$\\
C\,{\sc iii}    & 112 &  62   &     891    & $n\le$2s8f$^1$F$^o$\\
C\,{\sc iv}     & 64  &  59   &    1446    & $n\le$30\\
N\,{\sc i}      & 104 &  44   &     855    & $n\le$5f$^2$F$^o$\\
N\,{\sc ii}     & 41  &  23   &     144    & $n\le$2p$^3$d$^1$P$_{1}$\\
O\,{\sc i}      & 51  &  19   &     214    & $n\le$2s$^2$2p$^3$ $^4$S 4f$\,^3$F$_{3}$\\
O\,{\sc ii}     & 111 &  30   &    1157    & $n\le$2s$^2$2p$^2$ $^3$P 4d$^2$D$_{5/2}$\\
O\,{\sc iii}    & 86  &  50   &     646    & $n\le$2p$^4$f$^1$D\\
O\,{\sc iv}     & 72  &  53   &     835    & $n\le$2p$^2$p$^3$p$^2$P\\
O\,{\sc v}      & 78  &  41   &     523    & $n\le$2s5f$^1$F$^o_{3}$\\
Na\,{\sc i}     & 71  &  22   &    1614    & $n\le$30w2W\\
Mg\,{\sc ii}    & 65  &  22   &    1452    & $n\le$30w2W\\
Si\,{\sc ii}    & 59  &  31   &     354    & $n\le$3s$^2$ $^1$S 7g$^2$G$_{7/2}$\\
Si\,{\sc iii}   & 61  &  33   &     310    & $n\le$3s5g$^1$G$^e_{4}$\\
Si\,{\sc iv}    & 48  &  37   &     405    & $n\le$10f$^2$F$^o$\\
S\,{\sc ii}     &324  &  56   &    8208    & $n\le$3s3p$^3$ $^5$S 4p$^6$P\\
S\,{\sc iii}    & 98  &  48   &     837    & $n\le$3s3p$^2$ $^2$D 3d$\,^3$P\\
S\,{\sc iv}     & 67  &  27   &     396    & $n\le$3s3p $^3$P 4p$^2$D$_{5/2}$\\
Ca\,{\sc ii}    & 77  &  21   &    1736    & $n\le$3p$^6$30w2W\\
Ti\,{\sc ii}    & 152 &  37   &    3134    & $n\le$3d$^2$ $^3$F 5p$^4$D$_{7/2}$\\
Ti\,{\sc iii}   & 206 &  33   &    4735    & $n\le$3d$^6$f$^3$H$^o_{6}$\\
Fe\,{\sc ii}    & 115 &  50   &    1437    & $n\le$3d$^6$ $^1$G1 4sd$^2$G$_{7/2}$\\
Fe\,{\sc iii}   & 477 &  61   &    6496    & $n\le$3d$^5$ $^4$F 5s$^5$F$^e_{1}$\\
Fe\,{\sc iv}    & 294 &  51   &    8068    & $n\le$3d$^4$ $^5$D 4d$^4$G$_{5/2}$\\
Fe\,{\sc v}     & 191 &  47   &    3977    & $n\le$3d$^3$ $^4$F 4d$^5$F$^e_{3}$\\
Fe\,{\sc vi}    & 433 &  44   &   14103    & $n\le$3p5 $^2$P 3d$^4$ $^1$S $^2$P$^c_{3/2}$\\
Fe\,{\sc vii}   & 153 &  29   &    1753    & $n\le$3p5 $^2$P 3d$^3$ b$^2$D $\,^1$P$_{1}$\\
Co\,{\sc ii}    & 144 &  34   &    2088    & $n\le$3d$^6$ $^5$D 4s4p$\,^7$D$^{o}_{1}$\\
Co\,{\sc iii}   & 361 &  37   &   10937    & $n\le$3d$^6$ $^5$D 5p$^4$P$_{3/2}$\\
Co\,{\sc iv}    & 314 &  37   &    8684    & $n\le$3d$^5$ $^2$P 4p$\,^3$P$_{1}$\\
Co\,{\sc v}     & 387 &  32   &   13605    & $n\le$3d$^4$ $^3$F 4d$^2$H$_{9/2}$\\
Co\,{\sc vi}    & 323 &  23   &    9608    & $n\le$3d$^3$ $^2$D 4d$\,^1$S$_{0}$\\
Co\,{\sc vii}   & 319 &  31   &    9096    & $n\le$3p5 $^2$P d$^4$ $^3$F $^2$D$_{3/2}$\\
Ni\,{\sc v}     & 183 &  46   &    3065    & $n\le$3d$^5$ $^2$D3 4p$\,^3$F$_{3}$\\
Ni\,{\sc vi}    & 314 &  37   &    9569    & $n\le$3d$^4$ $^5$D 4d$^4$F$_{9/2}$\\
Ni\,{\sc vii}   & 308 &  37   &    9225    & $n\le$3d$^3$ $^2$D 4d$\,^3$P$_{2}$\\
\hline
                & 6426&       &   143054   &     \\
\hline
\end{tabular}
\end{center}
\end{table}

\begin{figure}
\epsfig{file=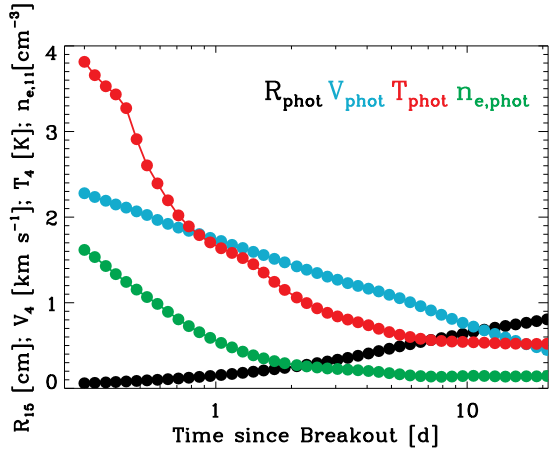,width=8.5cm}
\caption{Evolution of the radius $R_{15}$ (black; $R_{15} = R/10^{15}$\,cm),
velocity $V_4$ (blue; $V_4 = V/10^4$\,\kms),
temperature $T_4$ (red; $T_4=T/10^4$\,K),
and electron density $n_{e,11}$ (green; $n_{e,11} = n_e/10^{11}$\,cm$^{-3}$) at the photosphere
as a function of the time since shock breakout
(we adopt for the photosphere location the point where the inward-integrated electron-scattering optical
depth is 2/3). We use a logarithmic scale for the abscissa to better
render the fast early evolution of the photospheric conditions.
The dots correspond to the actual times at which the computations were performed.
\label{fig_phot}
}
\end{figure}

   To step in time, we adopt a fixed logarithmic increment, i.e., $\Delta t/t= 0.1$. Hence, to
   cover from the initial time of the Woosley model of 0.27\,d, to the final time of 20.8\,d, we
   computed $\log(20.8/0.27)/\log(1.1)\approx$46 consecutive timesteps.
   Each consecutive computation taking on average 1-2 days, the whole time sequence
   takes about 2--3 months.\footnote{It is tempting to try to step in time with larger increments (e.g., 20 to 30\% of the current time)
   but tests we have performed show that the results then differ slightly from those obtained with a choice of 10\%, or less,
   and are thus not converged. Furthermore, the convergence speed of the code is also much poorer since the changes in radiation
   and gas properties are much larger. For computing the long-term evolution of a SN ejecta, one can expedite the simulation
   by starting the time sequence at a few days after explosion.}
   It thus represents a considerable computational effort.
   For the first timestep in the sequence, we solve the radiative-transfer problem by fixing the temperature
   everywhere to what it was in the hydrodynamical input. This is suitable at depth, but the two solutions differ sizably
   in the surface layers where the assumption of LTE made by {\sc kepler}  is invalid.
   Hence, for the second timestep in the sequence, when we let the temperature
   vary and relax as needed, we observe a sizable jump in surface properties. Below, when describing our
   results,  we thus exclude that original relaxation step, and start our discussion with the timestep at 0.3\,d,
   the second epoch in our sequence.

   The spatial grid used for the radiative-transfer solution is set to have a constant increment in optical-depth, with  typically
   5 points per optical-depth decade, and a maximum of 140 points. This is much less than the original 523
   data points of the hydrodynamical-input grid. However, our grid is much finer for the regions where radiation decouples
   from matter, the transition from Eddington factors of 1/3 to 1 occurs over a few tens of points, while in the hydro
   input this transition takes place over merely a few points. Furthermore, as the ejecta recombine, our grid adjusts
   to track the recombination fronts (across which the optical depth varies steeply). In SN1987A, we have three
   main fronts, associated with the recombination to He\,{\sc ii}, He\,{\sc i}, and H\,{\sc i}. Within these fronts,
   the stiff and complicated coupling between the populations and the temperature is one of the main factors
   slowing down code convergence.\footnote{We encounter this severe problem
   for the present model for SN1987A as well as in Type Ib SN ejecta models, while for Type II-Plateau SN
   ejecta conditions, recombination fronts develop but never become very steep (Dessart \& Hillier 2010, in prep).}
   However, regions of sharp
   discontinuities in the hydrodynamical input (in composition in particular) are not resolved well and our grid
   then tends to smooth the corresponding variations. Given that these gradients are likely overestimated by the
   1D nature of the hydrodynamical model, this may not be such an issue.

\begin{table}
\caption{Summary of properties for the non-LTE time-dependent calculations presented in this work. For each epoch,
we give the total ejecta electron-scattering optical depth (in $10^3$),
the photospheric radius (in $10^{15}$\,cm), the photospheric velocity (in $10^{4}$\,\kms), the
photospheric temperature (in $10^{4}$\,K), as well as the emergent bolometric luminosity (in $10^7$\,L$_{\odot}$).
\label{tab_prop}}
\begin{tabular}{rccccc}
\hline
 Age       & $\tau_{\rm base,es}$ &    $R_{\rm phot}$      &   $V_{\rm phot}$   &   $T_{\rm phot}$   & $L_{\rm bol}$  \\
\hline
  [d]      &      [$10^3$]      &     [$10^{15}$\,cm]   &  [$10^{4}$\,\kms]  &   [$10^{4}$\,K]  &  [$10^7$\,L$_{\odot}$] \\
\hline
    0.30      &    22886.0    &    0.06      &     2.28    &    3.81  &   134.99  \\
    0.33      &    18884.8    &    0.06      &     2.24    &    3.66  &   104.37  \\
    0.37      &    15425.6    &    0.07      &     2.19    &    3.53  &    81.28  \\
    0.40      &    12851.8    &    0.07      &     2.15    &    3.43  &    66.47  \\
    0.44      &    10625.1    &    0.08      &     2.11    &    3.27  &    55.97  \\
    0.48      &     8781.2    &    0.09      &     2.07    &    2.91  &    48.82  \\
    0.53      &     7256.6    &    0.09      &     2.02    &    2.60  &    43.35  \\
    0.59      &     5988.2    &    0.10      &     1.97    &    2.39  &    38.20  \\
    0.64      &     4941.5    &    0.11      &     1.92    &    2.20  &    33.28  \\
    0.71      &     4077.2    &    0.12      &     1.88    &    2.02  &    29.71  \\
    0.78      &     3369.1    &    0.12      &     1.84    &    1.89  &    27.34  \\
    0.86      &     2778.3    &    0.13      &     1.80    &    1.79  &    24.98  \\
    0.95      &     2276.8    &    0.14      &     1.76    &    1.70  &    22.47  \\
    1.05      &     1863.7    &    0.16      &     1.72    &    1.64  &    20.29  \\
    1.16      &     1527.0    &    0.17      &     1.68    &    1.58  &    18.48  \\
    1.28      &     1254.3    &    0.18      &     1.64    &    1.52  &    16.94  \\
    1.41      &     1033.8    &    0.19      &     1.59    &    1.45  &    15.69  \\
    1.55      &      855.9    &    0.21      &     1.56    &    1.35  &    14.61  \\
    1.71      &      703.6    &    0.22      &     1.51    &    1.24  &    13.50  \\
    1.88      &      582.6    &    0.24      &     1.47    &    1.15  &    12.46  \\
    2.10      &      467.4    &    0.26      &     1.42    &    1.06  &    11.32  \\
    2.30      &      390.1    &    0.28      &     1.39    &    1.00  &    10.19  \\
    2.53      &      322.3    &    0.29      &     1.35    &    0.93  &     9.60  \\
    2.78      &      267.1    &    0.31      &     1.30    &    0.88  &     8.83  \\
    3.05      &      222.0    &    0.33      &     1.27    &    0.84  &     8.18  \\
    3.35      &      184.0    &    0.36      &     1.23    &    0.80  &     7.61  \\
    3.68      &      152.5    &    0.38      &     1.20    &    0.77  &     7.09  \\
    4.05      &      125.8    &    0.41      &     1.16    &    0.74  &     6.60  \\
    4.45      &      104.1    &    0.44      &     1.13    &    0.70  &     6.16  \\
    4.90      &       85.7    &    0.46      &     1.09    &    0.66  &     5.81  \\
    5.39      &       70.5    &    0.49      &     1.05    &    0.63  &     5.54  \\
    5.93      &       57.8    &    0.52      &     1.01    &    0.60  &     5.36  \\
    6.53      &       46.9    &    0.54      &     0.96    &    0.58  &     5.25  \\
    7.18      &       38.1    &    0.56      &     0.91    &    0.56  &     5.21  \\
    7.90      &       30.9    &    0.59      &     0.86    &    0.55  &     5.25  \\
    8.70      &       25.1    &    0.61      &     0.81    &    0.55  &     5.35  \\
    9.60      &       20.3    &    0.63      &     0.77    &    0.54  &     5.48  \\
   10.60      &       16.5    &    0.66      &     0.72    &    0.54  &     5.59  \\
   11.70      &       13.4    &    0.69      &     0.68    &    0.53  &     5.66  \\
   12.90      &       10.9    &    0.71      &     0.64    &    0.53  &     5.69  \\
   14.20      &        8.8    &    0.73      &     0.59    &    0.52  &     5.74  \\
   15.60      &        7.2    &    0.75      &     0.55    &    0.52  &     5.85  \\
   17.20      &        5.7    &    0.77      &     0.52    &    0.52  &     5.86  \\
   18.90      &        4.6    &    0.79      &     0.48    &    0.52  &     6.13  \\
   20.80      &        3.7    &    0.81      &     0.45    &    0.52  &     6.41  \\
\hline
\end{tabular}
\end{table}

\section{Ejecta evolution}
\label{sect_results}

    The simulations we present here were performed over a relatively short time, starting at
    0.27\,d after breakout and ending at 20.8\,d. While it may seem a short time compared
    to the bolometric evolution (with, for example, a light-curve peak at $\sim$80\,d), this corresponds
    to an increase in radius by a factor of 77 for all mass shells. Hence, over the 46 timesteps computed,
    the SN ejecta expanded enormously, and the associated changes at the photosphere, as well
    as above and underneath it, are large. Throughout the discussion,
    we define the SN photosphere as the ejecta location where the inward-integrated electron-scattering optical
    depth is 2/3.
    In the following section, we first present  the evolution
    of the ejecta properties, before presenting synthetic spectra and light curves.

\begin{figure*}
\epsfig{file=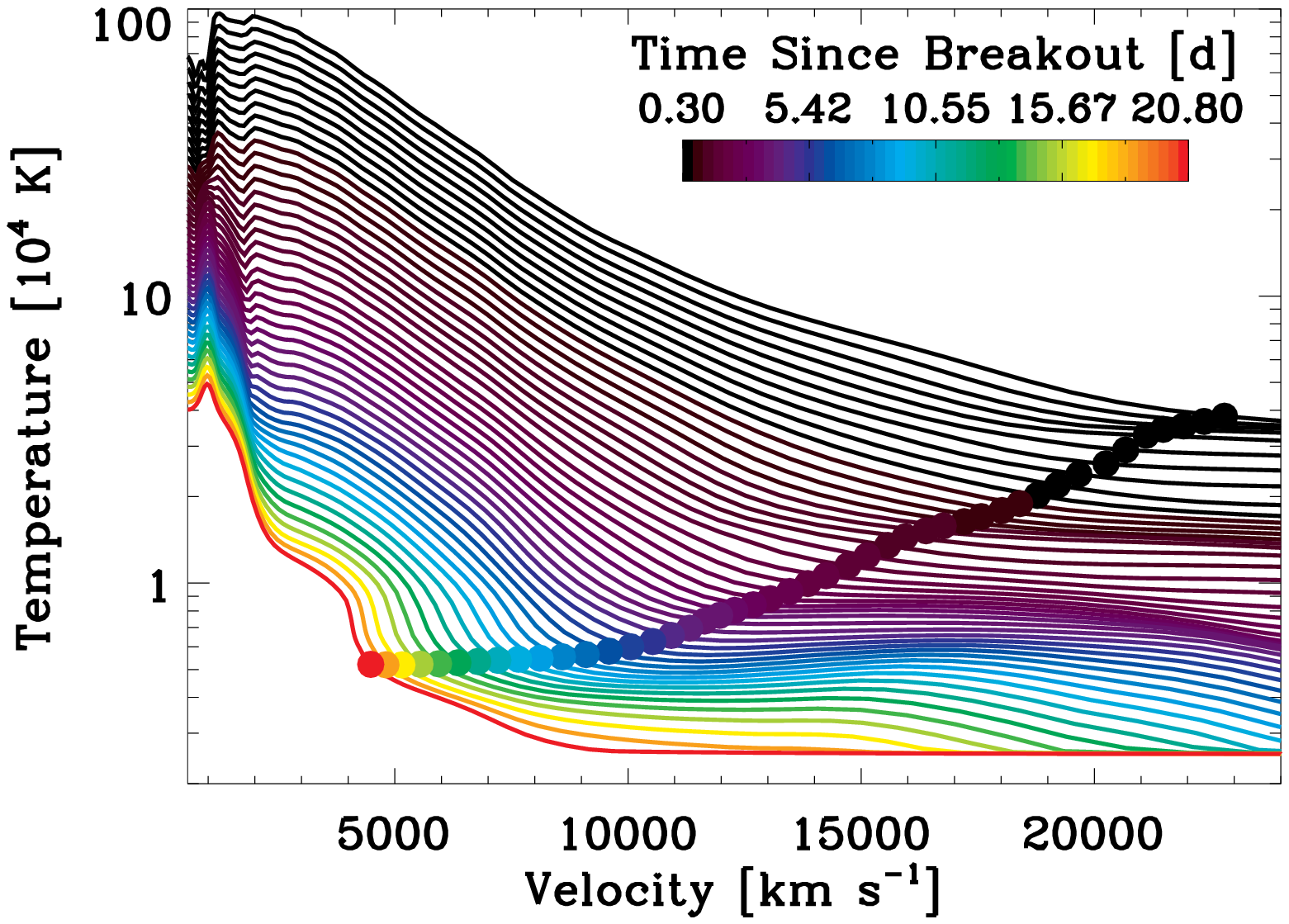,width=8.5cm}
\epsfig{file=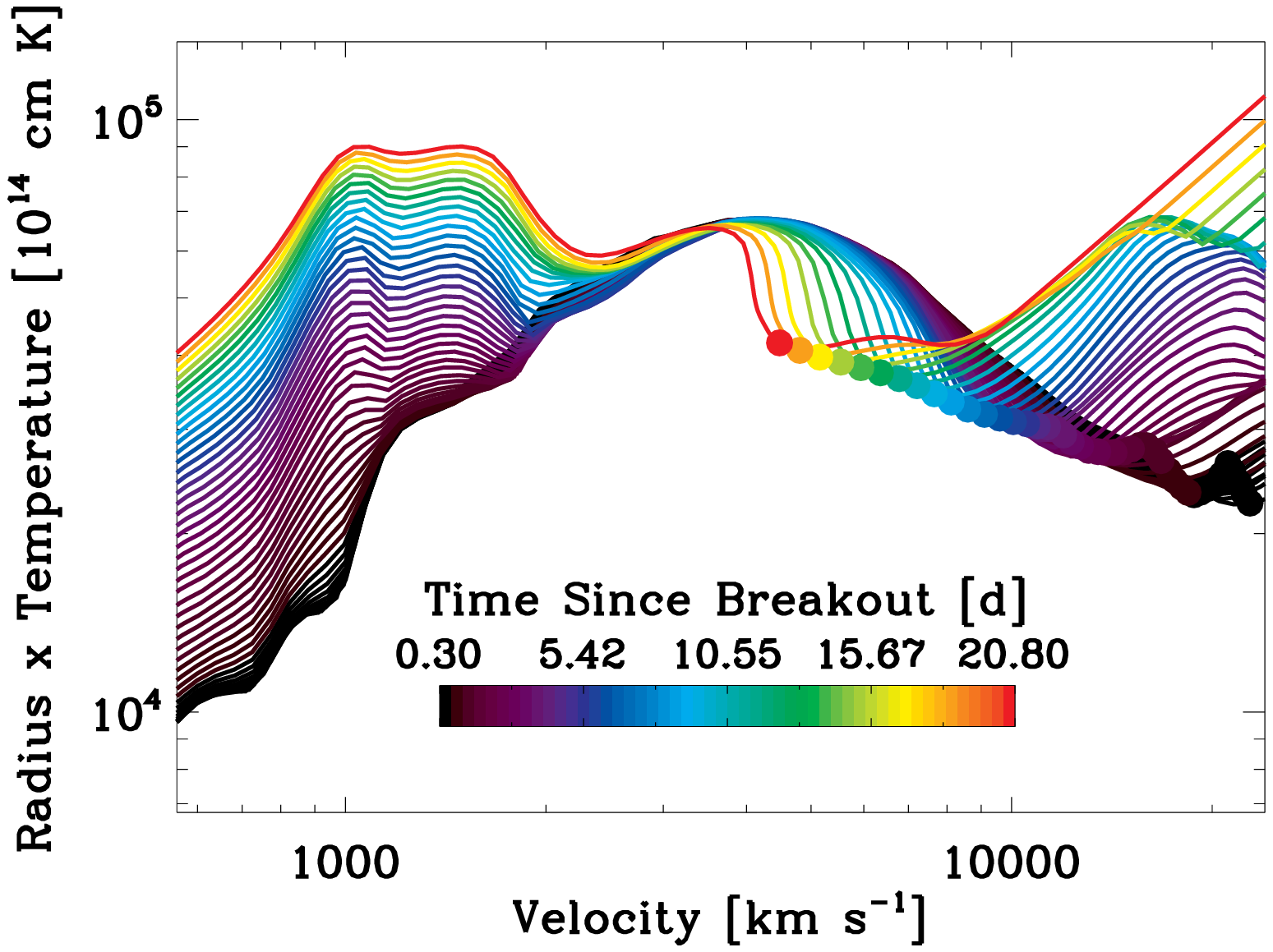,width=8.5cm}
\caption{{\it Left:} Temperature distribution as a function of ejecta velocity and time (a color-coding is used to
differentiate the epochs). We overplot with a dot the corresponding location of the photosphere.
From left to right, the two steep drops in the temperature curve at the last time in the sequence (red curve) correspond
to the He\,{\sc i} and H\,{\sc i} recombination fronts.
{\it Right:} Same as left, but now for the quantity $R \times T$, which should be a constant of time
for an adiabatically-expanding ejecta (we use a logarithmic abscissa scale for better visibility).
Deviations from the initial curve stem from energy gain due to radioactive decay at depth (below 2000\,\kms)
and from energy losses (radiation leakage) in the photosphere region and above.
The straight line we obtain at large velocities and late times results from the setting of a floor temperature
of 5000\,K up to 0.45\,d, then of 3000\,K up to 2.8\,d, and finally of 2500\,K for all later times.
\label{fig_temp}
}
\end{figure*}
\begin{figure*}
\epsfig{file=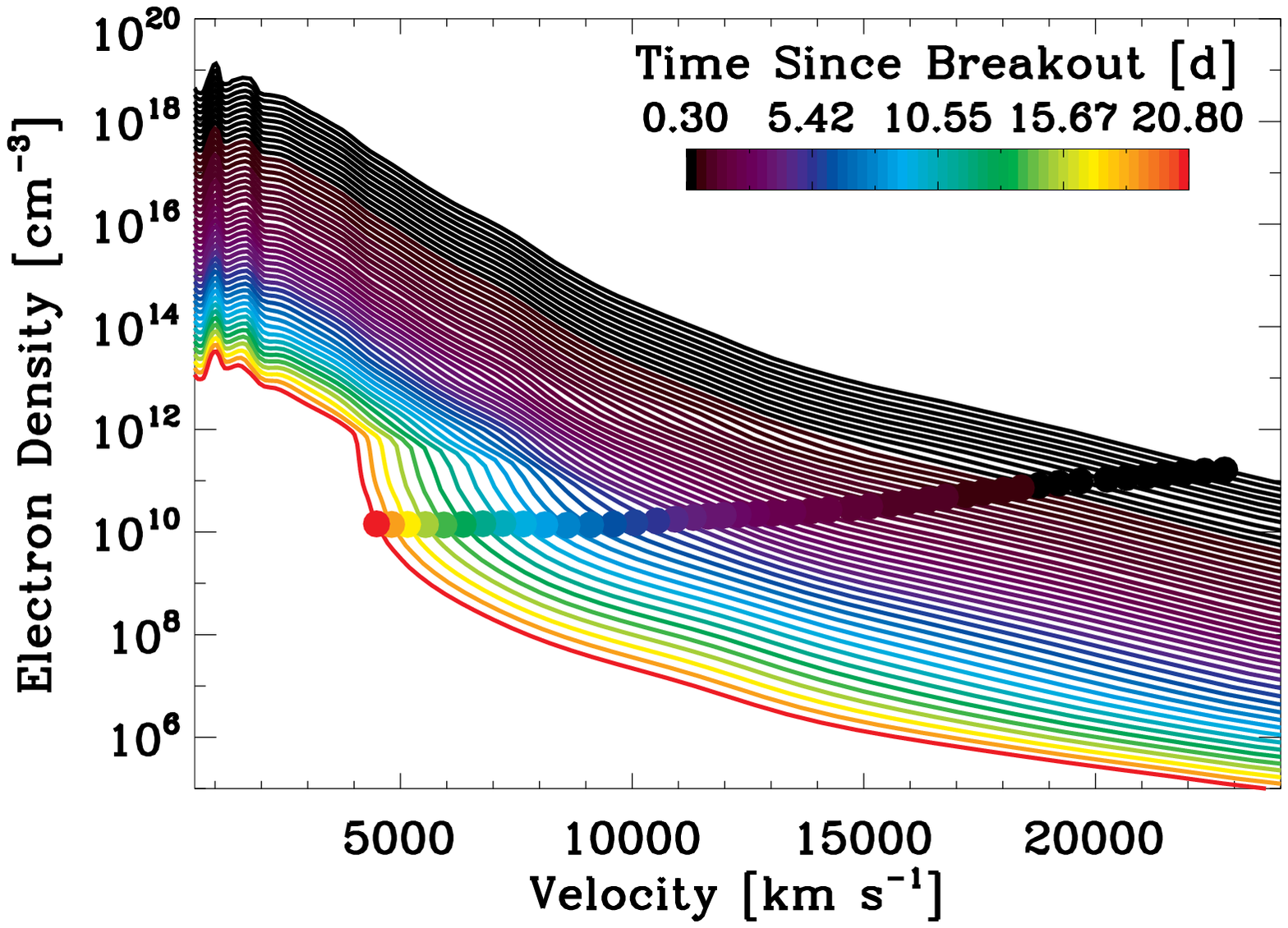,width=8.5cm}
\epsfig{file=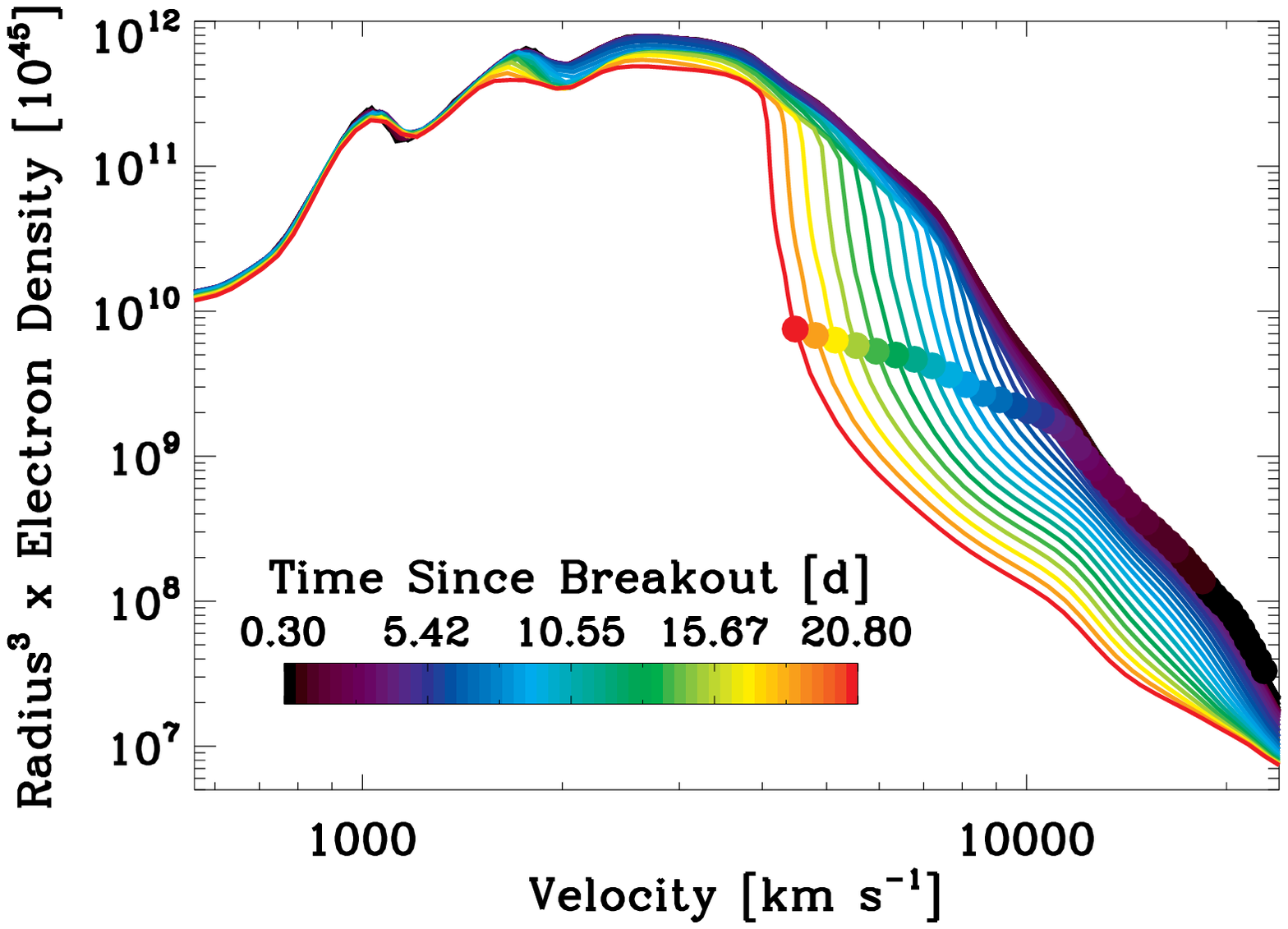,width=8.5cm}
\caption{{\it Left:} Same as for Fig.~\ref{fig_temp}, but now for the electron density.
Notice the steady decrease of the electron density with time $t$, following a general $1/t^3$ evolution,
but steepening at late times and at intermediate velocities as hydrogen recombines in those regions.
{\it Right:} Same as left, but now for the quantity $R^3 \times n_{\rm e}$, a quantity that is
a constant of time under fixed ionization. While the curves overlap at early times, several dips
develop at later times, associated with the recombination of hydrogen (one recombination front)
as well as of helium (two recombination fronts).
\label{fig_ed}
}
\end{figure*}

\begin{figure*}
\epsfig{file=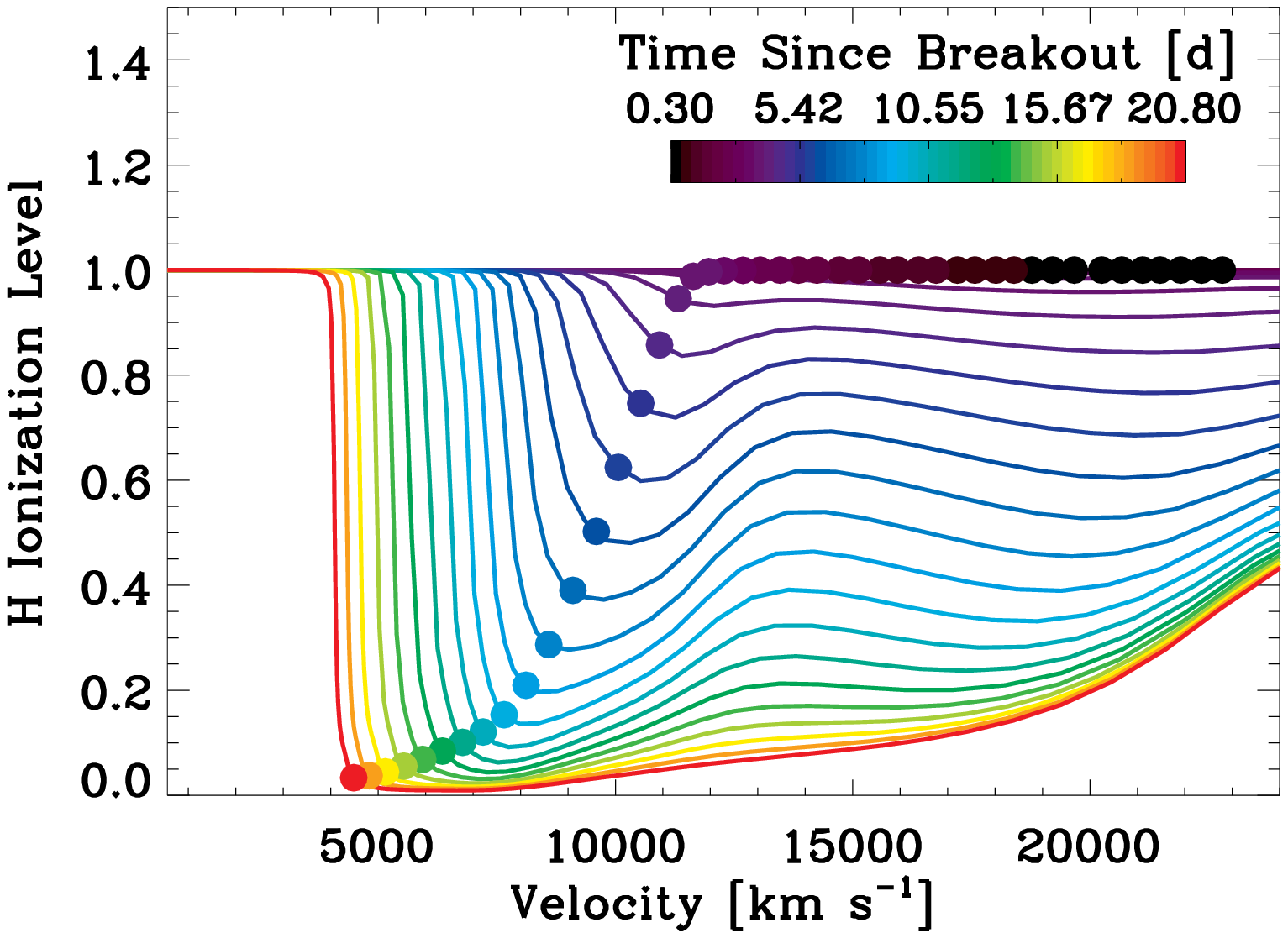,width=8.5cm}
\epsfig{file=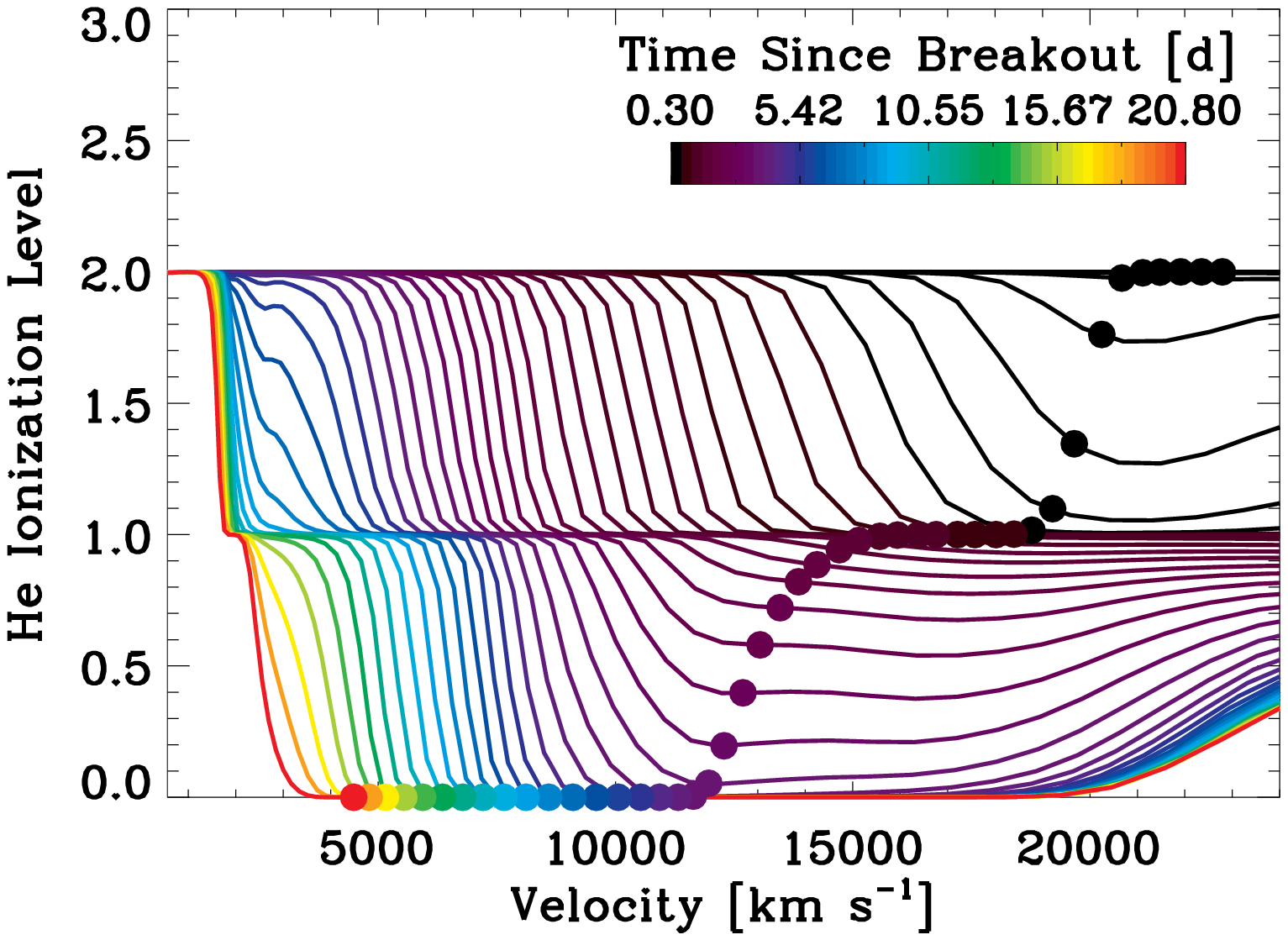,width=8.5cm}
\epsfig{file=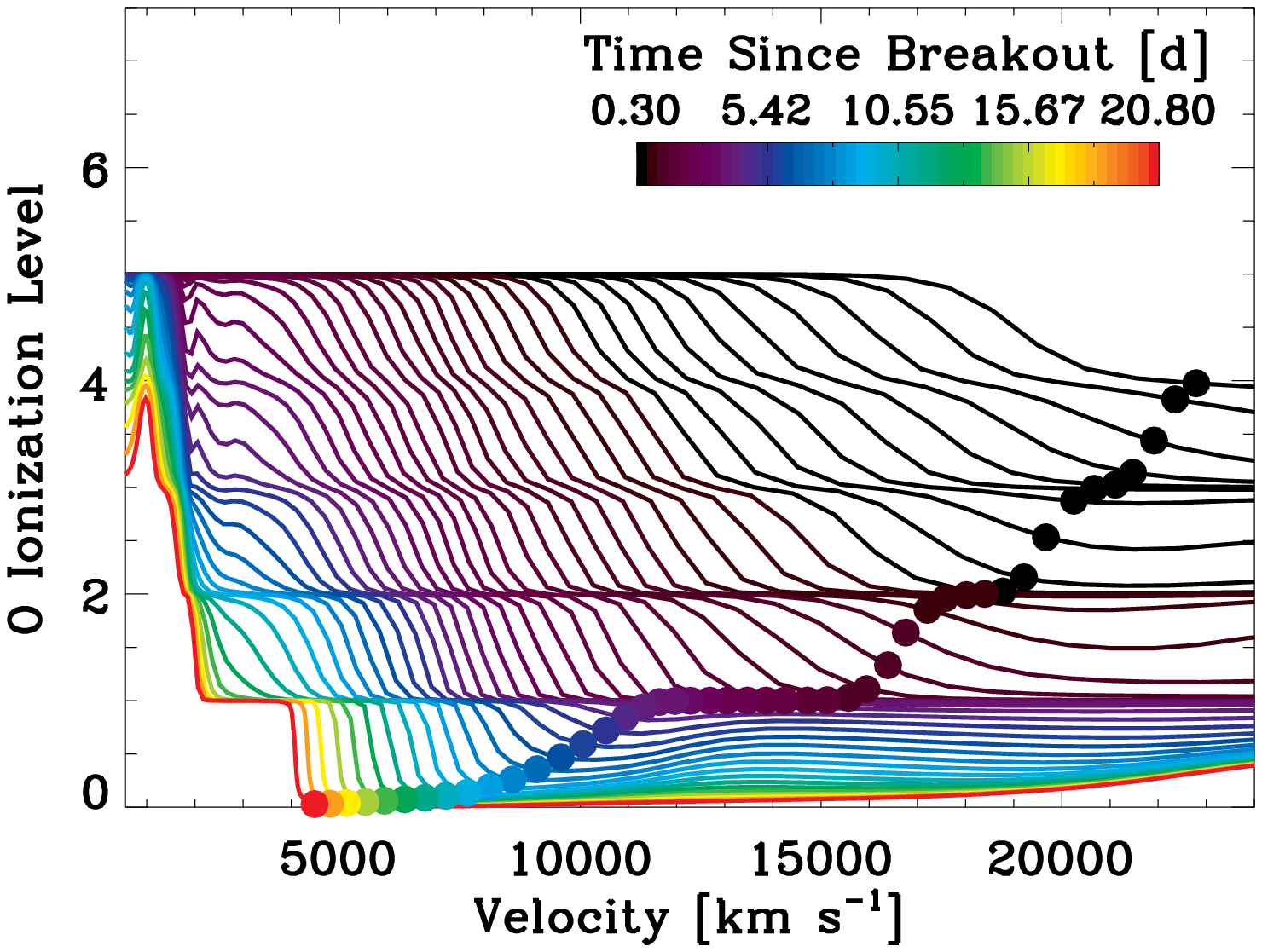,width=8.5cm}
\epsfig{file=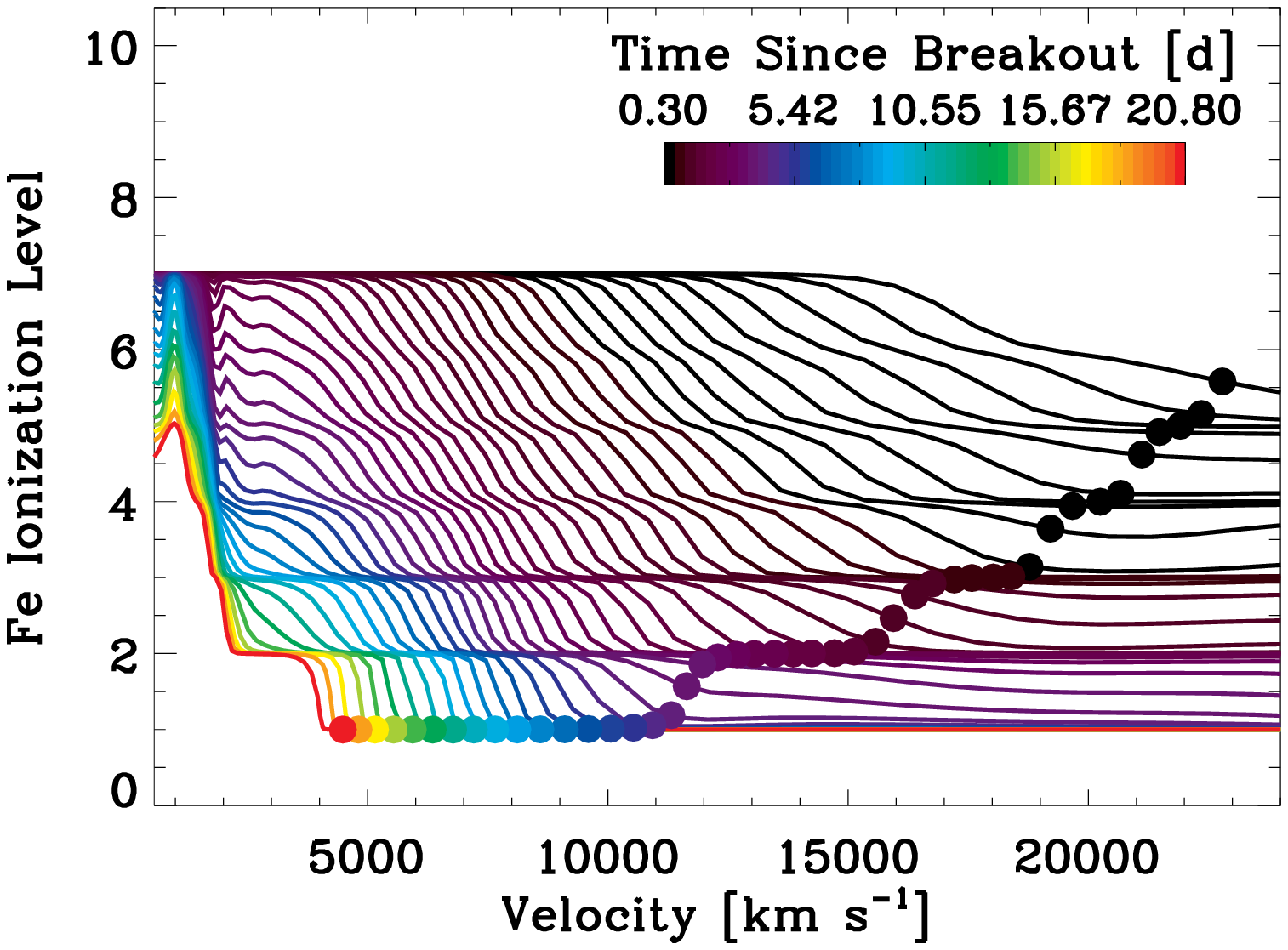,width=8.5cm}
\caption{
Same as Fig.~\ref{fig_temp}, but now for the level of ionization of hydrogen (top left), helium (top right),
oxygen (bottom left), and iron (bottom right). An ionization level $n$, where $n$ is a positive real number,
for a species X means that this species X is found at that location primarily in its $n^{\rm th}$-time ionized
form (i.e., the ionization level is equal to $\sum_i i {\rm X}^{i+}/ \sum_i {\rm X}^{i+})$
(see text for discussion)
\label{fig_ionization}
}
\end{figure*}

To illustrate the drastic evolution of the ejecta conditions over this $\sim$21-day span, we show
in Fig.~\ref{fig_phot} the radius, the velocity, the temperature, and the electron density at the
photosphere ($R_{\rm phot}$, $V_{\rm phot}$, $T_{\rm phot}$, and $n_{\rm e,phot}$) and as a function of
time after shock breakout. These quantities have been scaled to fit within
the same plotting range. They are also given in Table~\ref{tab_prop}, together with the base ejecta optical depth
(taking electron-scattering as the only opacity source for the purpose of that table)
and the emergent bolometric luminosity.

During this 21-day timespan the photospheric radius increases monotonically by over a factor of 10 from 6.0$\times$10$^{13}$ to 8.1$\times$10$^{14}$\,cm,
while the photospheric velocity drops from 23\,000 to 4500\,\kms. The decrease in photospheric velocity testifies
for the recession of the photosphere in mass. The photospheric temperature drops precipitously from 38\,100\,K\,
due to the fast expansion, but plateaus after 5--10 days at 5200\,K, which corresponds
to the hydrogen-recombination temperature - it will then remain at this value until the nebular phase.
Note that our adoption of a floor temperature (originally 5000\,K, and then decreased to 3000\,K after 0.45\,d,
and down to 2500\,K after 2.8\,d), used to avoid exponential overflow
for evaluations involving very high ionization species, is well below the temperature of the
photosphere at any time, and does not affect its properties.

The electron density at the photosphere varies by about a factor of ten during the 21-day span (green curve
in Fig.~\ref{fig_phot}).
This is because the optical depth at the photosphere is fixed by $\int_{R_{\rm phot}}^{\infty} \kappa \rho \, dr \equiv 2/3$,
where $\kappa$ is the continuum mass-absorption coefficient and $\rho$ is the mass density.
Its position adjusts as the ejecta expand, scaling as the inverse of the photospheric radius.
In the meantime, the total ejecta optical depth has decreased from 2.3$\times$10$^7$ to 3700 (we quote the
values for the electron-scattering optical depth, which differs little at such times and in such hydrogen-rich ejecta
from that associated with the Rosseland-mean opacity).

The photosphere location lies, in velocity space, above 4500\,\kms\, at all times, and thus
the composition at the photosphere has not changed through the 21-day evolution (see Fig.~\ref{fig_comp_87A_prog}).
At the photosphere, we have the following mass fractions :
$X_{\rm H}=0.478$, $X_{\rm He}= 0.517$, $X_{\rm C}= 4.2\times10^{-5}$, $X_{\rm N}= 2.5\times10^{-3}$,
$X_{\rm O}= 4.9\times10^{-4}$, $X_{\rm Fe}= 7.9\times10^{-4}$, $X_{\rm Ni}= 3\times10^{-5}$ (note that
the Nickel is primarily in its stable form, the corresponding $^{56}$Ni abundance being
initially 3$\times10^{-8}$).

Despite the relatively small time span of 21 days of our time sequence,
it covers a fast and very rich evolution for SN1987A.


\subsection{Ejecta temperature}

    The evolution of the ejecta temperature distribution echoes what was shown above for the photospheric
temperature. In the innermost part of the ejecta, below $\sim$2000\,\kms, the temperature drops
from about 10$^6$\, to 5$\times$10$^4$\,K, while above that, it is on the order of 1.5$\times$10$^4$\,K or less. This is a huge
change from the early very hot ejecta conditions we started with (Fig.~\ref{fig_temp}; dots give the location
of the photosphere at each time).
Notice the steepening of the temperature distribution just below the photosphere, associated
with the hydrogen recombination front. Two other fronts also develop, a strong one for the He\,{\sc i}--He\,{\sc ii} transition
and a weaker one for the He\,{\sc ii}--He\,{\sc iii} transition, both of which occur at large optical depths.

To reveal the ingredients that control the temperature we integrate Eq.~\ref{eq_zero_mom} over frequency to
give
\begin{equation}
  {1 \over cr^4}  {D(r^4 J )  \over Dt} + {1 \over r^2} {\partial (r^2 H)  \over \partial r}
  = \int_0^\infty \left(\eta_\nu - \chi_\nu J_\nu \right)  \,d\nu
 \end{equation}
Since at depth, $J \approx B$, $J \gg H$,  and the integral on the right-hand side is approximately zero (ignoring radioactive heating)
we have that $r^4J$ is constant, and hence  $r \times T$ is a constant of time. The same results follow from
Eq.~\ref{eq_energy} if we treat the radiation as a photon-gas with $e \equiv a T^4/ \rho$ and $P\equiv aT^4/3$.

Deviations from that constant $r \times T$ curve will reflect departures from adiabaticity,
associated here with the energy decay (a gain in regions where $^{56}$Ni
is present) and radiation losses (near the photosphere where escaping
photons extract energy from the thermal pool).
But regions lacking $^{56}$Ni and situated at large optical depth should evolve at constant $r \times T$.

In the right panel of Fig.~\ref{fig_temp}, we show the time evolution of that quantity $r \times T$
obtained in our simulations, plotted for
all ejecta mass shells as a function of time (we use a logarithmic scale for the abscissa for visibility).
At early times, all the curves overlap, but within a few days, a pronounced ``bump" forms below 2000\,\kms,
an excess which is caused by the energy contribution from radioactive decay.
Importantly, we note that it does not extend beyond 2000\,\kms, implying that the
locally-deposited energy from decay has not had time to diffuse out, even at 20.8 days after explosion.
As the photosphere is located at (or above) 4500\,\kms, radioactive decay has no effect on our synthetic spectra
during these initial 20.8.

At large velocities, the radiation that escapes
from the photosphere makes the curves dip. Note that the deficit extends below the photosphere, in regions
that ``feel'' the energy leakage taking place immediately above.
While Fig.~\ref{fig_temp} highlights some key properties governing the energy budget of SN ejecta in general,
and of SN1987A in particular, it also
serves as an important check for the numerics. Indeed, in regions where the $^{56}$Ni mass fraction
is small or negligible, and located well below the photosphere, the expansion is adiabatic and we find that
the temperature scales as the inverse of the radius, as it should.

\subsection{Ejecta ionization}

We now turn to the evolution of the electron-density distribution as a function of time (Fig.~\ref{fig_ed}).
Due to the homologous expansion, the mass-continuity equation leads to a $1/t^3$ evolution
of the mass density at a fixed velocity. Furthermore, if the evolution is at constant ionization, the quantity $r^3 \times n_{\rm e}$ is
a constant of time. We plot this quantity $r^3 \times n_{\rm e}$ in the right panel of Fig.~\ref{fig_ed}. While the curves overlap
at early times, there is a systematic and growing deviation at large velocity, starting just below the photospheric velocity at each time.
This deviation starts when the conditions of full ionization cease to prevail. It is caused by hydrogen recombination, which severely
depletes the density of free electrons. At depth, a similar but more modest deviation occurs,
and is associated with the recombination of helium.

While hydrogen and to a lesser extent helium recombination impact the electron density visibly,
all species present in the ejecta, no matter how relatively under-abundant, will recombine from their
high ionization states at the time of breakout to a neutral or once-ionized state after a few days/weeks.
We illustrate this systematic recombination in Fig.~\ref{fig_ionization}
for hydrogen (top left), helium (top right), oxygen (bottom left), and iron (bottom right). As before, a color coding
is used to differentiate the epochs - recall that the time increment is constant in the log (it increases by $\sim$10\% at each time
step). The key characteristics that apply to all is that the ionization is high at low velocity, reaches a minimum
around the photosphere, but increases again at high velocity (in particular for the dominant species hydrogen and helium).
This behavior reflects the ionization freeze-out in fast-expanding low-density SN ejecta,
and stems from incorporating time-dependent terms in the
statistical-equilibrium and energy equations \citep{DH08_time}.
This ionization freeze-out cannot prevent recombination altogether, but the effect is large enough
to dramatically alter the strength of numerous strong lines, such as H\,{\sc i} Balmer lines or Na\,{\sc i}D
(see \S\ref{sect_comp_to_obs}).


Initially, the material at depth is highly-ionized for all species, i.e.,  H$^+$, He$^{2+}$, O$^{5+}$, Fe$^{7+}$, while
at the photosphere we have H$^+$, He$^{2+}$, O$^{4+}$, Fe$^{5.5+}$.
These are highly-ionized conditions
that were not seen at the photosphere of SN1987A because observations started ``too late'',
at one day after explosion, when the ions that prevail at the photosphere are: H$^+$, He$^{+}$, O$^{2+}$, Fe$^{3+}$.
At the end of the simulation at 20.8\,d, the material at depth is still hot and optically-thick, and we find
H$^+$, He$^+$--He$^{2+}$, O$^+$--O$^{4+}$, Fe$^{+}$--Fe$^{5+}$, while at the photosphere
the ions are now close to their neutral state: H$^{0+}$, He$^{0+}$, O$^{0+}$, Fe$^{+}$.
This strong contrast in ionization between the optically-thick hot inner layers
and the optically-thin cool outer layers is what forces us to retain a broad range of ionization
stages for all important ions. In our current brute-force approach, any ion important at some
depth must be included at all depths.

The photosphere location tends to track the ionization minimum, bracketed between the fully ionized
hotter layers at large optical depth and the cool outer optically-thin regions subject to ionization freeze-out.
So, in the new approach followed, we retrieve the properties identified in that former work,
but now within the context of a calculation based on a physical hydrodynamical input of the explosion
and pre-SN progenitor evolution, and without any prescribed boundary conditions for the flux.
The evolution we see here further implies that events like SN1987A experience a faster change in
ejecta ionization within their first day, motivating the collection of multiple spectra {\it within}
the first day/night after explosion.

  \section{Synthetic Spectra and Bolometric Luminosity}
\label{sect_synthetic_rad}

In this section, we present the radiative properties of SN1987A, focusing exclusively on the
results from our non-LTE time-dependent radiative-transfer simulations.
 We scale our synthetic spectra adopting a distance of 50\,kpc to the LMC
 \citep{LMC_dist1,LMC_dist2,LMC_dist3,LMC_dist4}.

  \begin{landscape}
\begin{figure}
\includegraphics[width=22cm]{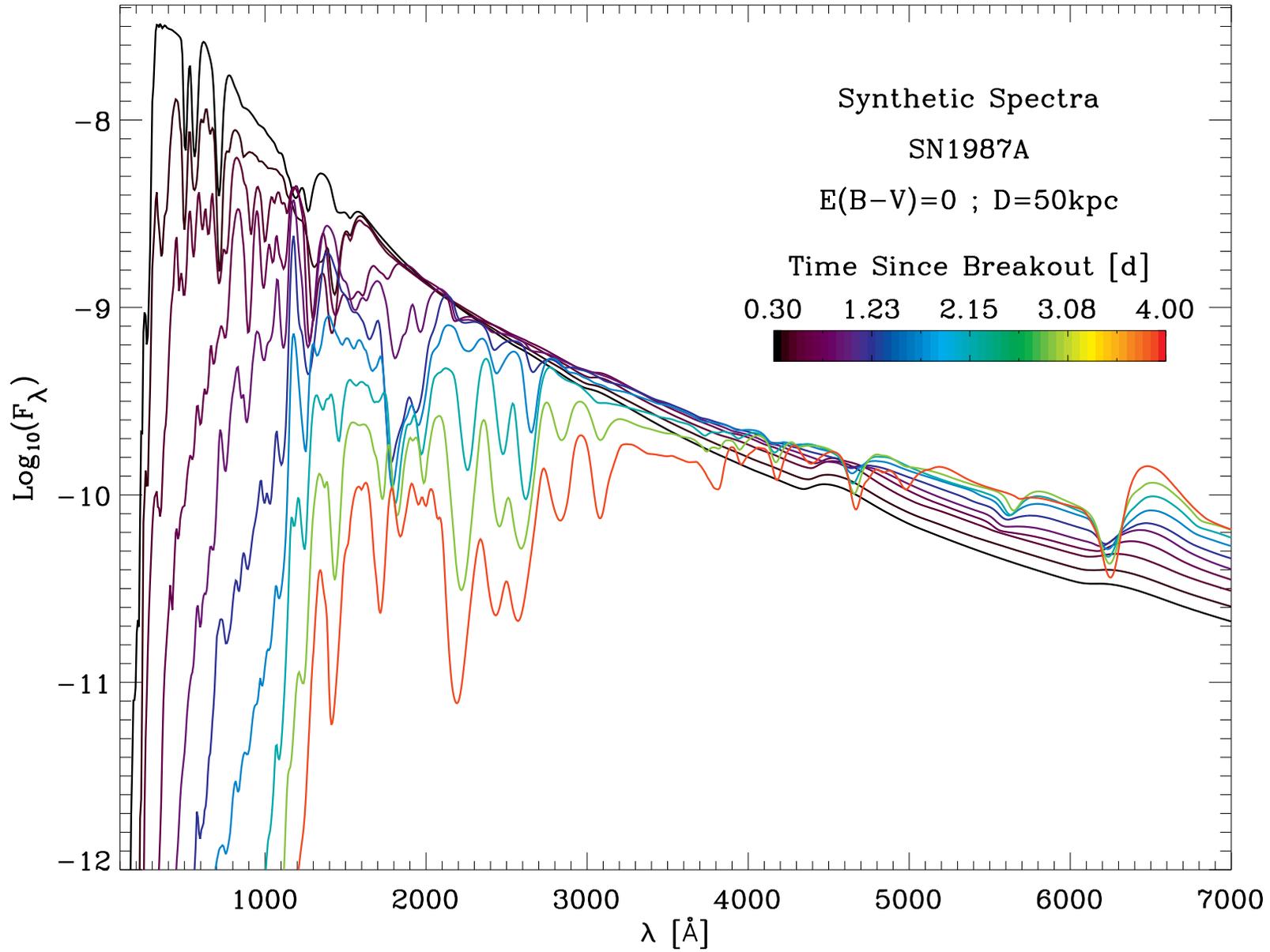}
\caption{Montage of synthetic spectra between 0.30\,d and 4\,d.
For better visibility, we interpolate in time between those epochs that were
computed, and show the spectra at the following times:
0.30, 0.40, 0.53, 0.71, 0.95, 1.26, 1.69, 2.25, 3.00, and 4.00\,d.
We used a constant increment in the $\log$ to better resolve the fast evolution at early times.
No reddening is applied, and the only scaling is for the distance, taken at 50\,kpc.
\label{fig_synthetic_spectra_early}
}
\end{figure}
\end{landscape}
\begin{landscape}
\begin{figure}
\includegraphics[width=22cm]{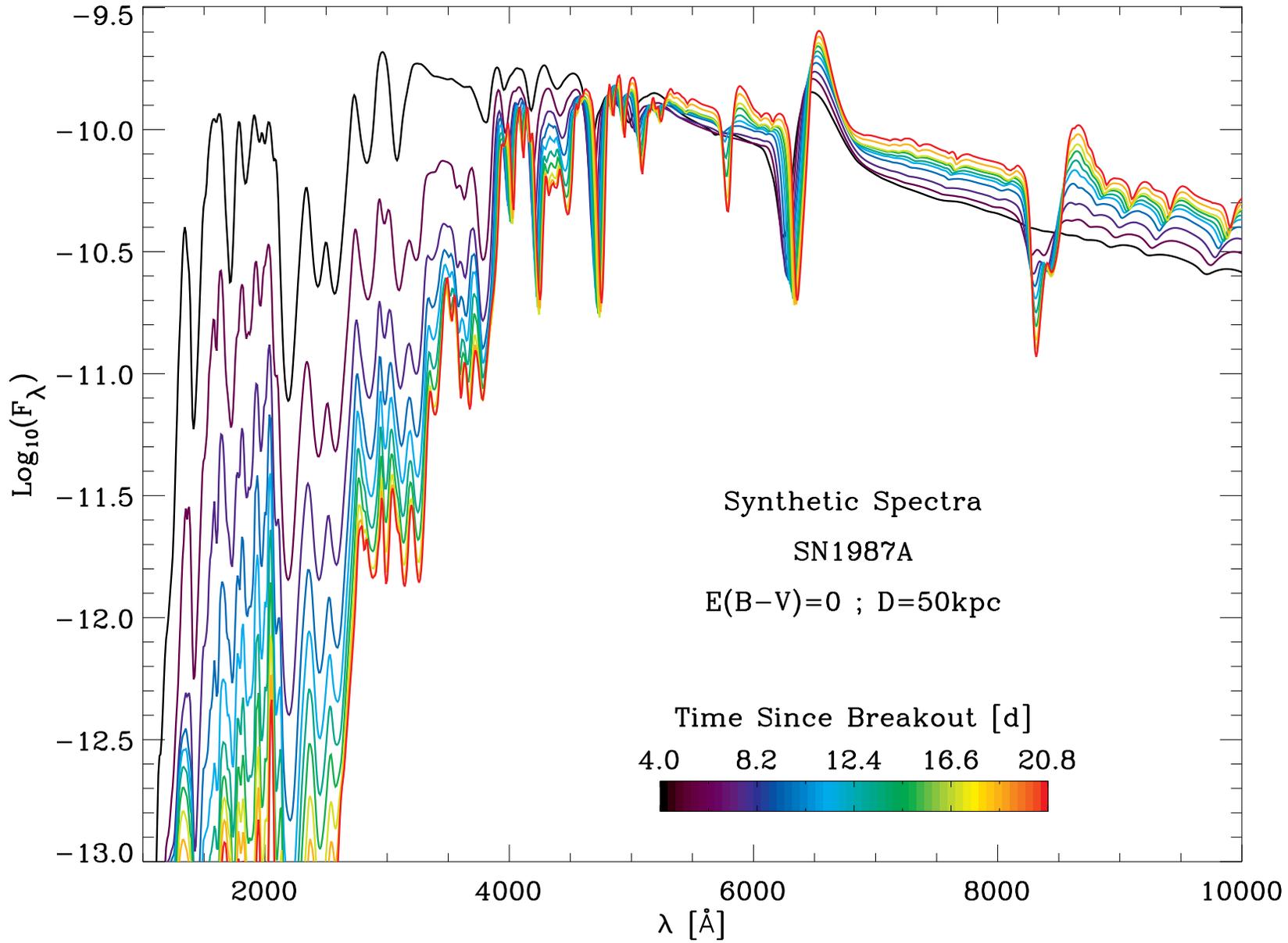}
\caption{Same as Fig.~\ref{fig_synthetic_spectra_early}, but now for later times.
The split in time between the previous figure and this one corresponds to the
epoch when the He\,{\sc i} 5879\AA\ line vanishes. This transition occurs prior to the appearance of
Na\,{\sc i}\,D. We use a constant time increment,
and we plot synthetic spectra at the following times:
4.00, 5.87, 7.73, 9.60, 11.47, 13.33, 15.20, 17.07, 18.93, and 20.8\,d.
No reddening is applied, and the only scaling is for the distance, taken at 50\,kpc.
Notice how H$\alpha$ strengthens with time despite the declining Lyman/Balmer continua
and the cooling of the photospheric layers down to the hydrogen-recombination temperature.
\label{fig_synthetic_spectra_late}}
\end{figure}
\end{landscape}

 \subsection{Spectroscopic evolution  and sources of line blanketing}

As our time sequence starts at 0.27\,d after explosion, we compute non-LTE time-dependent
spectra and light curves starting at epochs that were not observed (i.e., before day 1)
and extending to 20.8\,d.
We show our synthetic spectra for the early/fast evolution in Fig.~\ref{fig_synthetic_spectra_early}
(day 0.3 to 4; we omit the first date in the sequence, which was computed at fixed temperature, as given
by the hydrodynamical input)
and the subsequent/slow evolution in Fig.~\ref{fig_synthetic_spectra_late} (day 4 to 20.8).
Given the BSG nature of the progenitor, and in particular the modest progenitor-star radius of $\sim$50\,\rsun\,
at explosion, the initial fast expansion of the shocked outer layers of the progenitor star induces a dramatic cooling.
This is the cause of the much faster spectral evolution of SN1987A  compared to Type II-P SNe which have
RSG progenitor stars with surface radii of 500-1500\,\rsun.

Our simulations give a photosphere whose temperature decreases from 38\,100\,K at 0.3\,d to
$\sim$11\,000\,K at 2.0\,d. This is then followed by a slow decrease to $\gtrsim$5000\,K over a number of days.
As expected, and as shown in Fig.~\ref{fig_synthetic_spectra_early},
we obtain synthetic spectra whose peak-flux shifts very rapidly during these initial two days,
from around 300\AA\  to 3000\AA. Over that time, the UV flux drops by about two orders of magnitude while
the optical flux rises by a factor of a few. The UV flux is initially highly wavelength-dependent,  caused
by strong line blanketing from multiple ions. In contrast, the optical suffers little from line blanketing at early times
and the corresponding flux is essentially featureless.

\begin{figure}
\epsfig{file=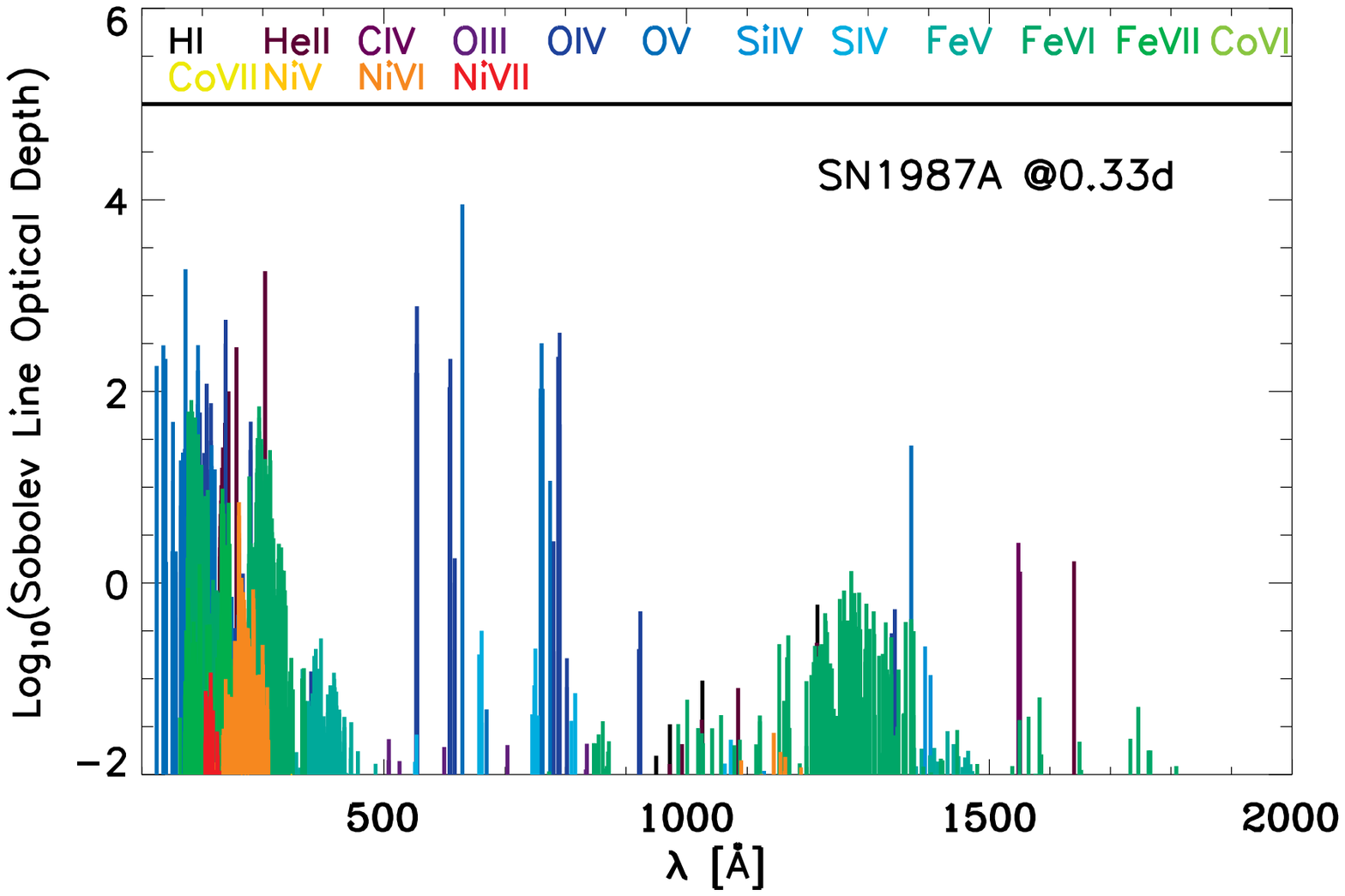,width=9cm}
\epsfig{file=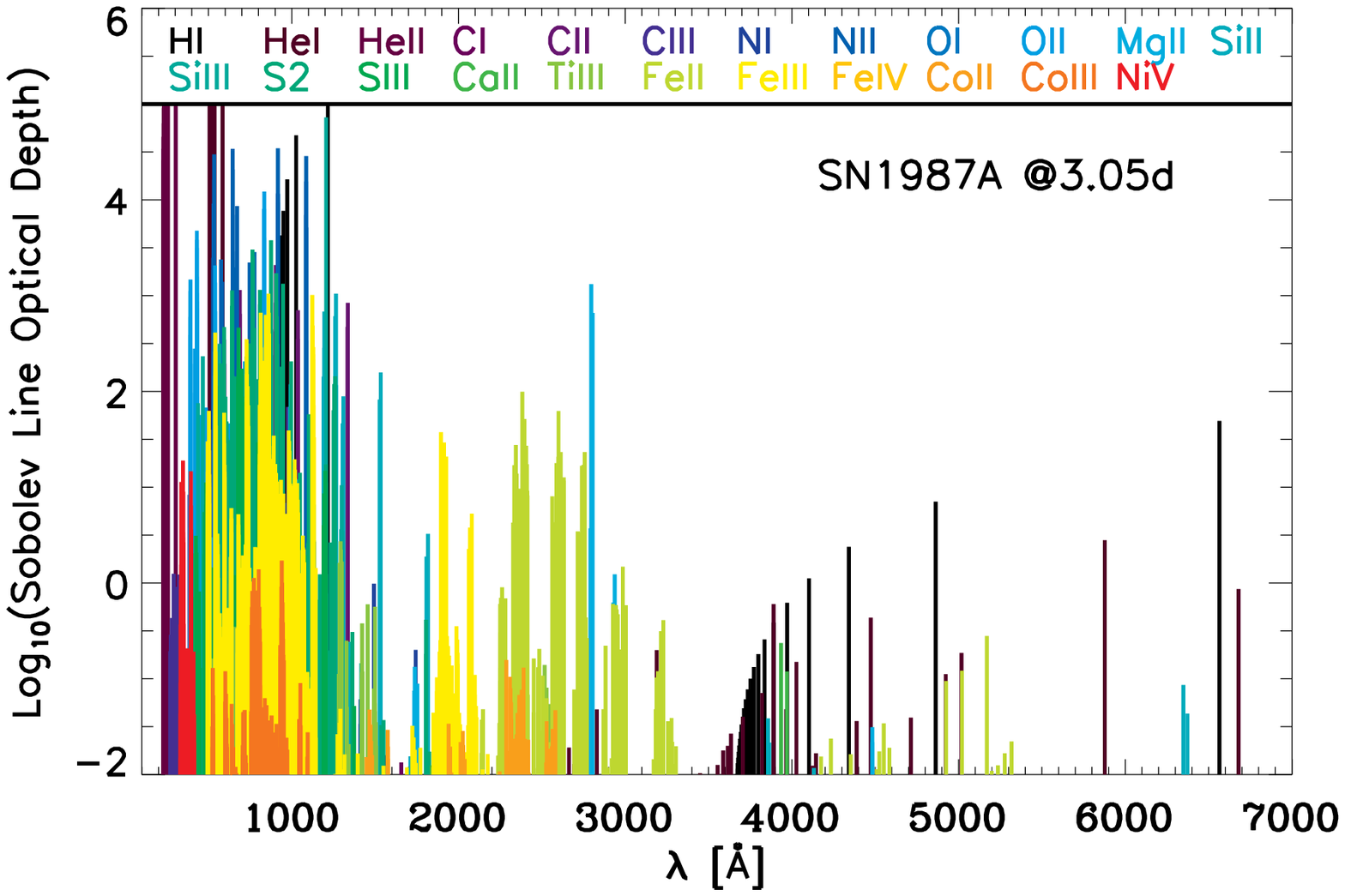,width=9cm}
\epsfig{file=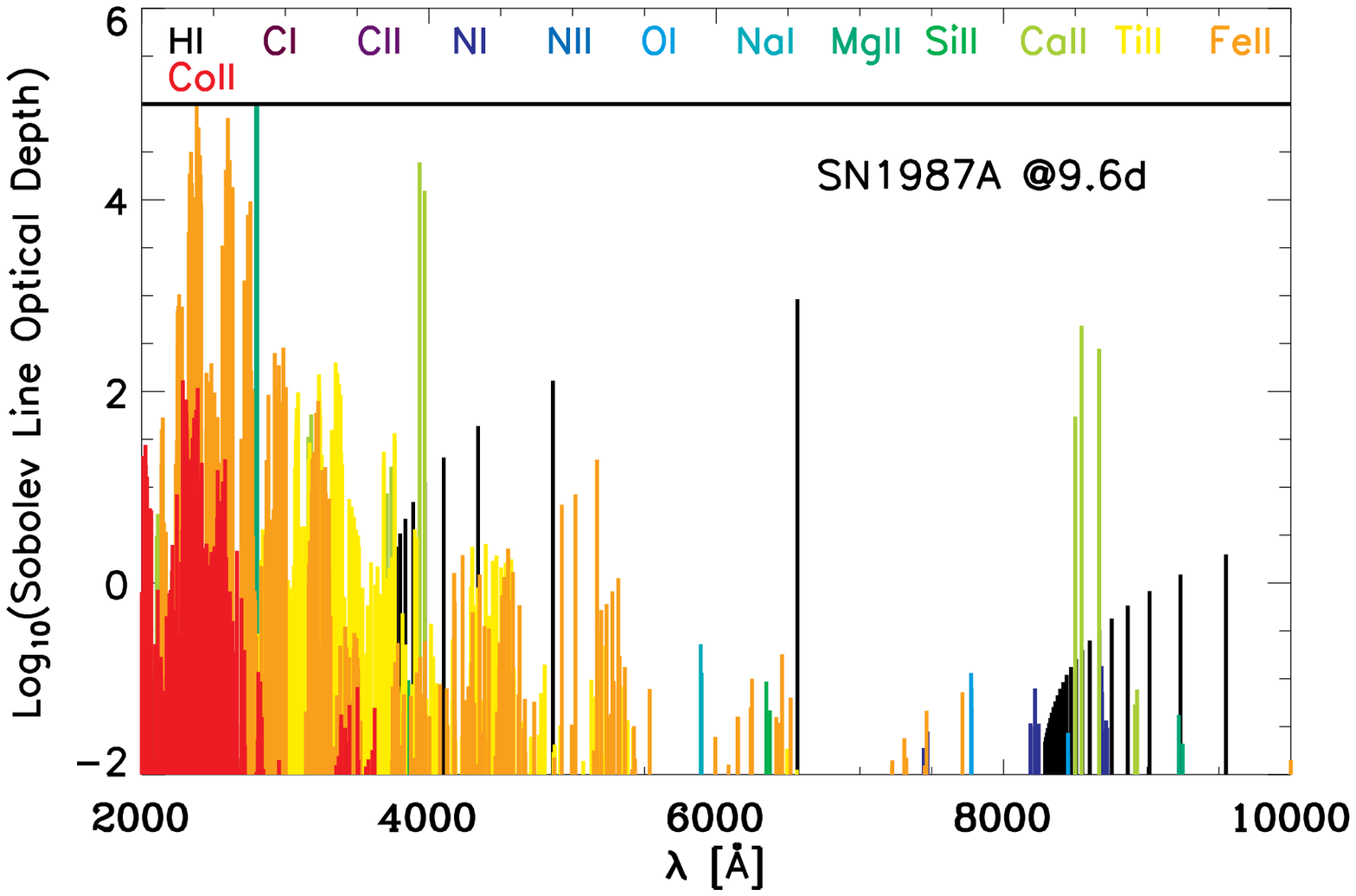,width=9cm}
\caption{{\it Top:} Illustration of the wavelength distribution of all lines having a Sobolev optical-depth $\tau_{\rm line}$
greater than 0.01 at the SN photosphere and 0.33 days after the explosion. We compute $\tau_{\rm line}$
in the co-moving frame using the Sobolev approximation. The abscissa range is 100-2000\AA.
This illustration does not intend to show the {\it effective} opacity felt by photons, but rather how
line opacity is distributed with wavelength and between ions (we overplot all line contributions rather
than add them).
{\it Middle:} Same as top, but now for the SN at 3.05\,d. The abscissa range is 100-7000\AA.
{\it Bottom:} Same as top, but now for the SN at 9.6\,d. The abscissa range is 2000-10\,000\AA.
\label{fig_ltau}
}
\end{figure}

Using the photosphere as a reference location,
we show in Fig.~\ref{fig_ltau} the spectral regions where important line blocking occurs at three epochs,
differentiating the contributing ions with a color coding.
At 0.33\,d, line blocking appears at short wavelengths
(200-2000\AA) and is associated with the highly-ionized ions that are present at the photosphere at those
early times, i.e., O\,{\sc iii}--{\sc v}, Si\,{\sc iv}, S\,{\sc iv}, Fe\,{\sc v}--\,{\sc vii}, Co\,{\sc vi}-{\sc vii}, Ni\,{\sc vi}--\,{\sc vii} 
(top panel of Fig.~\ref{fig_ltau}). Hence, considerable blanketing takes place at such times and under highly-ionized
conditions, although this takes place in the UV and is rarely seen.
Lines in the optical region are extremely weak (in part because of the very large photospheric velocity at those early times), 
and are associated with H\,{\sc i} (Balmer lines), He\,{\sc i} and He\,{\sc ii}.

At 3.05\,d (middle panel of Fig.~\ref{fig_ltau}),
the spectral regions with strong line blocking are still found shortward of 3000\AA, but  the associated ions
are of a lower ionization state (e.g., O\,{\sc ii}, Si\,{\sc ii}--\,{\sc iii}, S\,{\sc ii}--\,{\sc iii}, Ti\,{\sc iii},
Fe\,{\sc ii}--\,{\sc iv}, Co\,{\sc ii}--\,{\sc iii}). Ni\,{\sc ii} and Ni\,{\sc iii} would also contribute but have not been
included in this calculation. In the optical,
a small number of lines (relative to the forest of metal lines) associated with H\,{\sc i} and He\,{\sc i}
make their conspicuous appearance. He\,{\sc ii} 4687\AA\ is gone, and lines of once-ionized CNO elements are
now present.

Beyond 3--4 days, the SED continues its dramatic fading in the UV region (see bottom panel
of Fig.~\ref{fig_ltau} for an example at 9.8\,d), which decreases by a total
of about five orders of magnitude from 0.3\,d to 20\,d. In contrast, past 3--4 days and after the disappearance of He\,{\sc i} 5877\AA,
the optical range evolves slowly with a strengthening of the hydrogen Balmer lines, Ca\,{\sc ii},
and metals of low-ionization state (Si\,{\sc ii}, Ti\,{\sc ii}, Fe\,{\sc ii}, Co\,{\sc ii}).
Our simulations do not include Ni\,{\sc ii}--{\sc iv} in the model atom, and have zero abundance for
Scandium and Barium. Hence, no microphysical processes associated with these species/ions are
predicted by our computations.

\begin{figure}
\epsfig{file=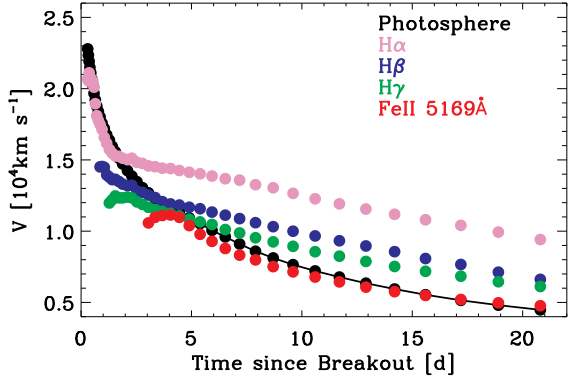,width=8.5cm}
\caption{Illustration of the evolution of the photospheric velocity (black) in comparison with
the velocity measured at maximum absorption in the P-Cygni profile troughs we associate
with H$\alpha$, H$\beta$, H$\gamma$, and Fe\,{\sc ii} 5169\AA.
Measurements are done on {\it synthetic spectra}. At times prior to 3-4 days, line overlap
tends to affect very weak lines. Hence, the measurement may be a hybrid of multiple contributions at earlier times.
However, notice how well Fe\,{\sc ii} 5169\AA, once it has become strong enough beyond a few days after explosion,
traces closely the photospheric velocity. Notice also how all lines underestimate the photospheric velocity
at early times, a property discussed in \citet{DH05_epm}.
\label{fig_vline}
}
\end{figure}

We wish to emphasize that at all times line blanketing severely depletes any overlapping flux. It achieves this very
effectively in the UV, in association with ions that may have a high or low ionization state. Line blanketing
is thus not born when conditions are cool; it is instead the case that as conditions cool, the spectral region where line
blanketing occurs shifts towards (observed) optical spectral regions. One must also recall that as this takes place,
the peak of the flux distribution shifts to longer wavelengths, so that there is less and less flux to block.
For example, at 20\,d, the line blanketing caused by Fe\,{\sc ii} is enormous at UV wavelengths, but there is hardly
any photons injected at the thermalization depth in the UV at such times.

\subsection{Ejecta kinematics from line profile morphology}

Inferring the photospheric velocity of the SN ejecta from spectroscopic observations at a given
epoch is important as it enters the determination of the total kinetic energy of the ejecta.
It is also a key component for the determination of distances to Type II SNe using the
expanding-photosphere method \citep{KK74_EPM,E96,DH05_epm}, the spectral-fitting
expanding-atmosphere method \citep{MBB02_87A}, or the standard-candle method \citep{HP_92}.
In Fig.~\ref{fig_vline}, and using the synthetic spectra for our present time sequence,
we show the Doppler velocity corresponding to the location of maximum profile absorption for several
optical lines (H$\alpha$: Magenta; H$\beta$: Blue; H$\gamma$: Green; Fe\,{\sc ii} 5169\AA: Red).
We find that at early times all
lines tend to show a maximum absorption at a Doppler velocity that underestimates the photospheric velocity
(shown in black), for a reason discussed in detail in \citet{DH05_epm}. However, once a given line strengthens,
that Doppler velocity then overestimates the photospheric velocity, by an amount that varies from
being large for H$\alpha$, to modest for H$\beta$ and H$\gamma$.
 For Fe\,{\sc ii} 5169\AA, this Doppler velocity underestimates the theoretical photospheric velocity
 at all times prior to $\sim$15\,d, and modestly overestimates it afterwards.

  In a separate test calculation, not shown here, we evolved the same hydrodynamical model
but with a three times larger iron abundance. In that case, the Fe{\sc ii} 5169\AA-line was stronger
at all times and the corresponding velocity location of maximum profile absorption matched more closely the
photospheric velocity. The accuracy of using the Fe{\sc ii} 5169\AA\ line for inferring the
photospheric velocity is thus conditioned by the environmental metallicity, and may therefore vary
with redshift or galaxy hosts.

\begin{figure}
\epsfig{file=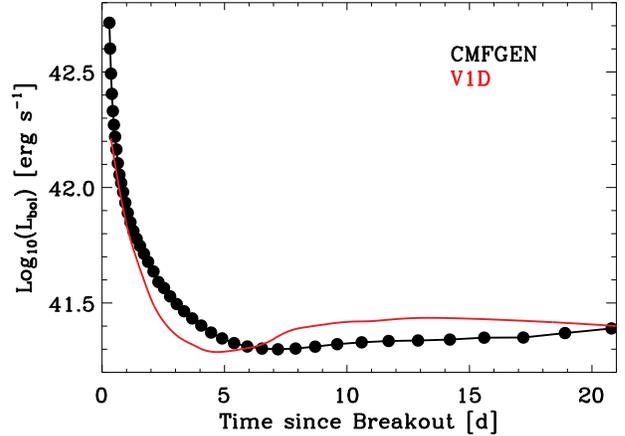,width=9cm}
\caption{Bolometric light curve for SN1987A based on our non-LTE time-dependent radiative-transfer
calculations with {\sc cmfgen} (black; dots refer to the epochs computed). We also overplot
the prediction, for the same ejecta, obtained with V1D \citep{DEW_09}, which agrees to within 10-20\%.
Both curves agree qualitatively and quantitatively with the results
of \citet[see their Fig.~8]{blinnikov_etal_2000}.
\label{fig_lbol}
}
\end{figure}

\subsection{Synthetic bolometric light curve}

To conclude this section on synthetic radiative properties,
we present the bolometric light curve for our time sequence in Fig.~\ref{fig_lbol}.
Starting at 0.3\,d, we miss the breakout phase, but capture the steep fall-off that immediately follows it until the
luminosity levels off after about 7 days. Beyond that, we then see a slow increase in bolometric luminosity.

In Fig.~\ref{fig_lbol}, we overplot the prediction of the one-dimensional radiation-hydrodynamics
V1D code,  as described in \citet{DEW_09}. This computation is done assuming gray transport
(i.e., one group), a mean opacity that neglects line contributions, and flux-limited diffusion.
The bolometric luminosity obtained with this much coarser approach agrees to within $\sim$20\%.
Furthermore, our non-LTE time-dependent line-blanketed treatment, shows an
evolution that agrees with the inferred observed bolometric light curve
as well as the radiation-hydrodynamics results obtained by \citet[see their Fig.~8 for comparison
with both their models and the inferred bolometric luminosity]{blinnikov_etal_2000}.

\begin{figure*}
\epsfig{file=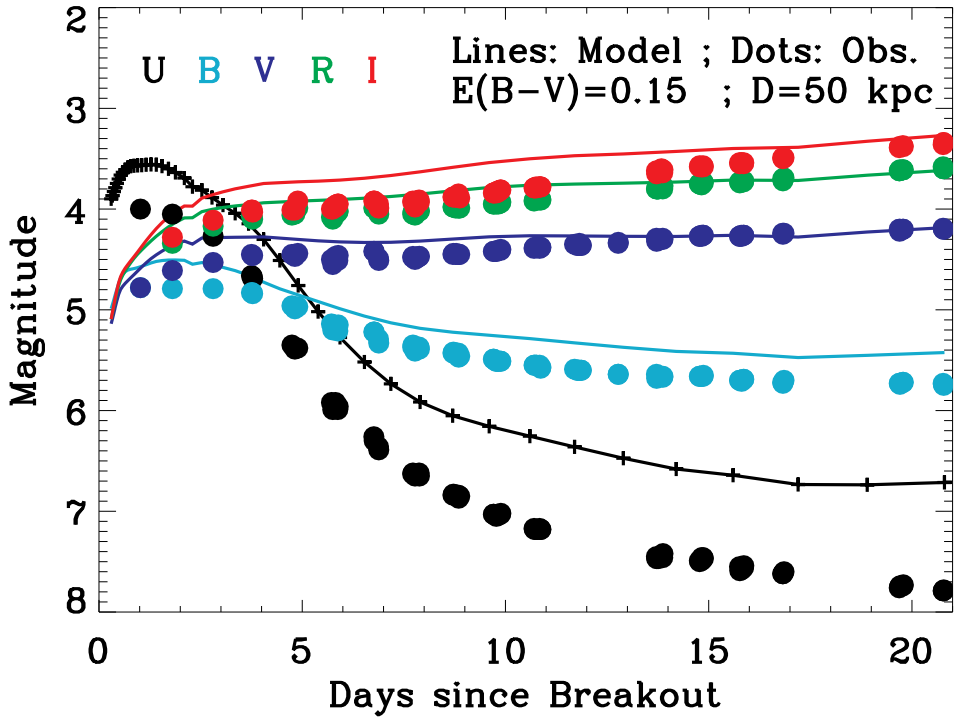,width=15cm}
\caption{Observed (filled circles) and theoretical (lines) $UBVRI$ light curves as a function of time since explosion.
We restrict the time baseline to 21 days to focus on that period over which we compute
our time sequence. The source of observational data is composite, coming from CTIO observations
and IAU circulars for earlier times.
The synthetic photometry was computed by integrating our synthetic spectra over the
Landolt filter bandpasses \citep{landolt_92}, and applying a reddening of
$E(B-V)=0.15$\,mag and scaling for a distance of 50\,kpc.
We plot crosses along the synthetic $U$-band light curve to indicate the epochs
at which our calculations are carried out. Notice the excellent match between theory and observations
for the $BVRI$ bands. For the $U$-band, our values lie above the observed ones at all times
(see text for discussion). \label{fig_LC}
}
\end{figure*}

The luminosity plateau and subsequent increase after $\sim$7\,d occurs at a time when the $^{56}$Ni energy
which, in this model, we deposit at depths below $\sim$2000\,\kms (there is a buffer of
about 10\,\msun\, between the photosphere and the $^{56}$Ni-rich regions at that time), has not had time
to diffuse out. The ``instantaneous" diffusion time for these inner and very dense layers is indeed
orders of magnitude larger than the age of the SN. Hence, this plateau and subsequent re-brightening
is not caused/triggered
by energy deposition from unstable isotopes, but results instead from the rate at which the photospheric
radius and temperature happen to evolve in the shock-heated envelope. Seven days after explosion
corresponds to the time in our simulations when the photospheric temperature becomes essentially constant
at $\gtrsim$5000\,K, which corresponds to conditions under which hydrogen recombines to its neutral state.
At the same time, the photospheric radius continues to grow and the flux has to go up (see
Fig.~\ref{fig_phot} for an illustration of those trajectories). The photospheric velocity at that time is $\sim$9000\,\kms
and there is no dramatic change in density gradient at that location. This point of inflection in the bolometric light
curve would likely be shifted for progenitor stars with different initial radii or for explosions of differing magnitude.

    \section{Comparison to observations}
\label{sect_comp_to_obs}

   Having discussed the radiative properties of the synthetic spectra of our SN1987A time sequence,
we now confront these to observations. We use the observational dataset presented in
\citet{phillips_etal_88}, complemented with early-time photometric data published in
IAU circulars \citep{87A_iauc1,87A_iauc2}, and the UV spectra obtained with IUE and
presented in \citet{pun_etal_95}.
We scale our synthetic spectra adopting a distance of 50\,kpc to the LMC
 \citep{LMC_dist1,LMC_dist2,LMC_dist3,LMC_dist4}.
When corrected for extinction, as when we compare to observed spectra and magnitudes, we employ
a reddening $E(B-V)=0.15$ and the reddening law of \citet{cardelli_etal_89}.
This seems adequate for obtaining a satisfactory match to observed
spectra, although this value could be altered by $\pm$0.05 and remain suitable.

    \subsection{Observed versus synthetic multi-band light curves}

In Fig.~\ref{fig_LC}, we show the observed (filled circles) and synthetic (lines and/or crosses; adopting a reddening of $E(B-V)=0.15$
and a distance of 50\,kpc) $UBVRI$ light curves for SN 1987A. The magnitudes are computed by
integrating the synthetic spectra over the Landolt filter-transmission functions \citep{landolt_92}.
We obtain an excellent agreement for the optical bands $BVRI$, but our predictions for the $U$-band magnitude
is offset, predicted too bright at all times. This discrepancy could be somewhat reduced, although not resolved,
by choosing a slightly higher reddening of 0.2\,mag.
The $U$-band magnitude decrease is also too slow in our simulations, and levels at a minimum that is
$\sim$1-1.5\,mag too bright. This mismatch may be physical, perhaps reflecting a slight offset in the density distribution
in the progenitor surface layers. The progenitor star model should perhaps have a slightly smaller radius, resulting in a faster
evolution of the SED to longer, optical, wavelengths.
It may also be associated with opacity issues, stemming
from under-abundant but nonetheless important species we presently ignore.
For example,  we are not including Al, which has strong resonance lines
in the UV, or, e.g., Cr, Ar, Ne which do contribute opacity there.
We may be missing some low ionization
stages, e.g., Ni\,{\sc ii} or Fe\,{\sc i}, although the $U$-band magnitude mismatch is present at all times, not just at late times
when the photospheric conditions are cool. Finally, we may be using a too small model atom for some species. For example,
we include only 115 levels of Fe\,{\sc ii}, which we find yields converged results at early times,
but is not quite sufficient when Fe\,{\sc ii} is the dominant Fe ion (we find that enhancing the number of Fe\,{\sc ii} levels
from 115 to 827 at the recombination epoch (past day 5) leads to a reduction of the already extremely small
flux in the 1000-2000\AA\ range, but leaves the spectrum at longer wavelengths unaffected).
Recall here that iron has a mass fraction of
less than 0.001 in our model, while hydrogen is the most abundant element and is the main donor of the
free-electrons that provide the bulk of the opacity.
In any case, the mismatch cannot be associated with something sizably inadequate in the progenitor model
or in the code since we obtain excellent agreement for the optical bands. It also intervenes in regions where the flux
is orders of magnitude smaller than in the optical, or orders of magnitude smaller than what it used to be at 0.3\,d.
But this goes to show that even in the case of SN1987A, for which we have tight constraints on the progenitor
star and the explosion properties, obtaining a complete agreement is a challenge.

  \subsection{Observed versus synthetic spectra}

\begin{landscape}
\begin{figure}
\includegraphics[width=11cm]{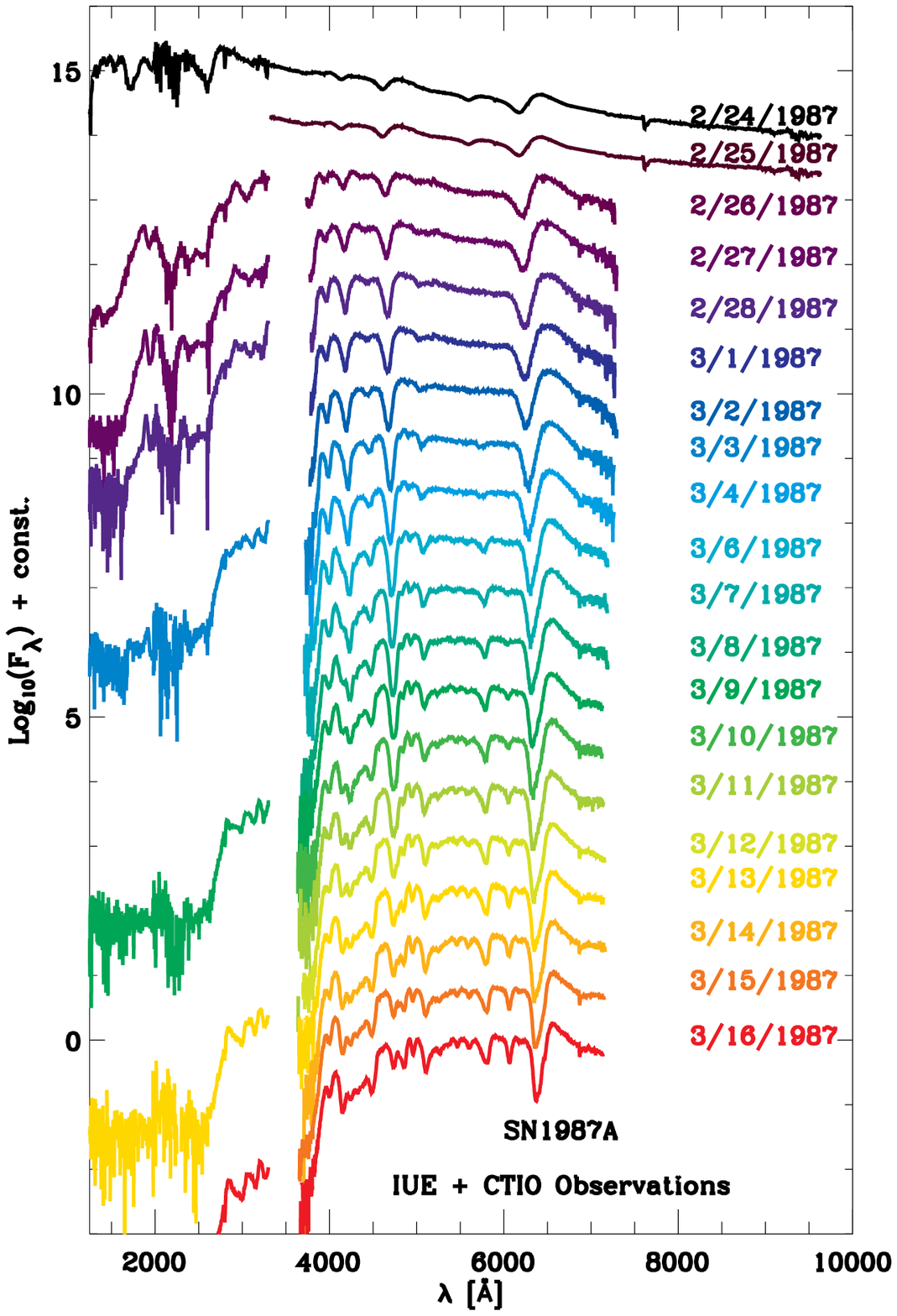}
\includegraphics[width=11cm]{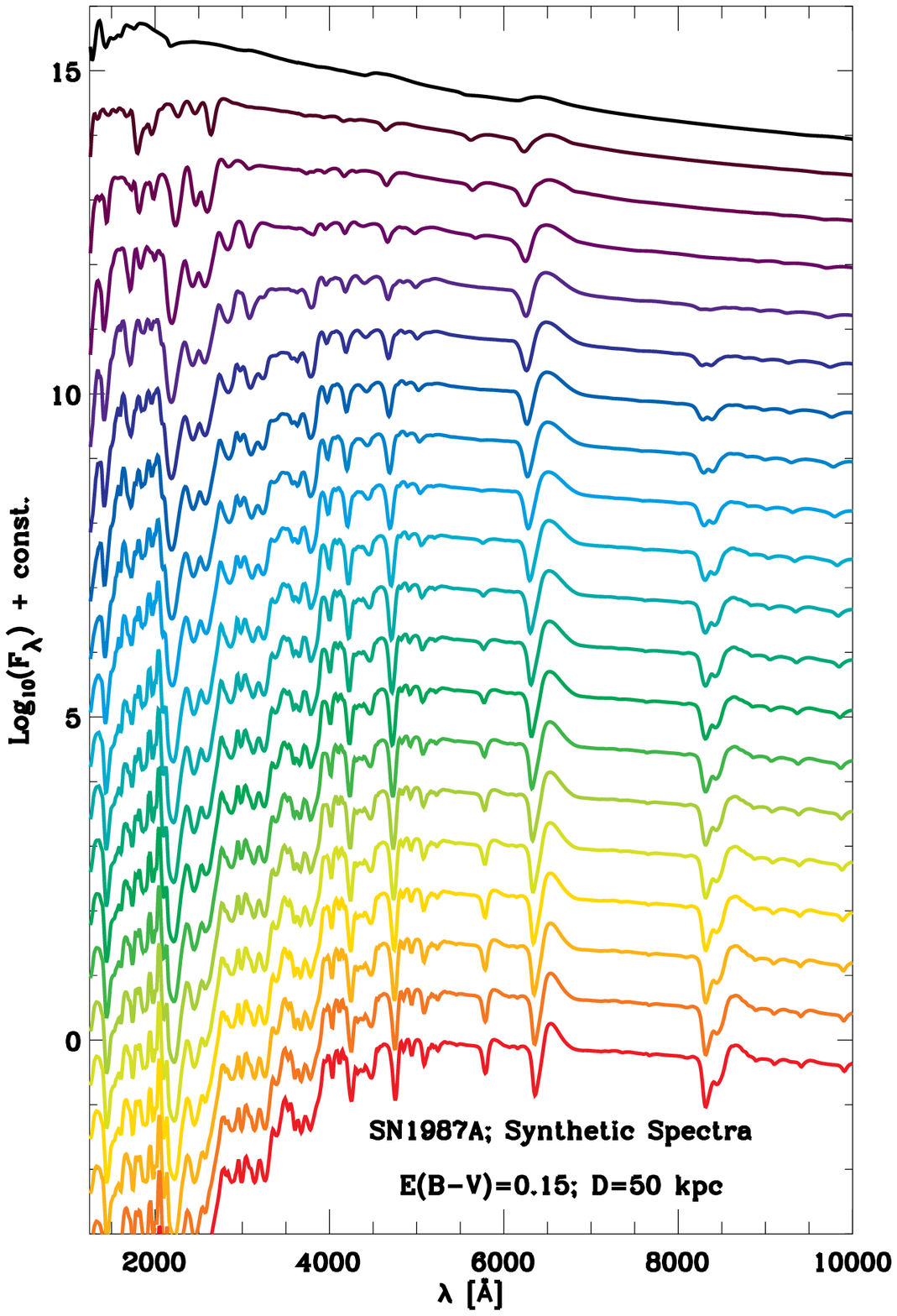}
\caption{{\it Left:} Montage of observed spectra for SN1987A, including the optical at all times
and the UV when available \citep{phillips_etal_88,pun_etal_95}. For visibility purposes, we apply a scaling between epochs,
ordering the spectra chronologically starting at the top. A color coding is used to differentiate each date.
{\it Right:} Same as left, but now showing our synthetic spectra, reddened
with $E(B-V)=0.15$ and scaled
for a distance of 50\,kpc. Since our time-sequence was not computed for the specific observed times, we interpolate
in between the computed epochs to make the correspondence exact in time between left and right panels.
Moreover, we apply the same final scaling, and the same color coding, as in the left panel to facilitate the visibility.
Importantly, apart from a small offset for the first epochs, the observed and theoretical fluxes agree
to within about 10\,\% in the optical (the agreement is not as good in the UV range),
which is quite remarkable given that we started from a hydrodynamical model of the
explosion and just let the ejecta evolve ``naturally'' from day 0.27 onwards.
\label{fig_spect_montage}}
\end{figure}
\end{landscape}

We present in the left panel of Fig.~\ref{fig_spect_montage} a montage of all observed UV and optical spectra
of SN1987A up to 20 days after explosion. These are stacked up vertically and shown on a logarithmic scale
for visibility. A color coding differentiates the epochs. In the right panel, we show a montage of synthetic
spectra (reddened using $E(B-V)=0.15$ and scaled for a distance of 50\,kpc), interpolated in time from our
time sequence so that the displayed dates match each observational epoch in the left panel. Importantly,
the same vertical shift is applied to both observed and synthetic  spectra for the same epoch.
Overall, the synthetic spectra shown in the right panel mirror with high fidelity the observed spectra shown
in the left panel.

For the first few epochs, echoing what was seen already in Fig.~\ref{fig_LC},
the synthetic fluxes are somewhat larger than observed. The color temperature is slightly too large,
as is the ionization state of the photosphere, so that all lines appear weaker above the continuum.
If physical, this mismatch could reflect the fact that the progenitor star used as an input in this simulation
has a slightly too large initial radius. A more compact progenitor star would induce a faster cooling
and a faster UV fading.

After a few days, the agreement between observations and theory is excellent, in particular
in the optical where both continuum and lines fit to within $\sim$10\% at all times. There are a few exceptions
to this, but H$\alpha$ is one line that embodies this agreement, as it is reproduced in both strength and width.
In the UV, the synthetic flux overestimates the
UV flux by a few tens of percent at most times. This echoes the flux mismatch in the $U$ band although the origin
may not be the same (opacity sources in the UV and around 3000-4000\AA\ differ; see Fig.~\ref{fig_ltau}).
Recall that the UV flux drops by 5 orders of magnitude between 0.3 and 20.8\,d, and that the UV flux
after a few days is a very small fraction of the total flux.
In any case, we will investigate this issue in the future, but here, given the complete lack of free parameters
after we start the time sequence, this agreement is impressive.

   We now present detailed fits to individual observed spectra. For this exercise, we allow ourselves to choose
a synthetic spectrum whose date may be offset by $\pm$0.5\,d from that of the observed date. This matters
at early epochs, when the evolution is fast, and for later dates when our photosphere seems to have somewhat
too high ionization/temperature (the $U$-band flux is visibly too high). Recall that we cannot tune the model
parameters to fit the observations since all the ejecta properties were inherited at the start of the sequence
from the adopted hydrodynamic explosion model.

\begin{figure*}
\epsfig{file=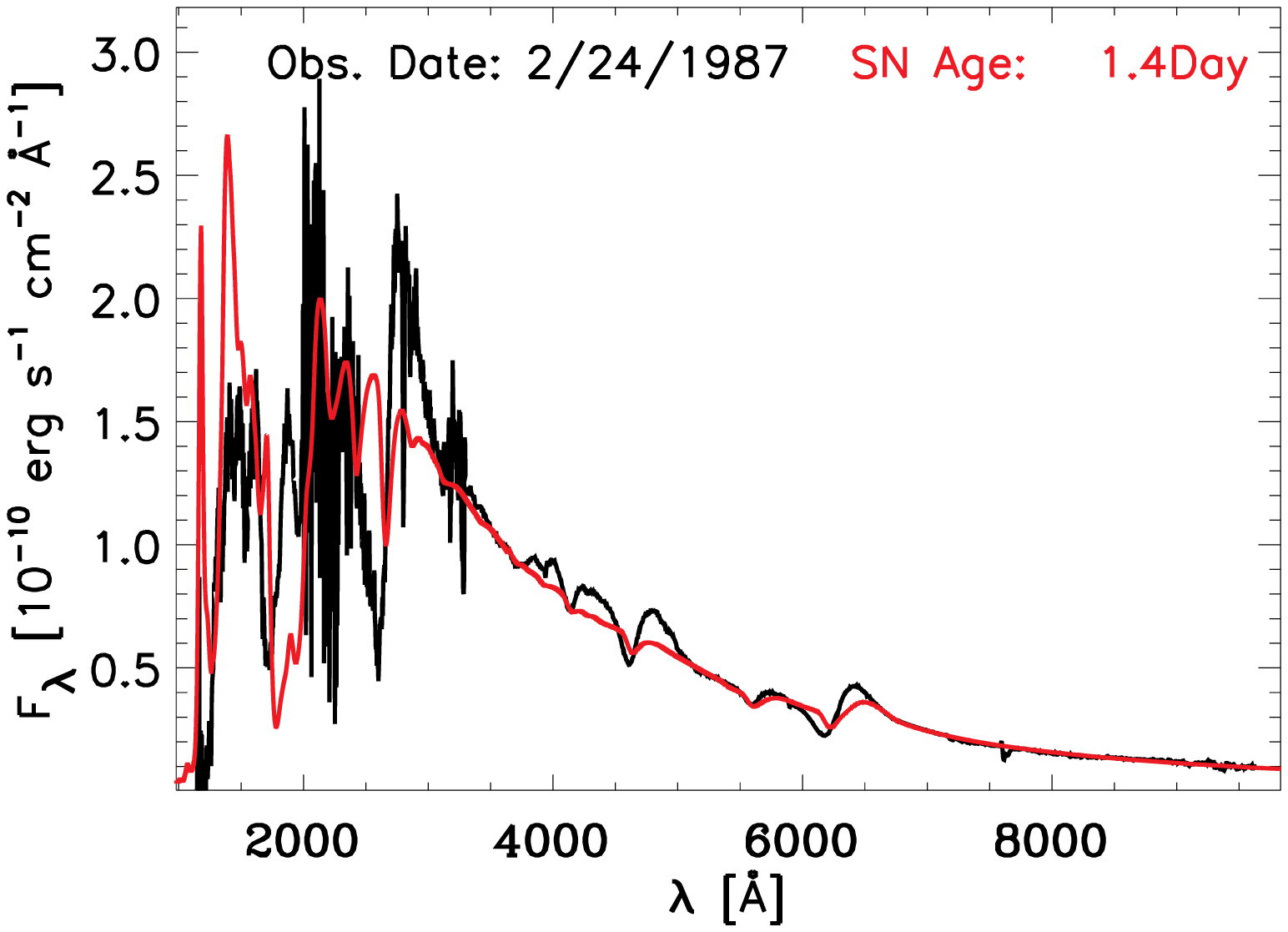,width=8.5cm}
\epsfig{file=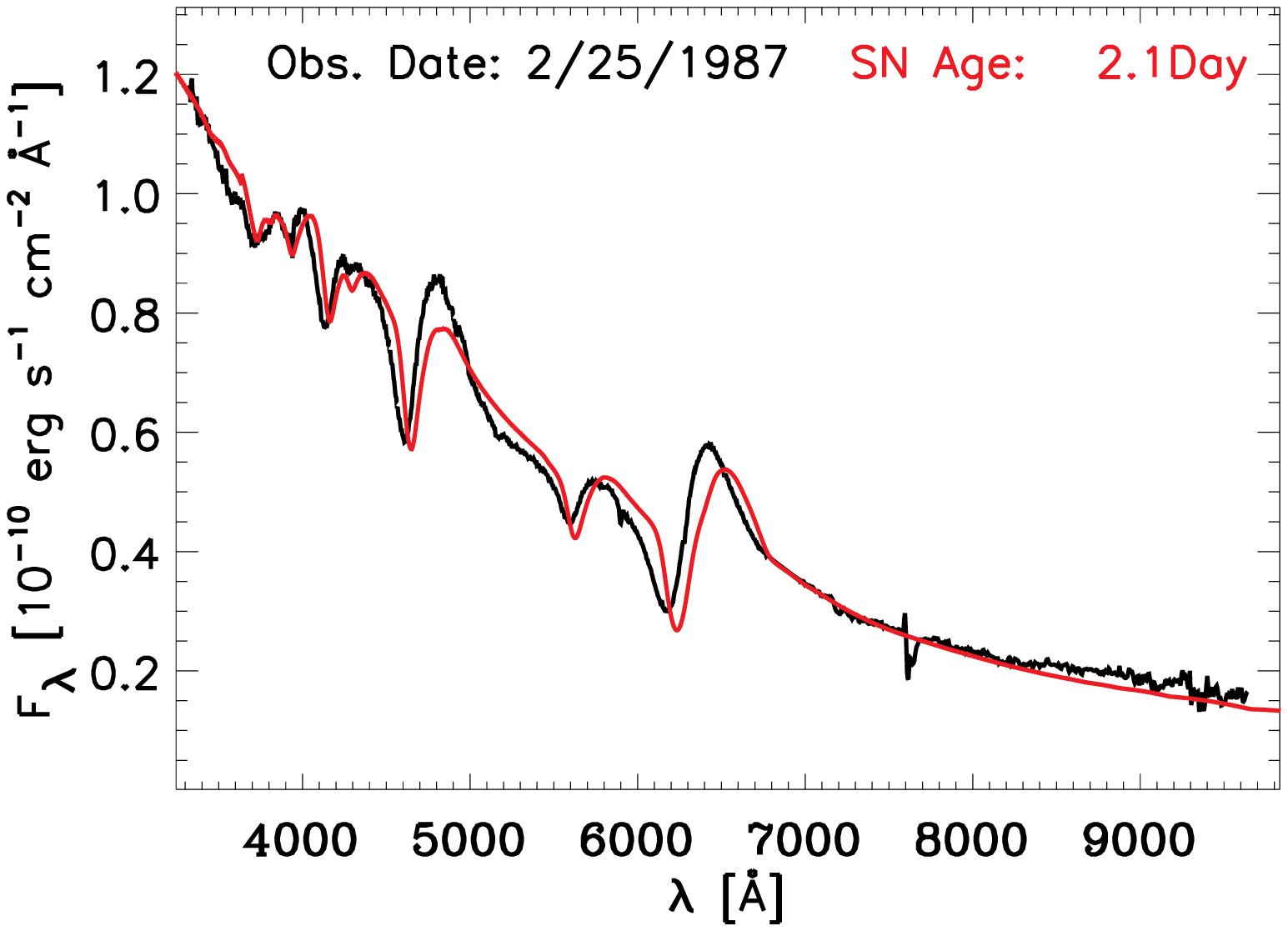,width=8.5cm}
\epsfig{file=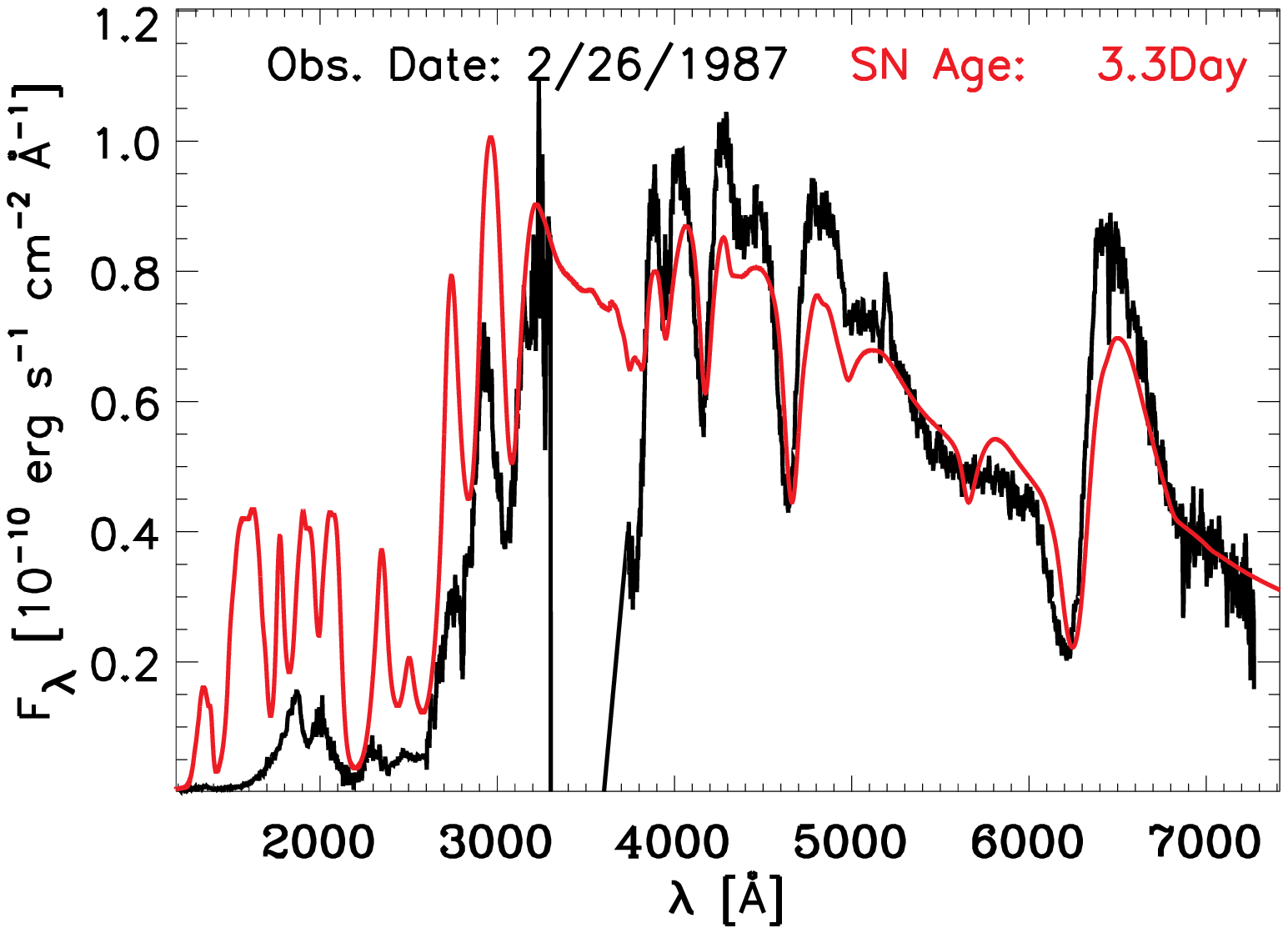,width=8.5cm}
\epsfig{file=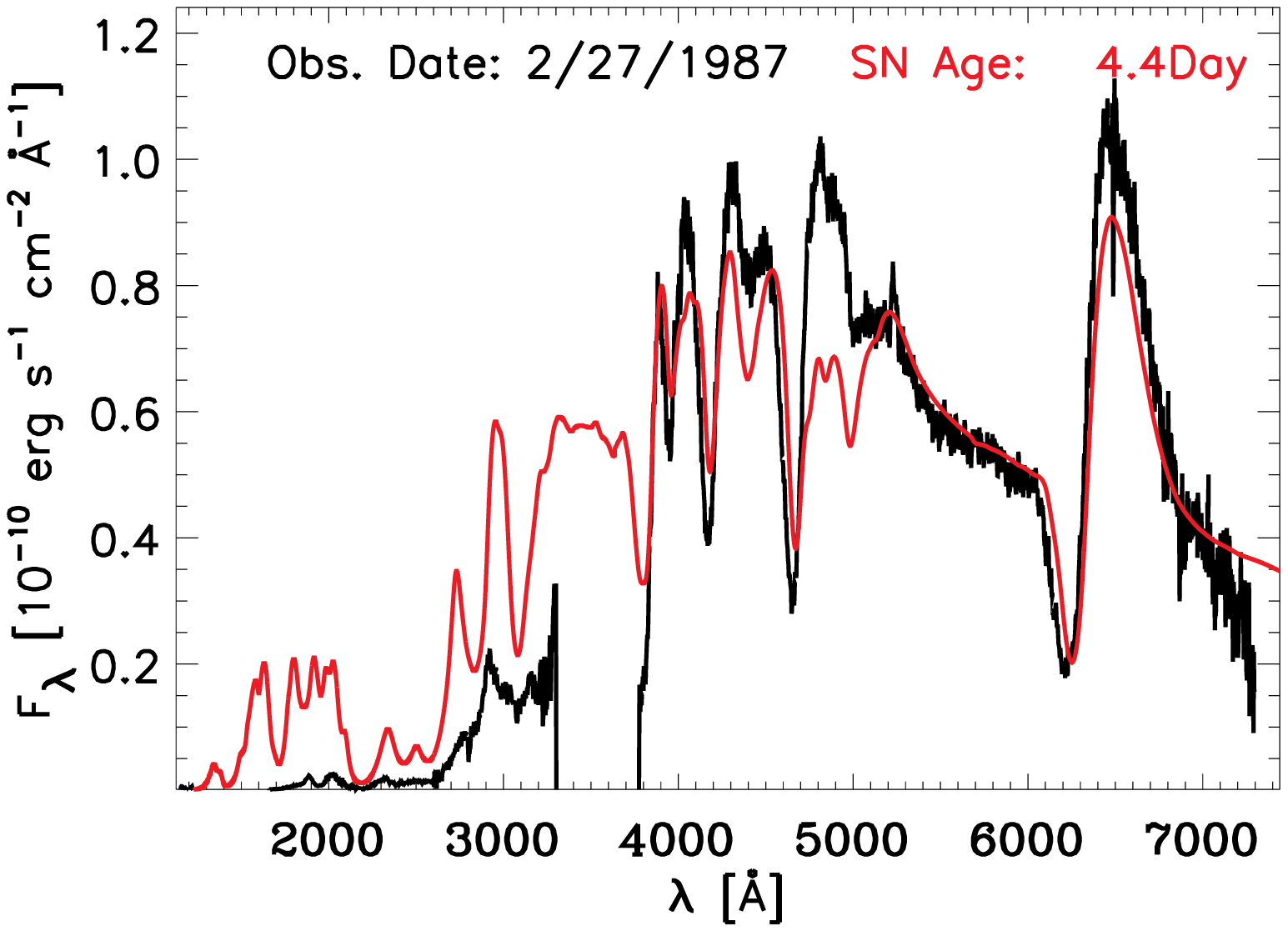,width=8.5cm}
\epsfig{file=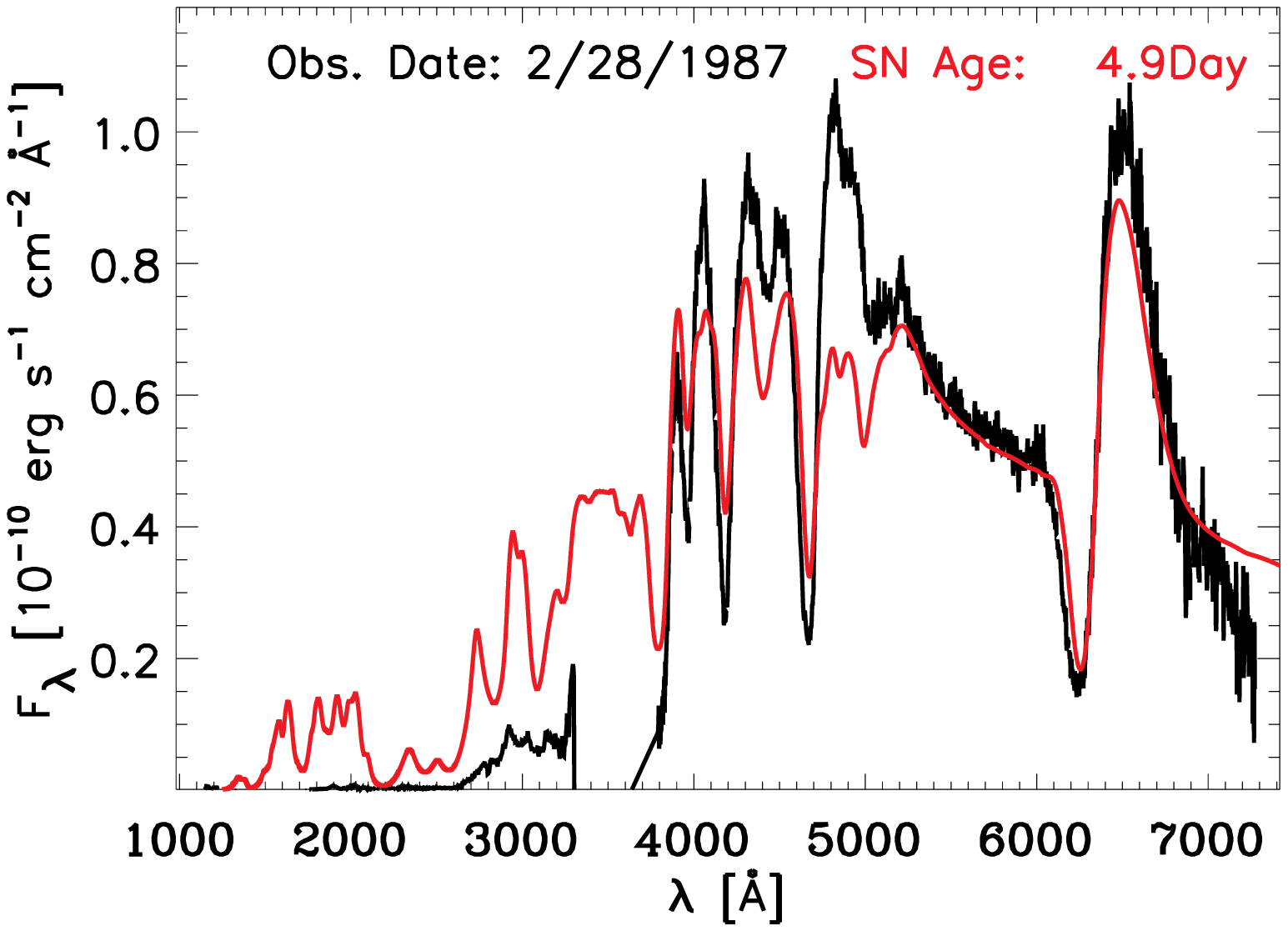,width=8.5cm}
\epsfig{file=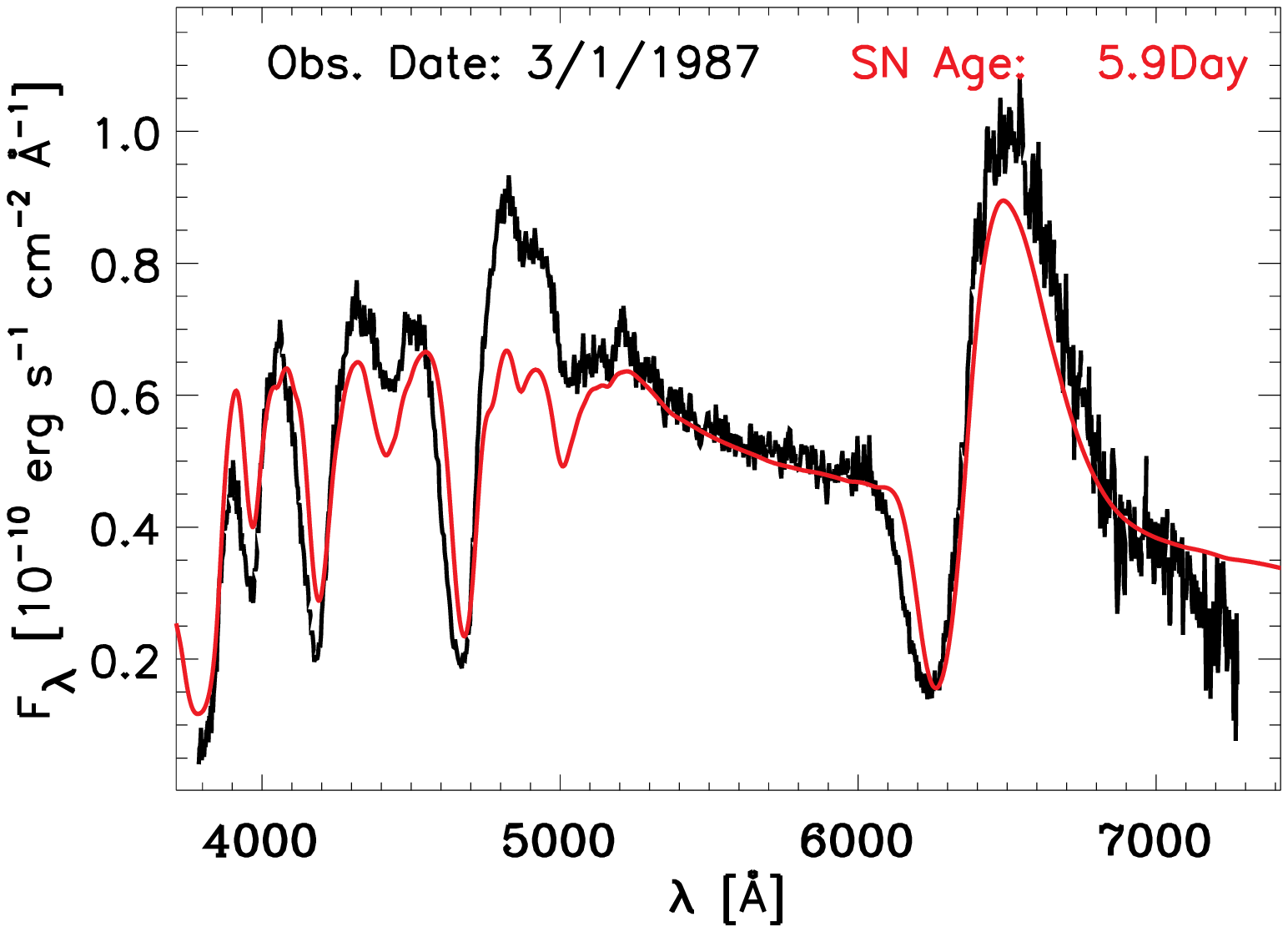,width=8.5cm}
\caption{Comparison between observed (black) spectra for SN1987A and our synthetic fits (red; reddened with $E(B-V)=0.15$ and
scaled for a distance of 50\,kpc) from our non-LTE time-dependent sequence, covering the epochs from the 24th of February
until the first of March 1987. A small absolute scaling of at most 30\% is applied to the synthetic spectra to bring them down to the level
of the observations (see text for discussion).
\label{fig_comp_spec1}
}
\end{figure*}

The first two spectra obtained for SN1987A on the 24th and the 25th of February 1987 show a blue continuum,
which represents a fossil of the conditions at breakout (top row of Fig.~\ref{fig_comp_spec1}). We see
broad and strong Balmer lines as well as the conspicuous presence of He\,{\sc i} 5877\AA, reproduced
here with the helium mass fraction prevailing at the surface of the BSG progenitor star. Hence, in contrast
with the previous results of \citet{EK89_87A,SAR90_87A, hoeflich_87,hoeflich_88}, it is indeed possible to reproduce this line
with an abundance typical of a BSG progenitor in the LMC.
While it is difficult to identify the specific ingredient(s) resolving this former discrepancy, this highlights
the need for realistic radiative-transfer models for an accurate reproduction of SN spectra, accounting
for non-LTE effects, line blanketing, etc.

The fit to observations is fair given the very fast evolution
of the photospheric conditions at these early times and the strong dependency on progenitor properties (in particular
radius). More frequent observations over these initial two days would have revealed further detail about this fast evolution
and provided additional constraints for the modeling.

  By the third observation on the 26th (middle-left panel in Fig.~\ref{fig_comp_spec1}),
  He\,{\sc i}\,5877\AA\  is gone and the first signs of metal lines in the optical
  appear (e.g., Fe\,{\sc ii}\,5169\AA). In our calculations, we still predict He\,{\sc i}\,5877\AA\
  (although it is no longer present in our time sequence one day later).
  Another sign that the ionization is too high at such times is the overestimated UV flux.
  The H$\alpha$ line profile is well fitted, although the emission part is somewhat underestimated.
  Recall that the early evolution is fast and that a very optically-thick line like H$\alpha$ is  sensitive
  to even slight modulations in ionization (ionization is changing rapidly at this epoch as hydrogen recombines
  in the photospheric layers). Interestingly, the H$\beta$ line is not as well fitted as H$\alpha$, although its
  formation process is analogous, and this discrepancy lasts for about a week.

  Continuing onwards, the ejecta recombine at the photosphere and the UV flux essentially vanishes.
  We fit the H$\alpha$ line profile and its evolution very well,
   both quantitatively in strength and width, but also qualitatively in its shape (i.e., how the flux
   varies from the blue edge to the red edge of the profile). This suggests that, together, our calculations and the initial
   model allow an accurate description of the H$\alpha$
   line-formation process; the density distribution and the kinetic energy of the original model are well suited.
   This contrasts with the former discrepancy in Balmer-line strengths encountered by \citet{SAR90_87A}, which
   at the time were associated with the potential role of clumping (note that no clumping is used here).
   The good reproduction of Balmer-line strengths or Na\,{\sc i}\,D supports the adequacy of our computed non-LTE time-dependent
   ionization structure.
   As pointed out by \citet{UC05_time_dep}, and independently confirmed by \citet{DH08_time},
   the competing recombination and expansion timescales
   for the ejecta lead to an ionization freeze-out, particularly visible in the optically-thin fast-expanding low-density regions.
   We capture this effect by including  time-dependent terms in the statistical-equilibrium and energy equations;
   this important aspect requires a full non-LTE treatment.

\begin{figure*}
\epsfig{file=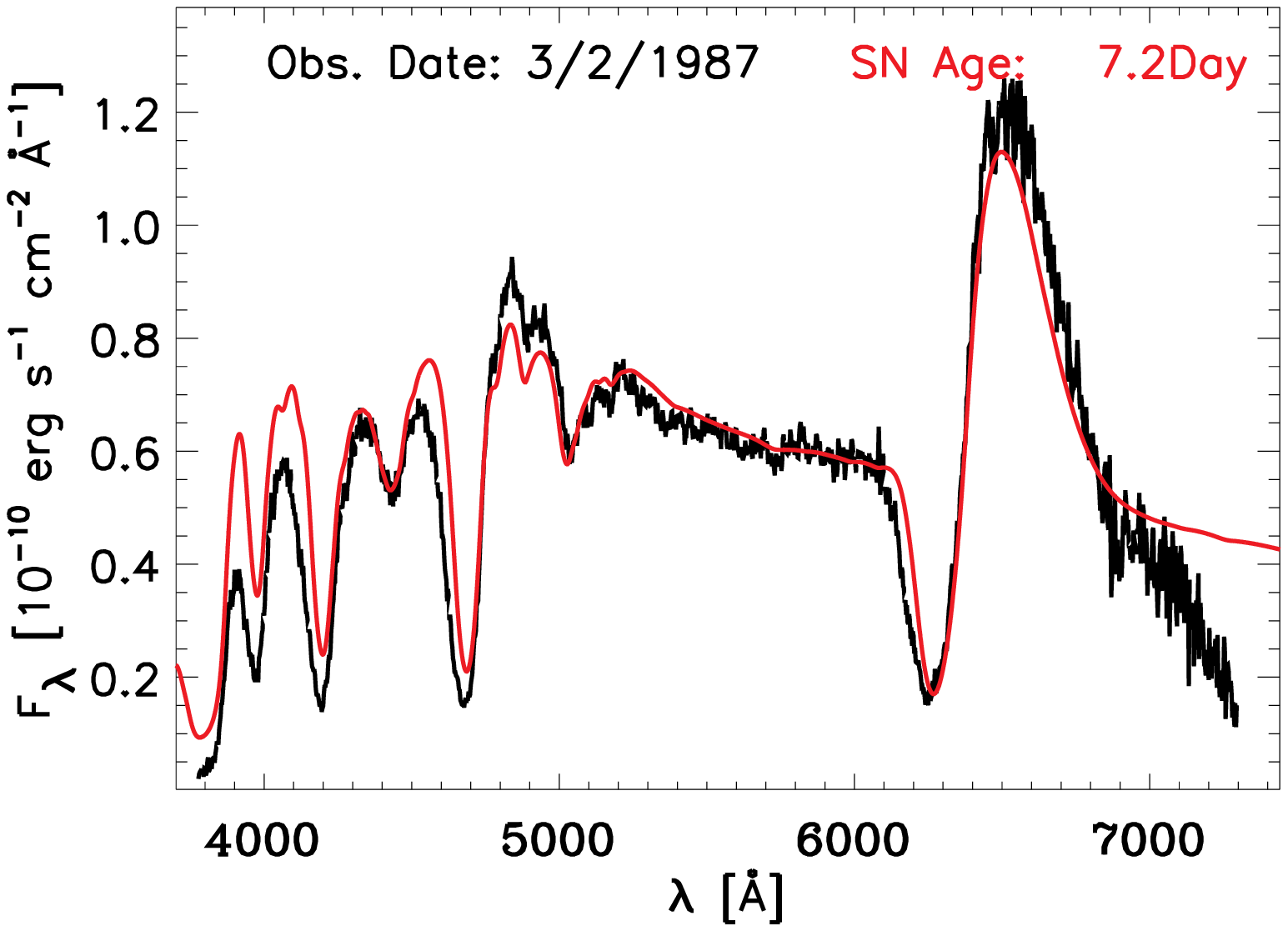,width=8.5cm}
\epsfig{file=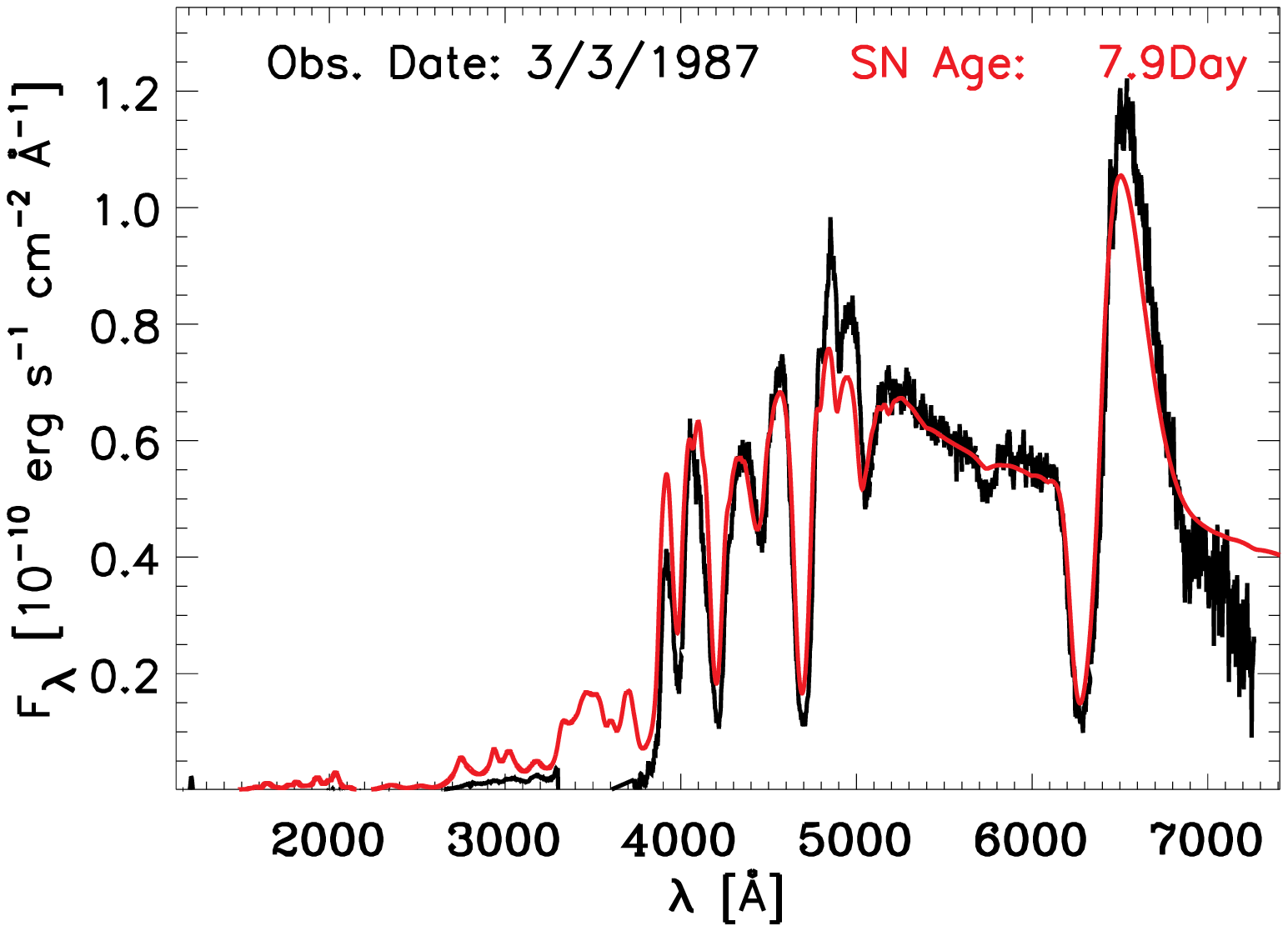,width=8.5cm}
\epsfig{file=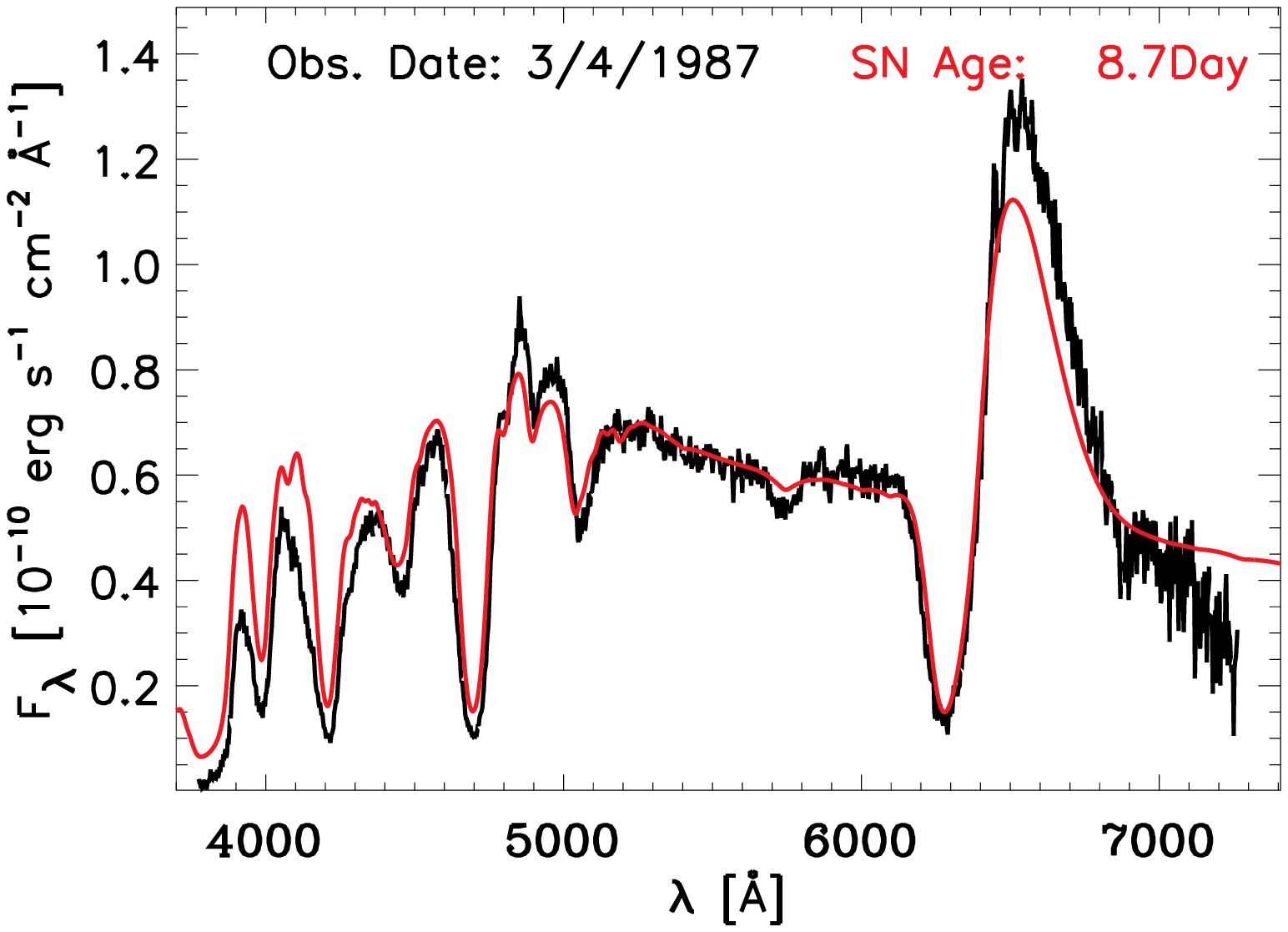,width=8.5cm}
\epsfig{file=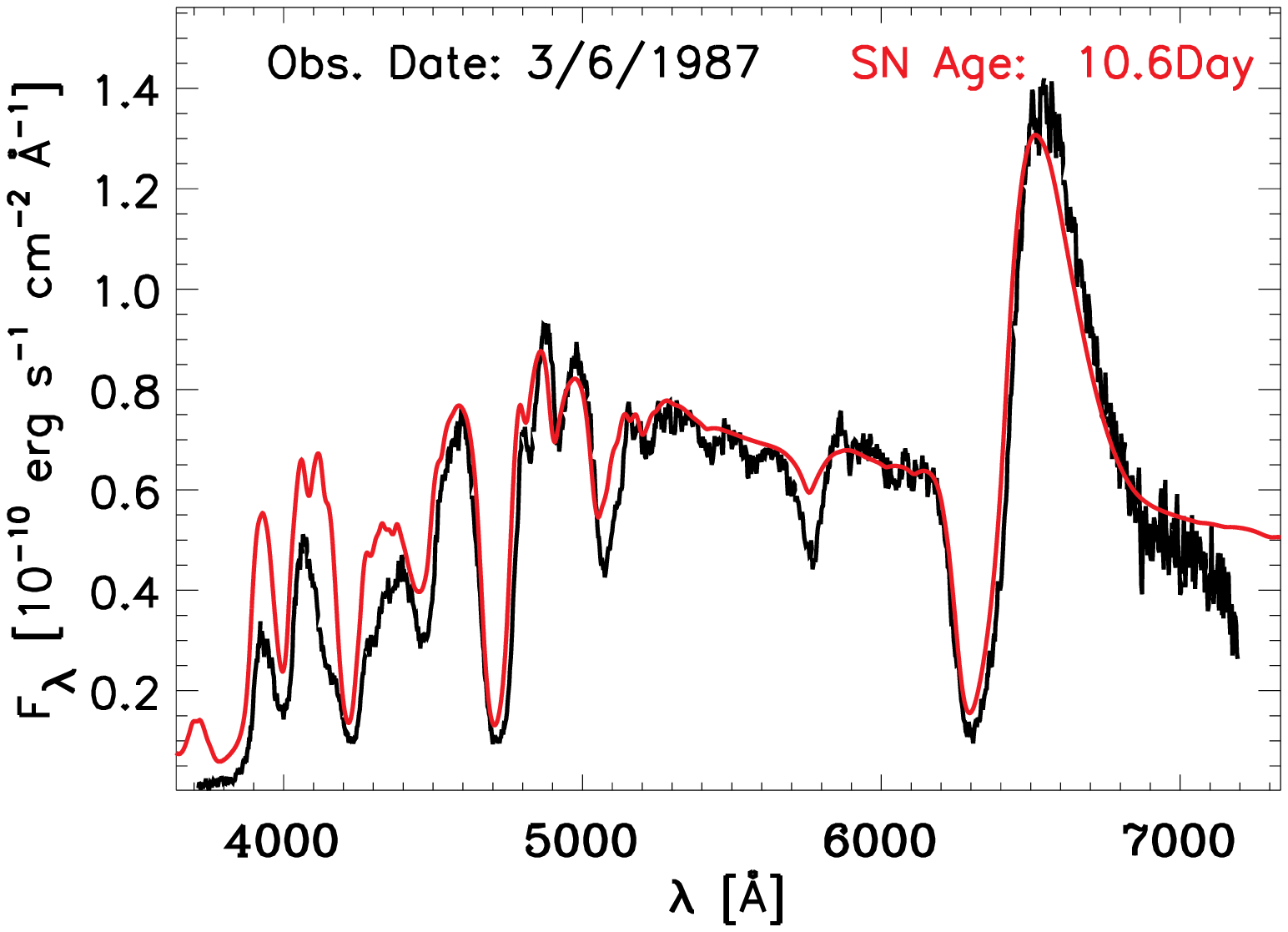,width=8.5cm}
\epsfig{file=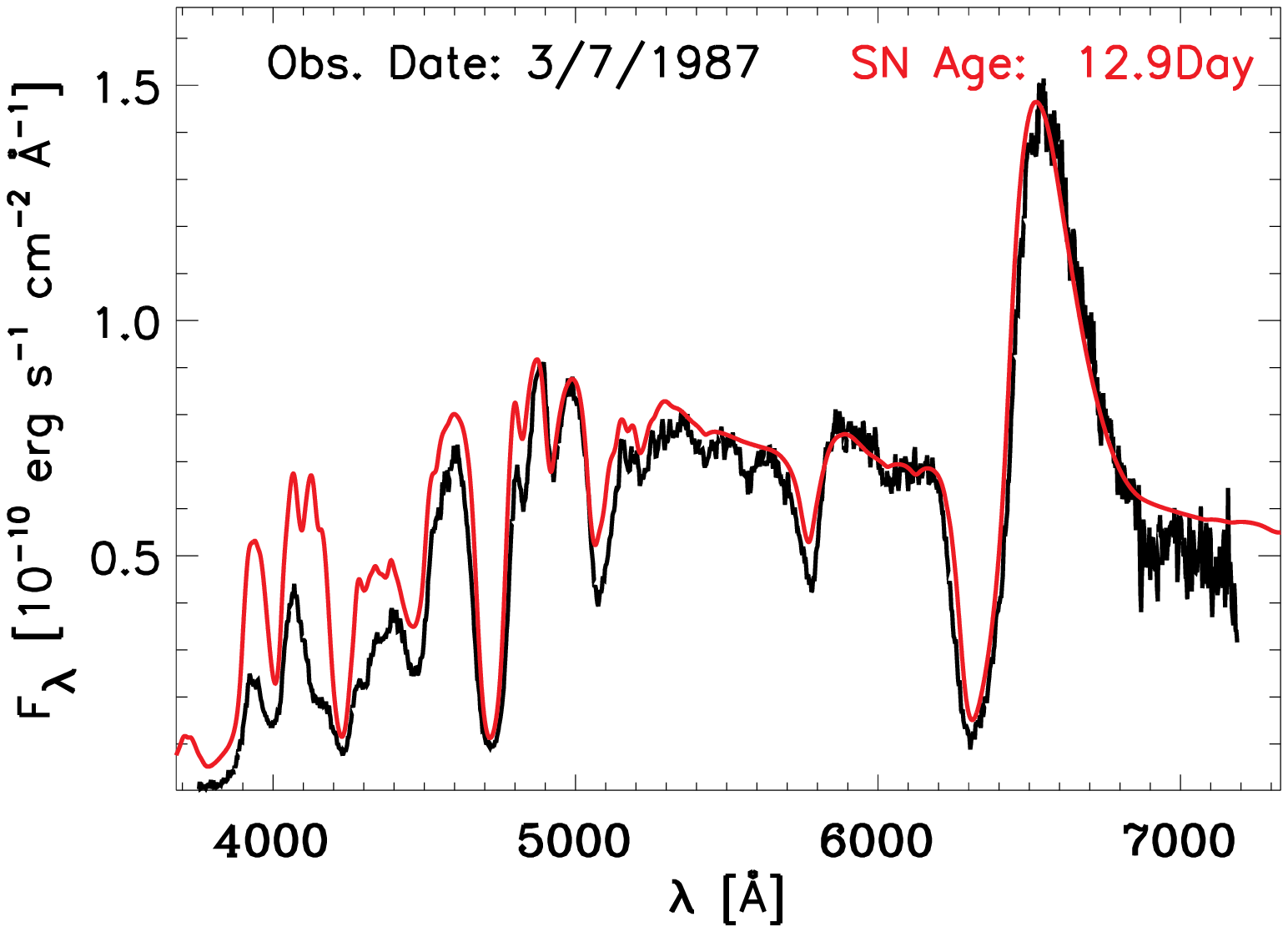,width=8.5cm}
\epsfig{file=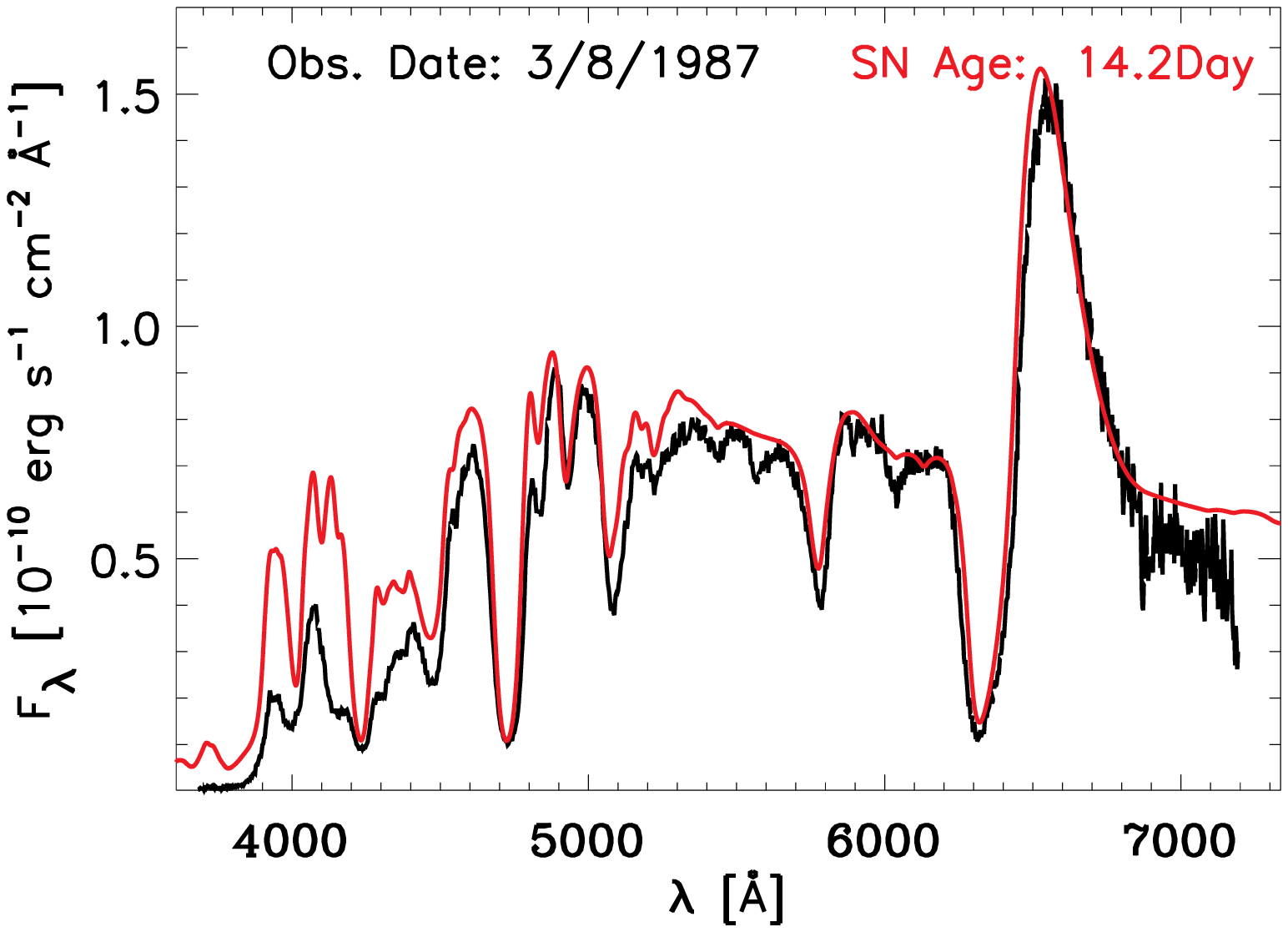,width=8.5cm}
\caption{Same as Fig.~\ref{fig_comp_spec1}, but showing dates between the 2nd and the 8th of March 1987.
The reddened and distance-scaled synthetic spectra are within a few percent at most of the observations.
\label{fig_comp_spec2}
}
\end{figure*}

\begin{figure*}
\epsfig{file=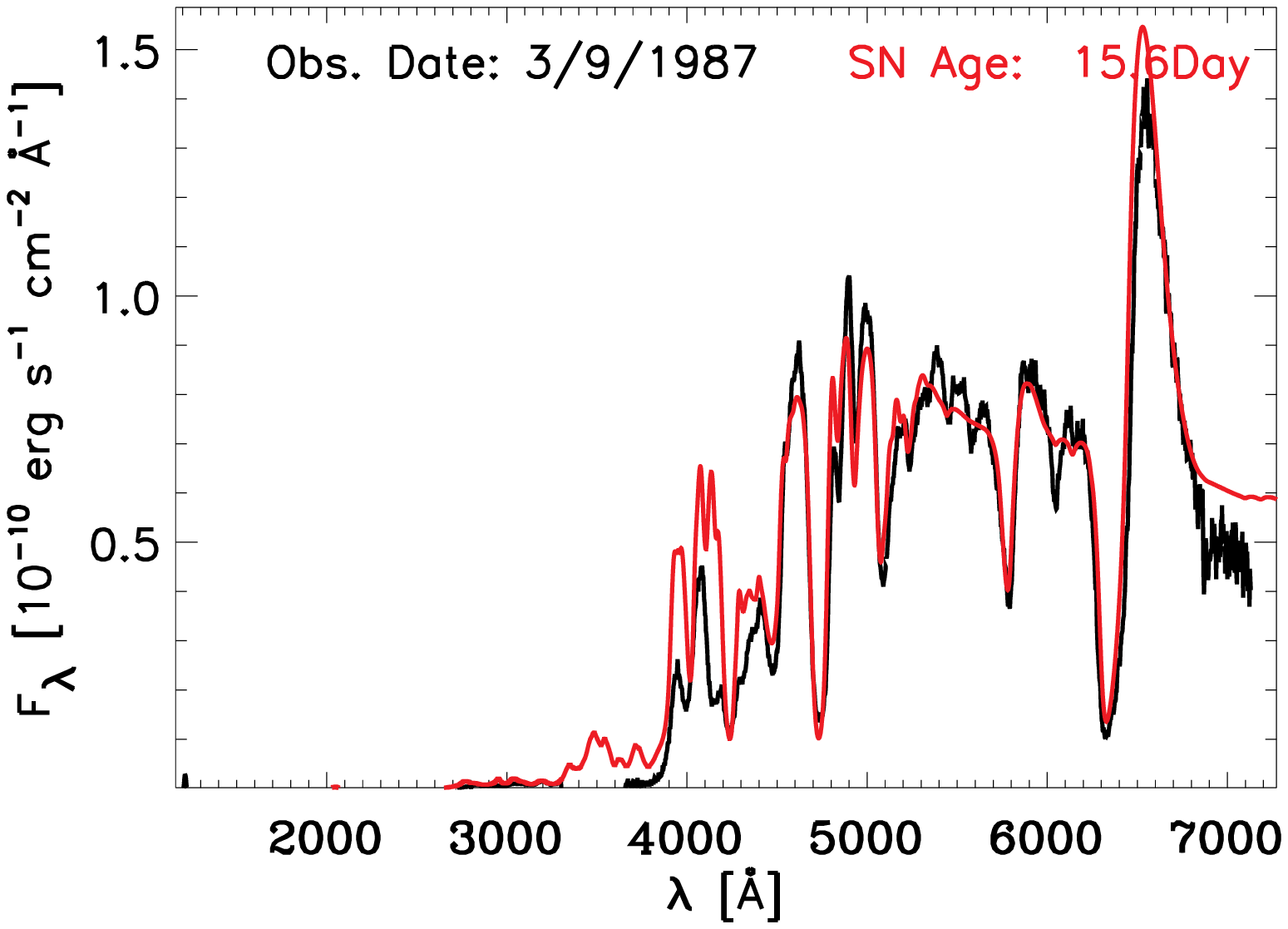,width=8.5cm}
\epsfig{file=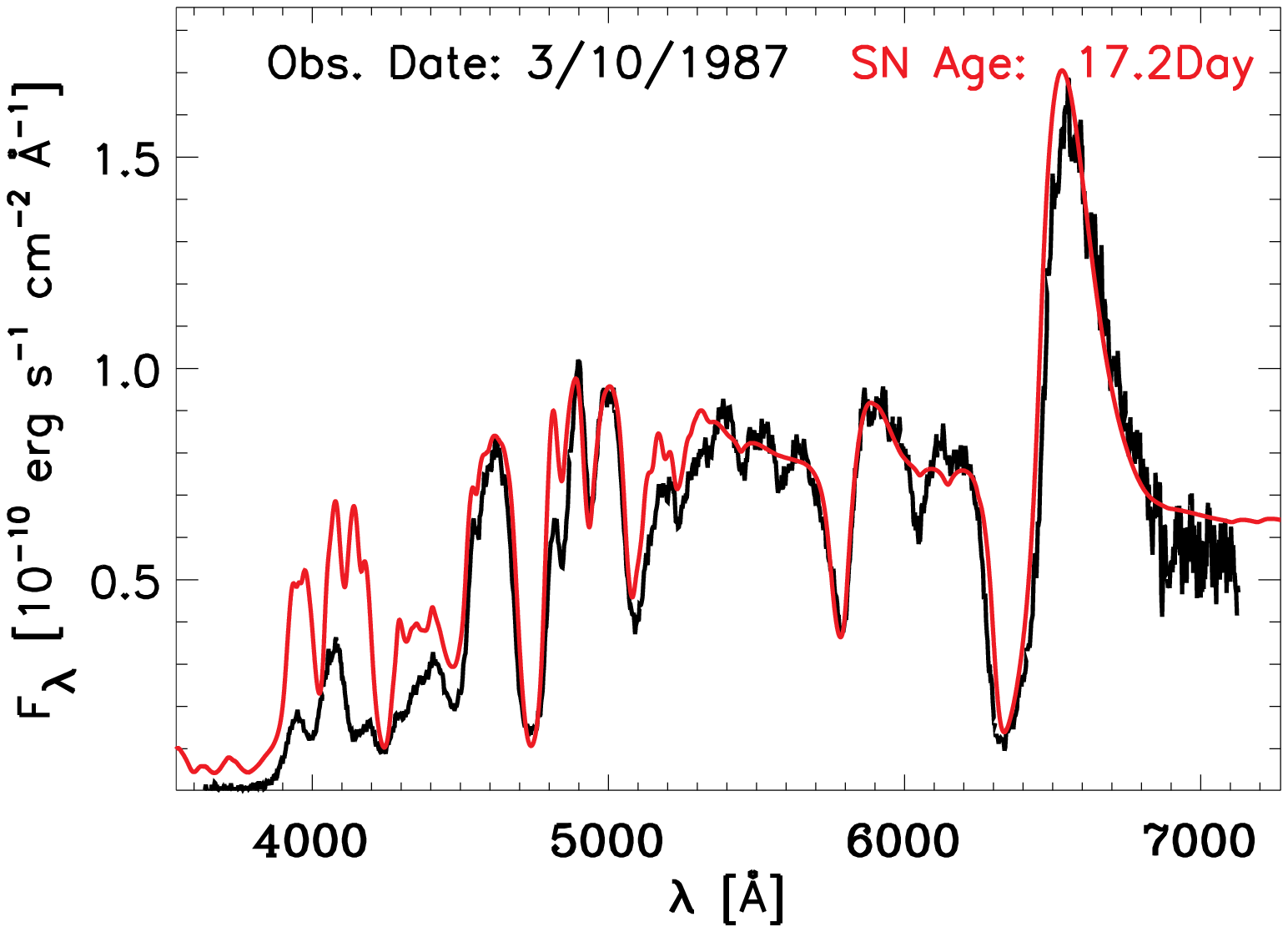,width=8.5cm}
\epsfig{file=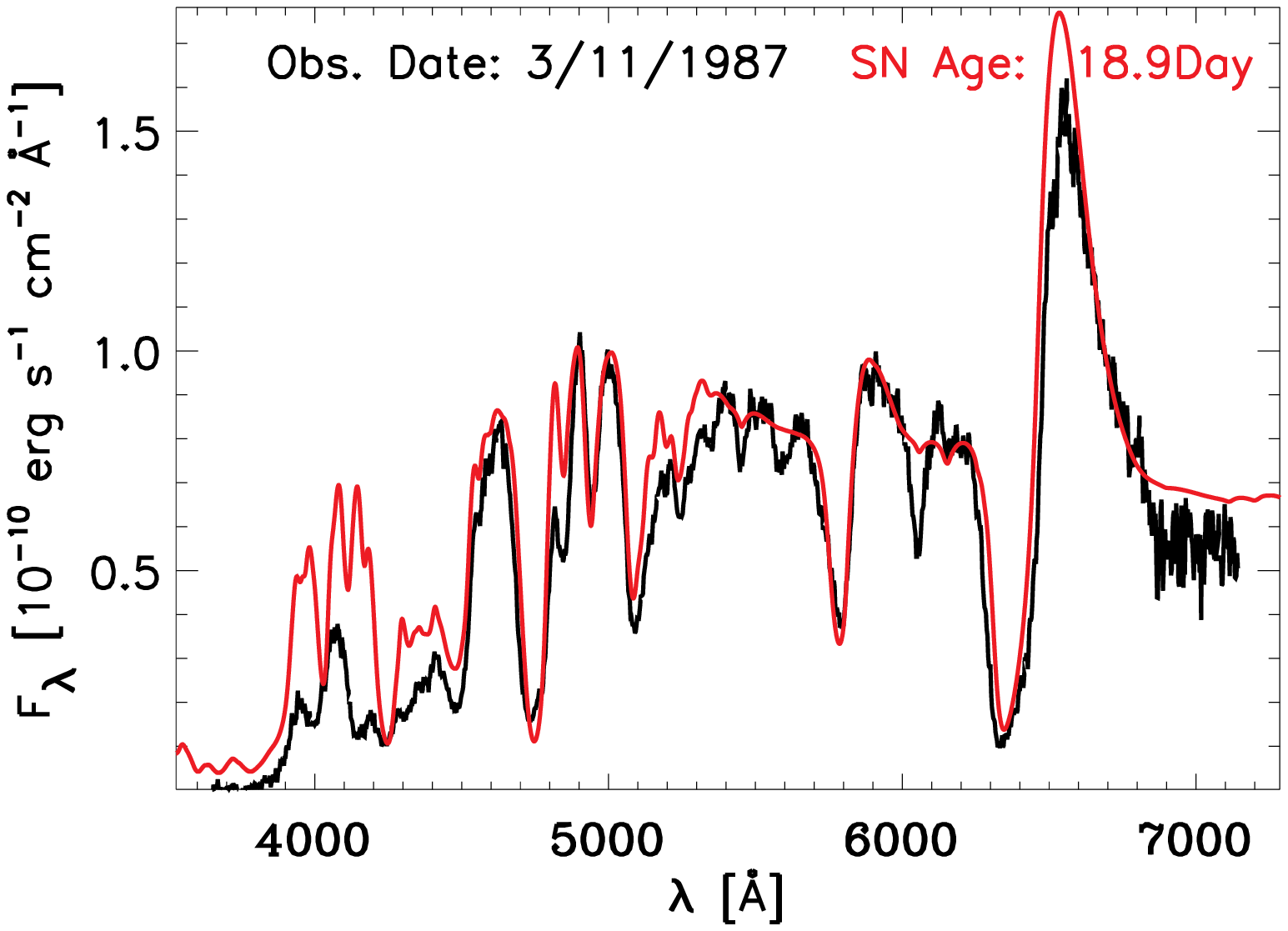,width=8.5cm}
\epsfig{file=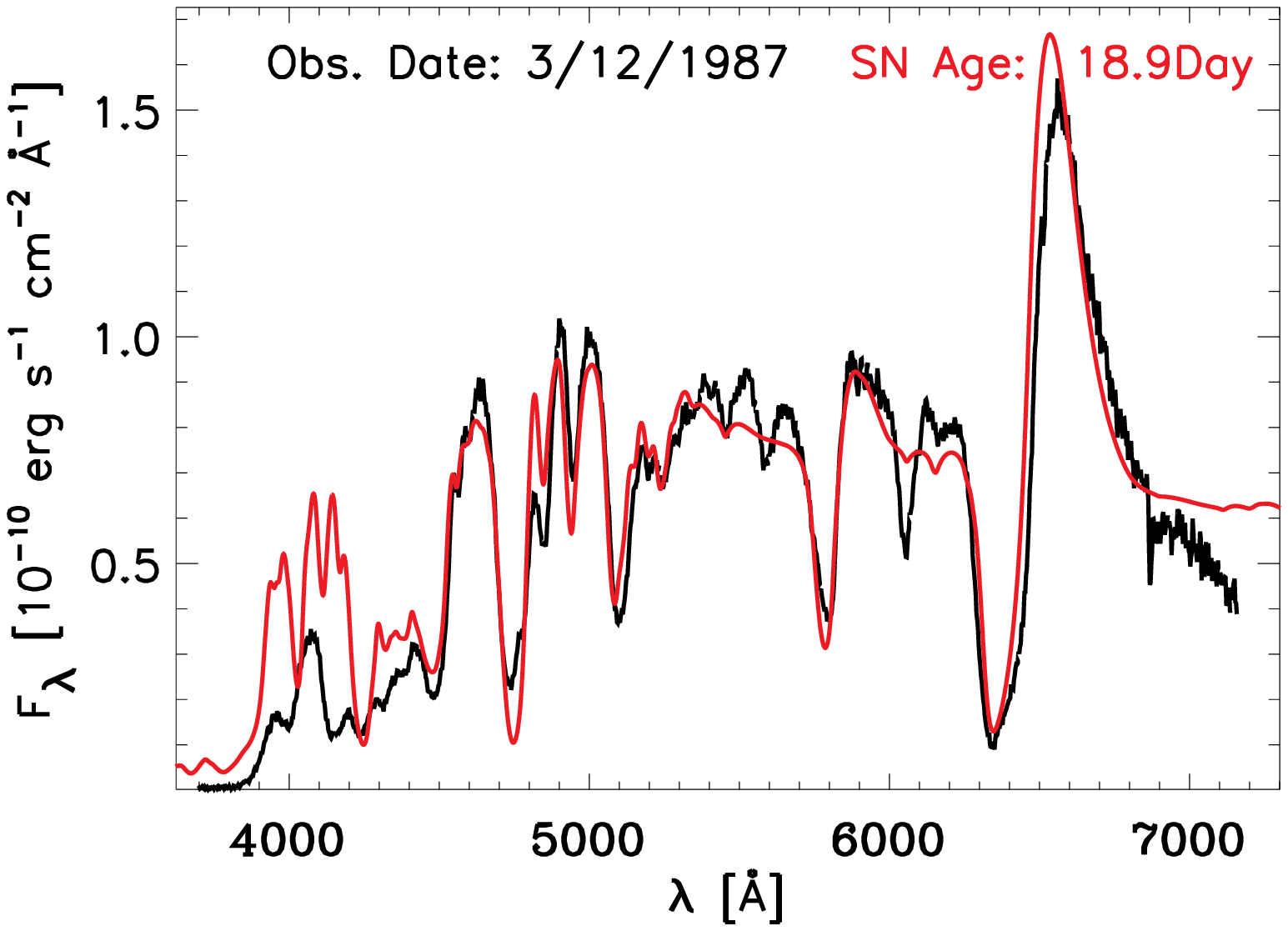,width=8.5cm}
\epsfig{file=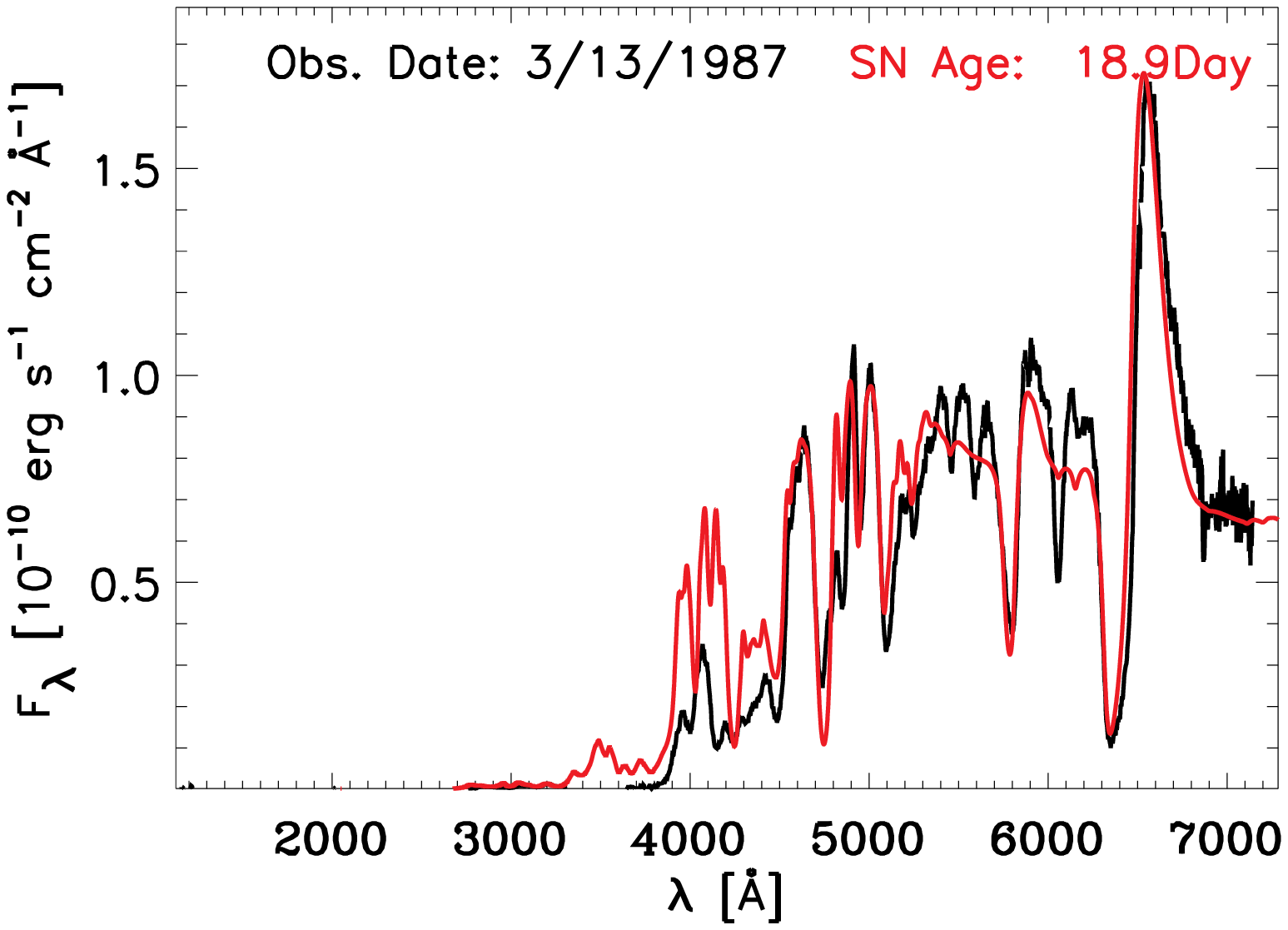,width=8.5cm}
\epsfig{file=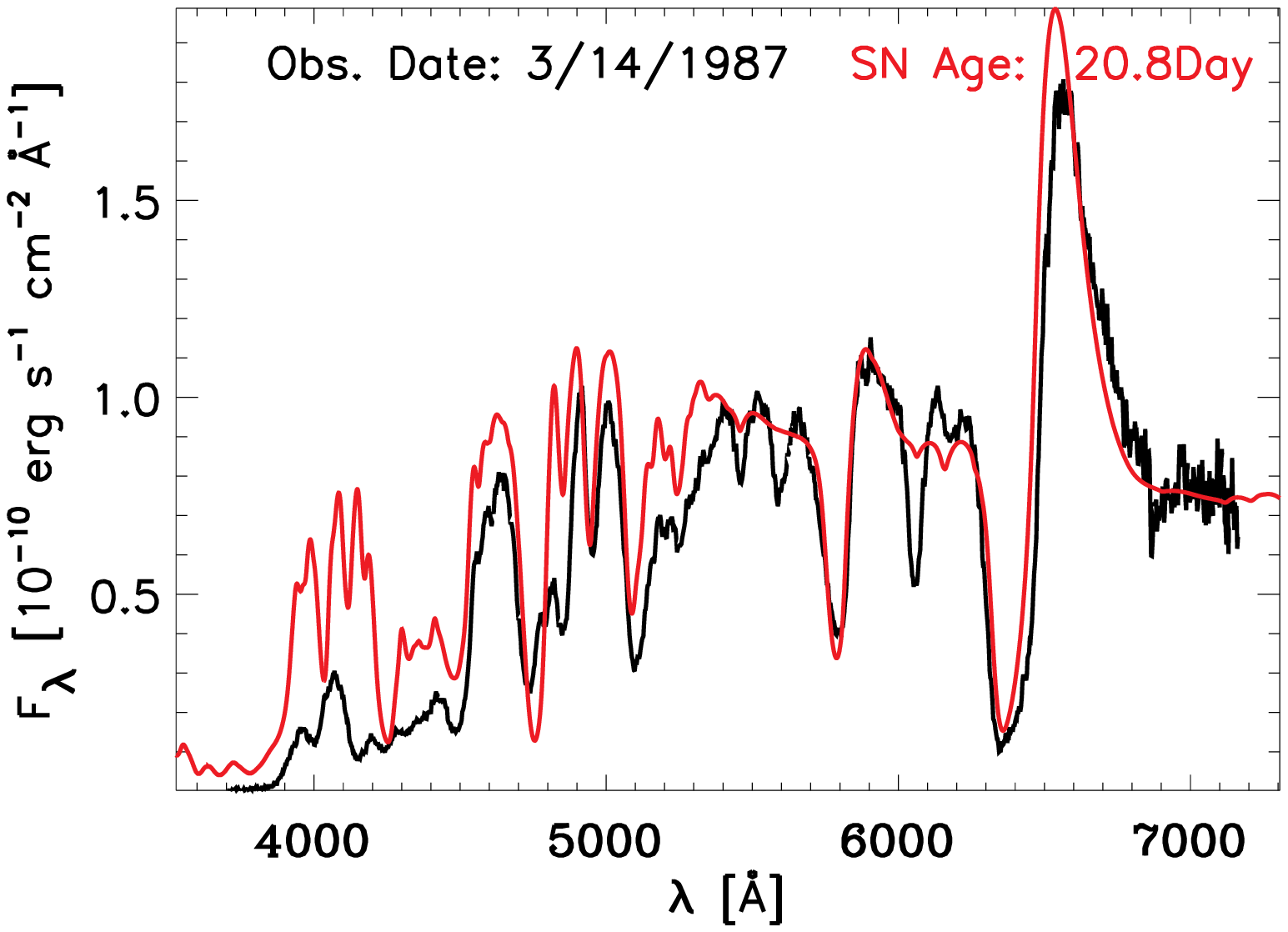,width=8.5cm}
\caption{Same as Fig.~\ref{fig_comp_spec1}, but showing dates between the 9th and the 14th of March 1987. Some of the prominent discrepancies are caused by missing lines of Ba and Sc (e.g.,Sc\,{\sc ii} $\lambda\lambda$ 5527, 5658, Ba\,{\sc ii} $\lambda$6142) which were  not included in the models. Sc\,{\sc ii} and Ba\,{\sc ii} were  previously identified in SN1987A by \citet{Wil87_1987A} (see also \citet{MBB02_87A} ).
\label{fig_comp_spec3}
}
\end{figure*}

   These findings have been challenged by \citet{de_etal_09} but their tests on SN1987A out to 20 days were performed
   using ejecta conditions that produce a strong Balmer continuum at late times  (their figures 8, 9) which is not seen in IUE observations
   (presented by  \citealt{pun_etal_95}, modeled in this paper, and shown, e.g., in Fig.~\ref{fig_spect_montage}--\ref{fig_comp_spec3}).
   This suggests that in their simulations the photospheric layers at such times are
   well above the recombination temperature for hydrogen and thus copiously ionized. In their model, ionization
   is likely controlled by Lyman/Balmer continuum photons, and since the photospheric layers are not recombining,
   the ionization freeze-out we predict cannot take place in their investigation. In contrast, time-dependent effects become important
   for determining the ionization structure of the ejecta when there are no Lyman/Balmer continuum photons
   to contribute, which inevitably occurs when hydrogen recombines, at around 7000\,K.

  After the 1st of March 1987 (Fig.~\ref{fig_comp_spec2}--\ref{fig_comp_spec3}),
  the optical flux becomes better fitted while we overestimate the very-faint UV flux.
  {The Na\,{\sc i}\,D doublet appears in our models on March 3rd, about a day later than in observations.
  It strengthens in time but matches the observations well after a few more days.}
  The fit to the H$\alpha$ line profile is quite spectacular. Recall that we do not treat Scandium and Barium
  and thus do not include their optical features.

\section{Discussion and conclusions}
\label{sect_ccl}

   We have presented a new approach for radiative-transfer modeling of SN ejecta, retaining the
key assets of our former approaches. By treating non-LTE explicitly and incorporating time-dependent terms in the statistical-equilibrium, energy, and moments of the radiative transfer equations, we improve the physical consistency of our computations. In particular, the simulations are performed on the entire ejecta, starting from a given hydrodynamical input at a given post-explosion time, making this an initial-value problem, while spatial boundaries are zero-flux
at the inner edge and free-streaming at the outer edge. As diffusion at the inner boundary does not need to
hold we can evolve the ejecta from its photospheric to its nebular phase.
Importantly, integrating our synthetic spectra over filter-transmission functions,
we compute non-LTE light-curves that account for the explicit role of line-blanketing,
without recourse to the standard expansion opacities and the assumption of LTE for the gas,
as is typically used (see, e.g., \citealt{blinnikov_etal_2000}).

Delaying a presentation of all the technical details of this new approach (Hillier \& Dessart 2010, in preparation),
we illustrate in this paper these modeling improvements with results for SN1987A, using as our starting conditions
the ejecta composition and structure of the hydrodynamical input model ``lm18a7Ad'' of Woosley (priv. comm.).
We evolved the ejecta gas and radiation from 0.27 to 20.8\,d after explosion, corresponding to a spatial expansion
of a factor of 77, and considerable cooling. At the photosphere, the ejecta conditions evolve
from fully-ionized to once-ionized and neutral. During the 21 day time sequence, the photosphere recedes in mass,
and thus in velocity, while its radius increases steadily. After a few days, the photospheric location is primarily
set by the location of the hydrogen ionization front.  Our computed temperature
and electron-density evolution reproduces predictions for a radiation-dominated homologously-expanding
gas, modulo the effect of cooling at the photosphere and heating from radioactive decay in the inner ejecta.
At the last time in the simulation, the photosphere is at 4500\,\kms, and is not influenced, either directly or
indirectly, by energy deposition from radioactive decay occurring below 2000\,\kms\ in our model.
The surface conditions computed here and prior to 20\,d are thus not affected by unstable nuclei.
Since the SN1987A model ejecta we employ are homogeneous above 4500\,\kms, the changes we observe in the spectrum
over this 21-day time span are conditioned by modulations in ionization rather than composition.

Adopting a distance of 50\,kpc and a reddening $E(B-V)=0.15$, we obtain very good agreement between
synthetic and observed light curves in the optical. Our $U$-band synthetic magnitudes are, however,
too bright by a fraction of a magnitude at early times, and this discrepancy slowly grows as the flux in the blue
ebbs.
Our synthetic spectra computed for the epoch 0.3--1\,d cannot be compared to observations, which started one day after explosion.
At such early times, we find an SED with a peak flux in the far-UV, where strong line blanketing occurs due
to 3-4-times ionized metal species (O, Fe, Ni), while the nearly-featureless optical range shows weak lines associated
with H\,{\sc i}, He\,{\sc ii} and He\,{\sc i}. After day 1 and until the 14th of March 1987, we compare our predictions to observations.
In general, the flux agreement in the optical is at the 10\% level (in an absolute sense, i.e., without any normalization), while
in the UV range it is within 30-50\%. In contrast with \citet{EK89_87A}, we reproduce the strength and morphology of
He\,{\sc i} lines at early times, under a photospheric composition that is compatible with a BSG progenitor star.
Throughout the time span of our sequence, we reproduce well the multi-band light curve,
the overall SED, and in particular the strength and shape of H$\alpha$. Since radioactive decay
does not influence the computed photospheric conditions in any way at such times, and given the
relatively good fit to observations (in particular if we focus on the bulk of the emergent luminosity,
which falls at optical wavelengths), it does not play a pivotal role for understanding the radiative properties
of SN1987A up to 20\,d, in contradiction with the proposition of \citet{MBB01_mixing,MBB02_87A}.
In contrast, our good reproduction of the Balmer line profiles at all times, and in particular when there are no longer any
photons in the Lyman and Balmer continua, gives strong support to our proper modeling of the time-dependent
ionization structure of the SN1987A ejecta, an important issue emphasized by \citet{UC05_time_dep} and \citet{DH08_time}.
\citet{de_etal_09} suggest that such time-dependent effects are much weaker, but their presentation for the recombination
phase of SN1987A is unconvincing since they use models that are fully-ionized and UV bright (in contradiction with observations).
Furthermore, despite the one-dimensionality of our approach, we achieve very competitive fits to observations. One may
argue the unresolved problems we encounter are caused by those neglected multi-dimensional effects, but we surmise this departure from
sphericity would also alter line profile shapes etc., which are fairly well reproduced (see, e.g., H$\alpha$).

We predict the fastest spectral evolution in SN1987A occurred prior to day 1. From 0.3\,d to 1\,d, most of the flux was emitted in the range
300-3000\AA\ which could have been captured by the IUE satellite. Prior to 0.27\,d, the photosphere would have shone
at shorter wavelengths, but the UV range would have revealed the long-wavelength tail of that SED,
testifying for the hotter conditions at earlier post-breakout times.
This suggests that obtaining multiple observations prior to day 1, across as broad a wavelength range as possible,
would have captured the phase of fastest evolution of SN1987A. While the UV shows the largest changes,
multiple spectra, rather than a unique spectrum, of SN1987A during both the first and the second night of its evolution would have been
valuable.\footnote{Covering the full optical range, from 3500\AA\ to 1$\mu$m allows us to gather information on Ca and CNO lines
in the red part of the spectrum, as well as constrain the reddening with the blue part.}
This applies to some extent to Type II-P SN as well, although their evolution is slower. In Type I SNe, the
SED likely evolves too fast after breakout to allow us to capture it,  but then,
the first few days after explosion would be very useful for constraining the surface composition of the progenitor.
Overall, observing frequently at early times would provide important constraints that are generally missing
in current observations.

The work presented here is an essential benchmarking of our code
since SN1987A is one of the best understood and best observed supernova. As it places numerous and tight observational
constraints, it offers {\it a valuable alternative, although not a replacement,}
to benchmarking against other codes.
The combined approach of using non-LTE, time dependence, and line blanketing
seems very promising. The future from here is to systematize such investigations by evolving a wide range of
ejecta for all supernova types and compare results with multi-epoch multi-wavelength observations. The aim is
then to use such models and observations to make quantitative inferences on the progenitor properties, pre-SN
star evolution, and the explosion mechanism. In the context of SN1987A, we now need to repeat the present
exploration with a variety of progenitor stars and explosion properties, in order to delineate the systematics
associated with core-collapse SN explosions of BSG stars. In parallel, it would be valuable to gather additional
and high quality observational data for similar Type II-peculiar events like SN1987A.

\section*{Acknowledgments}

We thank Stan Woosley for providing the input hydrodynamical model that was used as initial
conditions for our radiative-transfer calculations. We would also like
to thank Bob Kurucz, the people involved with the NIST Atomic Data Base, and the participants of the Opacity
and Iron Projects (with special thanks to Keith Butler, Sultana Nahar, Anil
Pradhan, and Peter Storey) for computing and  supplying atomic data to the astrophysical
community ---  without their work these calculations would not be possible.
LD acknowledges financial support from the European Community through an
International Re-integration Grant, under grant number PIRG04-GA-2008-239184.
LD also acknowledges financial support in the early stages of this work from the Scientific Discovery
through Advanced Computing (SciDAC) program of the DOE, under grant number DOE-FC02-06ER41452.
DJH acknowledges support from STScI grant HST-AR-11756.01.A.


\begin{thebibliography}{65}
\expandafter\ifx\csname natexlab\endcsname\relax\def\natexlab#1{#1}\fi

\bibitem[{A~collaboration of researchers~at Auburn~University \& the
  University~of Strathclyde(????)}]{ARS_auto}
A~collaboration of researchers~at Auburn~University, R.~C. \& the University~of
  Strathclyde. ????, Data Collections: Auburn-Rollins-Strathclyde Dielectronic
  Recombination Data

\bibitem[{{Arnett} {et~al.}(1989){Arnett}, {Bahcall}, {Kirshner}, \&
  {Woosley}}]{ABK89_rev}
{Arnett}, W.~D., {Bahcall}, J.~N., {Kirshner}, R.~P., \& {Woosley}, S.~E. 1989,
  \araa, 27, 629

\bibitem[{{Becker} \& {Butler}(1992)}]{BB92_FeV}
{Becker}, S.~R. \& {Butler}, K. 1992, \aap, 265, 647

\bibitem[{{Becker} \& {Butler}(1995{\natexlab{a}})}]{BB95_FeVI}
---. 1995{\natexlab{a}}, \aap, 294, 215

\bibitem[{{Becker} \& {Butler}(1995{\natexlab{b}})}]{BB95_FeIV}
---. 1995{\natexlab{b}}, \aap, 301, 187

\bibitem[{{Berrington} {et~al.}(1985){Berrington}, {Burke}, {Dufton}, \&
  {Kingston}}]{BBD85_col}
{Berrington}, K.~A., {Burke}, P.~G., {Dufton}, P.~L., \& {Kingston}, A.~E.
  1985, Atomic Data and Nuclear Data Tables, 33, 195

\bibitem[{{Blinnikov} {et~al.}(2000){Blinnikov}, {Lundqvist}, {Bartunov},
  {Nomoto}, \& {Iwamoto}}]{blinnikov_etal_2000}
{Blinnikov}, S., {Lundqvist}, P., {Bartunov}, O., {Nomoto}, K., \& {Iwamoto},
  K. 2000, \apj, 532, 1132

\bibitem[{{Brown} {et~al.}(2007){Brown}, {Dessart}, {Holland}, {Immler},
  {Landsman}, {Blondin}, {Blustin}, {Breeveld}, {Dewangan}, {Gehrels},
  {Hutchins}, {Kirshner}, {Mason}, {Mazzali}, {Milne}, {Modjaz}, \&
  {Roming}}]{BDH07_SN2005cs}
{Brown}, P.~J., {Dessart}, L., {Holland}, S.~T., {Immler}, S., {Landsman}, W.,
  {Blondin}, S., {Blustin}, A.~J., {Breeveld}, A., {Dewangan}, G.~C.,
  {Gehrels}, N., {Hutchins}, R.~B., {Kirshner}, R.~P., {Mason}, K.~O.,
  {Mazzali}, P.~A., {Milne}, P., {Modjaz}, M., \& {Roming}, P.~W.~A. 2007,
  \apj, 659, 1488

\bibitem[{{Busche} \& {Hillier}(2005)}]{BH05_2D}
{Busche}, J.~R. \& {Hillier}, D.~J. 2005, \aj, 129, 454

\bibitem[{{Cardelli} {et~al.}(1989){Cardelli}, {Clayton}, \&
  {Mathis}}]{cardelli_etal_89}
{Cardelli}, J.~A., {Clayton}, G.~C., \& {Mathis}, J.~S. 1989, \apj, 345, 245

\bibitem[{{Clement} {et~al.}(2008){Clement}, {Xu}, \& {Muzzin}}]{LMC_dist3}
{Clement}, C.~M., {Xu}, X., \& {Muzzin}, A.~V. 2008, \aj, 135, 83

\bibitem[{{Dall'Ora} {et~al.}(2004){Dall'Ora}, {Storm}, {Bono}, {Ripepi},
  {Monelli}, {Testa}, {Andreuzzi}, {Buonanno}, {Caputo}, {Castellani}, {Corsi},
  {Marconi}, {Marconi}, {Pulone}, \& {Stetson}}]{LMC_dist2}
{Dall'Ora}, M., {Storm}, J., {Bono}, G., {Ripepi}, V., {Monelli}, M., {Testa},
  V., {Andreuzzi}, G., {Buonanno}, R., {Caputo}, F., {Castellani}, V., {Corsi},
  C.~E., {Marconi}, G., {Marconi}, M., {Pulone}, L., \& {Stetson}, P.~B. 2004,
  \apj, 610, 269

\bibitem[{{De} {et~al.}(2009){De}, {Baron}, \& {Hauschildt}}]{de_etal_09}
{De}, S., {Baron}, E., \& {Hauschildt}, P.~H. 2009, \mnras, 1740

\bibitem[{{Dessart} {et~al.}(2008){Dessart}, {Blondin}, {Brown}, {Hicken},
  {Hillier}, {Holland}, {Immler}, {Kirshner}, {Milne}, {Modjaz}, \&
  {Roming}}]{DBB08_SN2005cs}
{Dessart}, L., {Blondin}, S., {Brown}, P.~J., {Hicken}, M., {Hillier}, D.~J.,
  {Holland}, S.~T., {Immler}, S., {Kirshner}, R.~P., {Milne}, P., {Modjaz}, M.,
  \& {Roming}, P.~W.~A. 2008, \apj, 675, 644

\bibitem[{{Dessart} \& {Hillier}(2005{\natexlab{a}})}]{DH05_epm}
{Dessart}, L. \& {Hillier}, D.~J. 2005{\natexlab{a}}, \aap, 439, 671

\bibitem[{{Dessart} \& {Hillier}(2005{\natexlab{b}})}]{DH05_qs_SN}
---. 2005{\natexlab{b}}, \aap, 437, 667

\bibitem[{{Dessart} \& {Hillier}(2006)}]{DH06_SN1999em}
---. 2006, \aap, 447, 691

\bibitem[{{Dessart} \& {Hillier}(2008)}]{DH08_time}
---. 2008, \mnras, 383, 57

\bibitem[{{Dessart} \& {Hillier}(2009)}]{DH09_boulder}
{Dessart}, L. \& {Hillier}, D.~J. 2009, in {Recent directions in astrophysical
  quantitative spectroscopy and radiation hydrodynamics}, American Institute of
  Physics

\bibitem[{{Dessart} {et~al.}(2009){Dessart}, {Livne}, \& {Waldman}}]{DEW_09}
{Dessart}, L., {Livne}, E., \& {Waldman}, R. 2009, ArXiv e-prints

\bibitem[{{Eastman} \& {Kirshner}(1989)}]{EK89_87A}
{Eastman}, R.~G. \& {Kirshner}, R.~P. 1989, \apj, 347, 771

\bibitem[{{Eastman} {et~al.}(1996){Eastman}, {Schmidt}, \& {Kirshner}}]{E96}
{Eastman}, R.~G., {Schmidt}, B.~P., \& {Kirshner}, R. 1996, \apj, 466, 911

\bibitem[{{Hamuy} \& {Pinto}(2002)}]{HP_92}
{Hamuy}, M. \& {Pinto}, P.~A. 2002, \apjl, 566, L63

\bibitem[{{Herald} {et~al.}(1987){Herald}, {McNaught}, {Morel}, {Madore},
  {Shelton}, {Claria}, {Davidsen}, {Kimble}, {Gregg}, \&
  {Kirshner}}]{87A_iauc2}
{Herald}, D., {McNaught}, R.~H., {Morel}, M., {Madore}, B., {Shelton}, I.,
  {Claria}, J., {Davidsen}, A., {Kimble}, R., {Gregg}, M., \& {Kirshner}, R.
  1987, \iaucirc, 4317, 1

\bibitem[{{Hillier} \& {Miller}(1998)}]{HM98_lb}
{Hillier}, D.~J. \& {Miller}, D.~L. 1998, \apj, 496, 407

\bibitem[{{H{\"o}flich}(1987)}]{hoeflich_87}
{H{\"o}flich}, P. 1987, Mitteilungen der Astronomischen Gesellschaft Hamburg,
  70, 192

\bibitem[{{H{\"o}flich}(1988)}]{hoeflich_88}
---. 1988, Proceedings of the Astronomical Society of Australia, 7, 434

\bibitem[{{Hummer} {et~al.}(1993){Hummer}, {Berrington}, {Eissner}, {Pradhan},
  {Saraph}, \& {Tully}}]{HBE93_IP}
{Hummer}, D.~G., {Berrington}, K.~A., {Eissner}, W., {Pradhan}, A.~K.,
  {Saraph}, H.~E., \& {Tully}, J.~A. 1993, \aap, 279, 298

\bibitem[{{Kingdon} \& {Ferland}(1996)}]{KF96_chg}
{Kingdon}, J.~B. \& {Ferland}, G.~J. 1996, \apjs, 106, 205

\bibitem[{{Kirshner} \& {Kwan}(1974)}]{KK74_EPM}
{Kirshner}, R.~P. \& {Kwan}, J. 1974, \apj, 193, 27

\bibitem[{{Kunkel} {et~al.}(1987){Kunkel}, {Madore}, {Shelton}, {Duhalde},
  {Bateson}, {Jones}, {Moreno}, {Walker}, {Garradd}, {Warner}, \&
  {Menzies}}]{87A_iauc1}
{Kunkel}, W., {Madore}, B., {Shelton}, I., {Duhalde}, O., {Bateson}, F.~M.,
  {Jones}, A., {Moreno}, B., {Walker}, S., {Garradd}, G., {Warner}, B., \&
  {Menzies}, J. 1987, \iaucirc, 4316, 1

\bibitem[{{Kurucz}(2009)}]{Kur09_ATD}
{Kurucz}, R.~L. 2009, in American Institute of Physics Conference Series, Vol.
  1171, American Institute of Physics Conference Series, ed. {I.~Hubeny,
  J.~M.~Stone, K.~MacGregor, \& K.~Werner}, 43--51

\bibitem[{{Kurucz}(2010)}]{Kur_web}
{Kurucz}, R.~L. 2010

\bibitem[{{Landolt}(1992)}]{landolt_92}
{Landolt}, A.~U. 1992, \aj, 104, 340

\bibitem[{{Lennon} \& {Burke}(1994)}]{LB94_N2}
{Lennon}, D.~J. \& {Burke}, V.~M. 1994, \aaps, 103, 273

\bibitem[{{Lennon} {et~al.}(1985){Lennon}, {Dufton}, {Hibbert}, \&
  {Kingston}}]{LDH85_CII_col}
{Lennon}, D.~J., {Dufton}, P.~L., {Hibbert}, A., \& {Kingston}, A.~E. 1985,
  \apj, 294, 200

\bibitem[{{Leonard} {et~al.}(2006){Leonard}, {Filippenko}, {Ganeshalingam},
  {Serduke}, {Li}, {Swift}, {Gal-Yam}, {Foley}, {Fox}, {Park}, {Hoffman}, \&
  {Wong}}]{leonard_etal_06}
{Leonard}, D.~C., {Filippenko}, A.~V., {Ganeshalingam}, M., {Serduke},
  F.~J.~D., {Li}, W., {Swift}, B.~J., {Gal-Yam}, A., {Foley}, R.~J., {Fox},
  D.~B., {Park}, S., {Hoffman}, J.~L., \& {Wong}, D.~S. 2006, \nat, 440, 505

\bibitem[{{Lucy}(1987)}]{lucy_87}
{Lucy}, L.~B. 1987, \aap, 182, L31

\bibitem[{{Mazzali} {et~al.}(1992){Mazzali}, {Lucy}, \&
  {Butler}}]{mazzali_etal_92}
{Mazzali}, P.~A., {Lucy}, L.~B., \& {Butler}, K. 1992, \aap, 258, 399

\bibitem[{{Mendoza}(1983)}]{Men83_col}
{Mendoza}, C. 1983, in IAU Symposium, Vol. 103, Planetary Nebulae, ed.
  {D.~R.~Flower}, 143--172

\bibitem[{{Mihalas} \& {Mihalas}(1984)}]{MM84_RH}
{Mihalas}, D. \& {Mihalas}, B.~W. 1984, {Foundations of radiation
  hydrodynamics}, ed. {Mihalas, D.~\& Mihalas, B.~W.}

\bibitem[{{Mitchell} {et~al.}(2002){Mitchell}, {Baron}, {Branch}, {Hauschildt},
  {Nugent}, {Lundqvist}, {Blinnikov}, \& {Pun}}]{MBB02_87A}
{Mitchell}, R.~C., {Baron}, E., {Branch}, D., {Hauschildt}, P.~H., {Nugent},
  P.~E., {Lundqvist}, P., {Blinnikov}, S., \& {Pun}, C.~S.~J. 2002, \apj, 574,
  293

\bibitem[{{Mitchell} {et~al.}(2001){Mitchell}, {Baron}, {Branch}, {Lundqvist},
  {Blinnikov}, {Hauschildt}, \& {Pun}}]{MBB01_mixing}
{Mitchell}, R.~C., {Baron}, E., {Branch}, D., {Lundqvist}, P., {Blinnikov}, S.,
  {Hauschildt}, P.~H., \& {Pun}, C.~S.~J. 2001, \apj, 556, 979

\bibitem[{{Nahar}(1995)}]{Nahar95_FeII}
{Nahar}, S.~N. 1995, \aap, 293, 967

\bibitem[{{Nahar}(2010)}]{Nahar_OSU}
---. 2010, NORAD-Atomic-Data

\bibitem[{{Nussbaumer} \& {Storey}(1983)}]{NS83_LTDR}
{Nussbaumer}, H. \& {Storey}, P.~J. 1983, \aap, 126, 75

\bibitem[{{Nussbaumer} \& {Storey}(1984)}]{NS84_CNO_LTDR}
---. 1984, \aaps, 56, 293

\bibitem[{{Phillips} {et~al.}(1988){Phillips}, {Heathcote}, {Hamuy}, \&
  {Navarrete}}]{phillips_etal_88}
{Phillips}, M.~M., {Heathcote}, S.~R., {Hamuy}, M., \& {Navarrete}, M. 1988,
  \aj, 95, 1087

\bibitem[{{Pietrzy{\'n}ski} {et~al.}(2009){Pietrzy{\'n}ski}, {Thompson},
  {Graczyk}, {Gieren}, {Udalski}, {Szewczyk}, {Minniti}, {Ko{\l}aczkowski},
  {Bresolin}, \& {Kudritzki}}]{LMC_dist4}
{Pietrzy{\'n}ski}, G., {Thompson}, I.~B., {Graczyk}, D., {Gieren}, W.,
  {Udalski}, A., {Szewczyk}, O., {Minniti}, D., {Ko{\l}aczkowski}, Z.,
  {Bresolin}, F., \& {Kudritzki}, R. 2009, \apj, 697, 862

\bibitem[{{Pinto} \& {Eastman}(2000)}]{PE00_Ia_anal}
{Pinto}, P.~A. \& {Eastman}, R.~G. 2000, \apj, 530, 744

\bibitem[{{Pun} {et~al.}(1995){Pun}, {Kirshner}, {Sonneborn}, {Challis},
  {Nassiopoulos}, {Arquilla}, {Crenshaw}, {Shrader}, {Teays}, {Cassatella},
  {Gilmozzi}, {Talavera}, {Wamsteker}, {Fransson}, \& {Panagia}}]{pun_etal_95}
{Pun}, C.~S.~J., {Kirshner}, R.~P., {Sonneborn}, G., {Challis}, P.,
  {Nassiopoulos}, G., {Arquilla}, R., {Crenshaw}, D.~M., {Shrader}, C.,
  {Teays}, T., {Cassatella}, A., {Gilmozzi}, R., {Talavera}, A., {Wamsteker},
  W., {Fransson}, C., \& {Panagia}, N. 1995, \apjs, 99, 223

\bibitem[{{Ribas} {et~al.}(2002){Ribas}, {Fitzpatrick}, {Maloney}, {Guinan}, \&
  {Udalski}}]{LMC_dist1}
{Ribas}, I., {Fitzpatrick}, E.~L., {Maloney}, F.~P., {Guinan}, E.~F., \&
  {Udalski}, A. 2002, \apj, 574, 771

\bibitem[{{Schmutz} {et~al.}(1990){Schmutz}, {Abbott}, {Russell}, {Hamann}, \&
  {Wessolowski}}]{SAR90_87A}
{Schmutz}, W., {Abbott}, D.~C., {Russell}, R.~S., {Hamann}, W.-R., \&
  {Wessolowski}, U. 1990, \apj, 355, 255

\bibitem[{{Seaton}(1987)}]{Sea87_OP}
{Seaton}, M.~J. 1987, Journal of Physics B Atomic Molecular Physics, 20, 6363

\bibitem[{{Shine} \& {Linsky}(1974)}]{SL74}
{Shine}, R.~A. \& {Linsky}, J.~L. 1974, \solphys, 39, 49

\bibitem[{{Smartt}(2009)}]{smartt_09}
{Smartt}, S.~J. 2009, ArXiv e-prints

\bibitem[{{Tayal}(1997{\natexlab{a}})}]{T97_SII_col}
{Tayal}, S.~S. 1997{\natexlab{a}}, \apjs, 111, 459

\bibitem[{{Tayal}(1997{\natexlab{b}})}]{T97_SIII_col}
---. 1997{\natexlab{b}}, \apj, 481, 550

\bibitem[{{Utrobin} \& {Chugai}(2005)}]{UC05_time_dep}
{Utrobin}, V.~P. \& {Chugai}, N.~N. 2005, \aap, 441, 271

\bibitem[{{Weaver} {et~al.}(1978){Weaver}, {Zimmerman}, \&
  {Woosley}}]{weaver_etal_78}
{Weaver}, T.~A., {Zimmerman}, G.~B., \& {Woosley}, S.~E. 1978, \apj, 225, 1021

\bibitem[{{Williams}(1987)}]{Wil87_1987A}
{Williams}, R.~E. 1987, \apjl, 320, L117

\bibitem[{{Woosley} \& {Weaver}(1995)}]{woosley_weaver_95}
{Woosley}, S.~E. \& {Weaver}, T.~A. 1995, \apjs, 101, 181

\bibitem[{{Zhang} \& {Pradhan}(1995{\natexlab{a}})}]{ZP95_FeII_col}
{Zhang}, H.~L. \& {Pradhan}, A.~K. 1995{\natexlab{a}}, \aap, 293, 953

\bibitem[{{Zhang} \& {Pradhan}(1995{\natexlab{b}})}]{ZP95_FeIII_col}
---. 1995{\natexlab{b}}, Journal of Physics B Atomic Molecular Physics, 28,
  3403

\bibitem[{{Zhang} \& {Pradhan}(1997)}]{ZP97_FeIV_col}
---. 1997, \aaps, 126, 373

\end{thebibliography}

\end{document}